\documentclass[10pt]{article}
\usepackage{multicol}
\usepackage{indentfirst}
\usepackage{latexsym}
\usepackage{bm}
\usepackage{graphicx}
\usepackage{subfigure}
\usepackage[top=1in, bottom=1in, left=1.25in, right=1.25in]{geometry}

\usepackage{epsfig,dsfont,amssymb,amsmath,amsthm,amsfonts,amsbsy,mathrsfs}
\usepackage{psfrag}
\usepackage{subfigure}
\usepackage{epstopdf}
\usepackage{hyperref}
\usepackage{float}
\usepackage{array}
\usepackage{color}
\usepackage{booktabs}
\usepackage{enumerate}
\usepackage{authblk}
\theoremstyle{plain}
\newtheorem{theorem}{Theorem}[section]

\theoremstyle{definition}
\newtheorem{definition}{Definition}[section]

\newtheorem*{problem*}{Problem}

\theoremstyle{remark}
\newtheorem{remark}{Remark}[section]

\newtheorem*{solution*}{Solution}
\newtheorem*{remark*}{Remark}

\DeclareMathOperator*{\argmin}{arg\,min}

\numberwithin{equation}{section}

\begin{document}

\title{Directed migration of microscale swimmers by an array of shaped obstacles: modeling and shape optimization}

\author[1]{Jiajun Tong\thanks{E-mail: jiajun@cims.nyu.edu}}
\author[1,2]{Michael J.\;Shelley}
\affil[1]{\small Applied Mathematics Laboratory, Courant Institute, New York University, USA}
\affil[2]{\small Flatiron Institute, Simons Foundation, USA}

\date{}
\maketitle

\begin{abstract}
Achieving macroscopic directed migration of microscale swimmers in a fluid is an important step towards utilizing their autonomous motion.
It has been experimentally shown that directed motion can be induced, without any external fields, by certain geometrically asymmetric obstacles due to interaction between their boundaries and the swimmers.
In this paper, we propose a kinetic-type model to study swimming and directional migration of microscale bimetallic rods in a periodic array of posts with non-circular cross-sections.
Both rod position and orientation are taken into account; rod trapping and release on the post boundaries are modeled 
by empirically characterizing curvature and orientational dependence of the boundary absorption and desorption.
Intensity of the directed rod migration, which we call the normalized net flux, is then defined and computed given the geometry of the post array.
We numerically study the effect of post spacings on the flux; we also apply shape optimization to find better post shapes that can induce stronger flux.
Inspired by preliminary numerical results on two candidate posts, we perform an approximate analysis on a simplified model to show the key geometric features a good post should have.
Based on that, three new candidate shapes are proposed which give rise to large fluxes.
This approach provides an effective tool and guidance for experimentally designing new devices that induce strong directed migration of microscale swimmers.
\end{abstract}

\section{Introduction}\label{section: introduction}
Microscale swimmers, such as bacteria or chemically active colloids, move autonomously in a fluid by converting energy in the local environment into mechanical work \cite{elgeti2015physics, ebbens2010pursuit}.
Possible applications of synthetic microswimmers include drug delivery \cite{patra2013intelligent, balasubramanian2011micromachine}, cargo transport \cite{sundararajan2008catalytic}, and environmental remediation \cite{soler2013self}.
One important task in manipulating microswimmers is to achieve their directed macroscopic motion, as opposed to their long-time isotropic motion, which results from a combination of ballistic swimming and angular diffusion \cite{howse2007self, palacci2010sedimentation}.
Directed migration can be easily induced by externally imposed fields, such as chemical gradient \cite{berg1972chemotaxis, mitchell2006bacterial} or electromagnetic fields \cite{anderson1989colloid, ghosh2009controlled, volpe2011microswimmers, tierno2008magnetically}.
A different approach is to place obstacles in the environment.
It has been demonstrated that obstacles can dramatically change the motion of microswimmers.
Due to hydrodynamic \cite{spagnolie2015geometric} or steric \cite{kantsler2013ciliary} interactions of swimmers with obstacles, swimmers can aggregate \cite{li2011accumulation, li2009accumulation, berke2008hydrodynamic}, slide \cite{volpe2011microswimmers, spagnolie2015geometric, takagi2014hydrodynamic}, hover \cite{uspal2015self}, or even reverse swimming direction \cite{cisneros2006reversal}.
There is a developing body of works investigating boundaries of obstacles guiding microswimmers, using flat walls \cite{volpe2011microswimmers}, v-shaped funnels \cite{galajda2007wall, lambert2010collective}, spherical obstacles \cite{spagnolie2015geometric, takagi2014hydrodynamic, simmchen2016topographical}, or teardrop-shaped posts \cite{wykes2017guiding}.

\begin{figure}
\centering
\includegraphics[height = 0.35\textwidth]{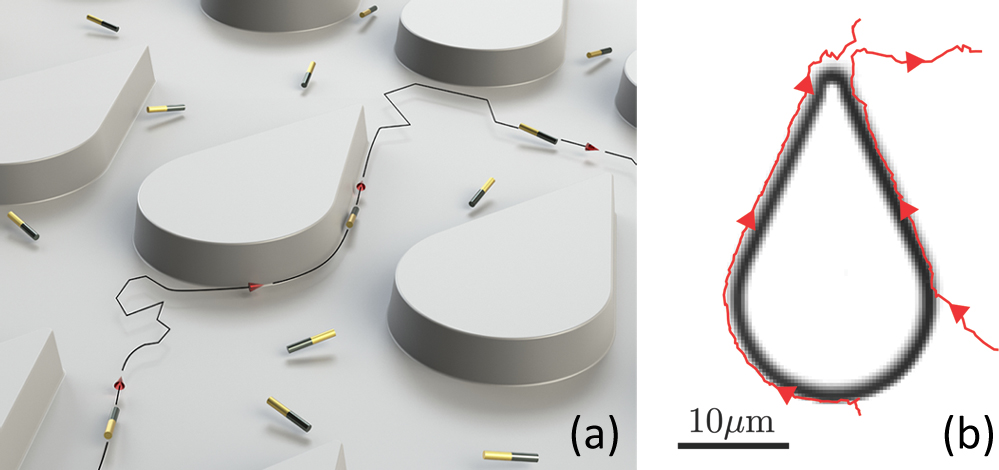}
\caption{(a) Sketch of bimetallic microscale Au-Pt rods swimming and interacting with an array of teardrop-shaped posts. These rods swim, due to self-electrophoresis, primarily along the bottom or the posts.
They move along their axes with the Pt-end leading; their positions and orientations are subject to random fluctuations.
After encountering a post, swimming rods tend to travel along the post boundary and preferentially depart from its tip. (b) Experimental observation of two typical trajectories of swimming rods when interacting with a teardrop shaped post. The pictures are adapted from our recent experimental paper \cite{wykes2017guiding}.}
\label{figure: Au-Pt rod swimming from our experiment paper}
\end{figure}

In a series of recent works, microscale bimetallic segmented rods composed of gold and platinum (Au-Pt) have been experimentally studied as a prototype of artificial microswimmers \cite{takagi2014hydrodynamic, paxton2004catalytic}.
These rods, typically $2\,\mu$m in length and $300$ nm in diameter, move autonomously in aqueous solutions of hydrogen peroxide ($\mathrm{H_2O_2}$), with the Pt end leading, due to self-electrophoresis which generates a slip flow along the rod surface \cite{moran2011electrokinetic, wang2006bipolar}.
They move with a constant speed along their axes while their positions and orientations are subject to random fluctuations.
As they are much denser than water, the Au-Pt rods swim primarily along the microscope coverslip or the obstacles.
It is demonstrated in our paper \cite{takagi2014hydrodynamic} that, these swimming rods can get captured by solid spheres resting on a horizontal plane, and orbit closely around them with little change in their speed, until they are released due to angular diffusion.
An uneven spatial distribution of the rods near the spheres and statistics of trapping time are obtained.
In a more recent study \cite{wykes2017guiding}, we show that when the rods swim in a periodic array of teardrop-shaped posts, they interact with the vertical walls of the posts in a similar way; yet the rods preferentially leave the posts at the post's sharp tips due to large boundary curvature there, rendering a statistically biased swimming over long times.
See Figure\;\ref{figure: Au-Pt rod swimming from our experiment paper} for a sketch of the rods swimming and interacting with an array of teardrop-shaped posts, as well as a picture from the experiment showing typical motion of rods when they encounter a post \cite{wykes2017guiding}.
It has been experimentally confirmed that the rods are most likely to migrate through the array in the direction pointed by the tips of the teardrop-shaped posts \cite{wykes2017guiding}.

In this paper, we shall present a kinetic-type model of Au-Pt rods swimming in a periodic array of posts with non-circular cross-sections, such as teardrop-shaped posts, and thus generating directed migration over long time.
Position and orientation of the rods are both taken into account, as well as the effect of thermal fluctuation.
Trapping and release of the rods on the post boundary are modeled via empirically defined rate functions and angular distributions, accounting for curvature and orientational dependence of the boundary absorption and desorption, respectively.
Distributions of the rods in the free-space and on the post boundary are found through numerical simulations; the intensity of the directed migration is then defined and calculated.

The degree of directed migration crucially relies on many features of the array, such as the spacings and shapes of the posts in it.
We study the effect of post spacing by numerical simulation.
We also perform numerical shape optimization to investigate how to choose the shape of posts judiciously so that stronger directed migration of the swimming rods is achieved.
General theory of shape optimization is well-established from the analysis point of view \cite{sokolowski1992introduction,delfour2011shapes,pironneau2012optimal}, while numerous applications can be found in many areas of applied mathematics and physics, such as problems involving swimming \cite{keaveny2013optimization,wilkening2008shape} and fluid motion \cite{walker2010shape,mohammadi2010applied}.
In our study, a mathematical derivation of the shape optimization is presented in the Supplementary Materials, together with an iterative optimization method based on an explicitly preconditioned steepest ascent method.
We apply shape optimization to two candidate post shapes, and observe a significant increase in directed migration.
To better understand the optimization results, we propose a simplified model that well explains the geometric features arising in the optimized shape.
This enables us to empirically determine the key geometric ingredients of designing posts.
We conclude by giving three new post shapes designed on these ingredients, which does give rise to a significantly stronger directed migration than naive choices.

The rest of the paper is organized as follows.
In Section \ref{section: modeling}, we present the model for Au-Pt rods swimming in a periodic rectangular array of posts and define the quantity that measures the intensity of the directed migration.
The numerical method to compute the ensemble distributions of rod positions and orientations as well as the intensity of the directed migration appear in the Supplementary Materials.
In Section \ref{section: shape optimization}, we formulate the optimization problem seeking better designs of the array so that it induces stronger directed migration.
A formal overview of the shape optimization theory, a full derivation of equations involved, and numerical methods for solving these equations and performing shape optimization, are also left to the Supplementary Materials.
Dimensionless parameters and rates are specified in Section \ref{section: model choices}; while numerical results are presented in Section \ref{section: numerical results} to study the effects of post spacings and optimization of post shape.
To understand key geometric features a good post should have, we perform an approximate analysis to a simplified model in Section \ref{section: approximate simplified model}.
Based on that, three new post designs inducing strong directed migration are proposed in Section \ref{section: explorations of other designs}.
We conclude the paper with a brief discussion in Section \ref{section: discussion}.

\section{Theory}
\subsection{Modeling microscale swimming rods in a periodic array of posts}\label{section: modeling}

With the typical swimming pattern of the rod described in Section \ref{section: introduction} and sketched in Figure\;\ref{figure: Au-Pt rod swimming from our experiment paper}, we shall build a kinetic-type model for rods swimming in a periodic array of posts.
The model will be presented in a dimensionless manner; the non-dimensionalization will be left to Section \ref{section: model choices}.
We start from modeling the environment in which the rods are swimming.

\begin{figure}
\centering
\subfigure[]{\includegraphics[height = 0.35\textwidth]{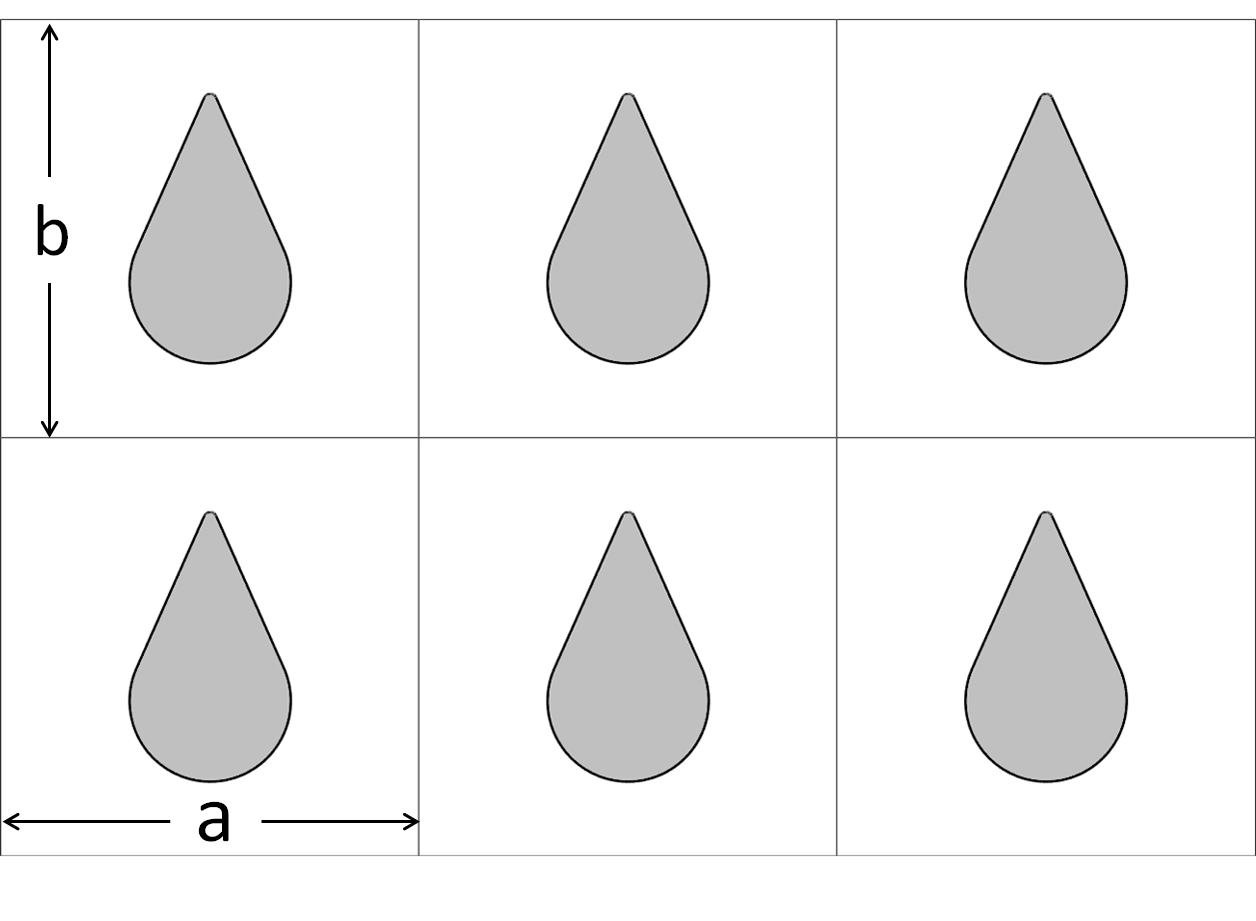}}\qquad
\subfigure[]{\includegraphics[height = 0.35\textwidth]{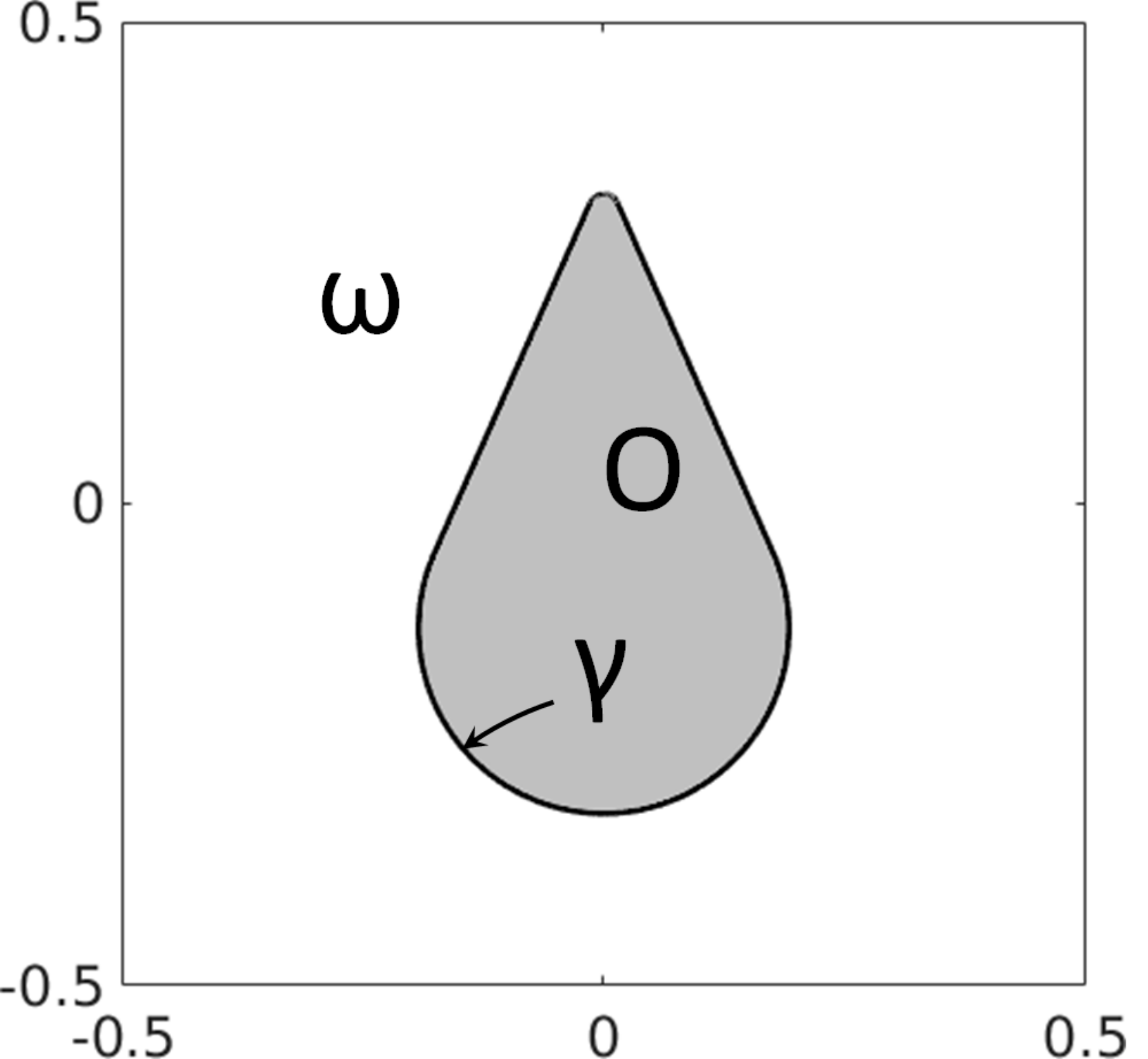}\label{figure: unit cell}}
\caption{(a) A periodic array of teardrop-shaped posts. Its unit cells, separated by solid lines, have width $a$ and height $b$; here $a=b=1$. Domains occupied by the posts are marked as grey. (b) An enlarged view of one unit cell in (a). $O$ denotes the grey domain occupied by the post; $\omega$ denotes the exterior domain filled with fluid; $\gamma$ denotes the boundary of $\omega$ between them.}
\label{figure: geometry of the domain}
\end{figure}

Consider a rectangular periodic array of posts printed on the microscope coverslip.
The posts are solid cylinders which neither fluid nor the rods can penetrate \cite{wykes2017guiding}.
A dilute suspension of the Au-Pt rods in the aqueous hydrogen peroxide solution is then placed on the coverslip, so that the rods can autonomously swim in the complex landscape.
Note that since the rods always swim in a quasi-two-dimensional fashion along the bottom, it suffices to consider the system in two dimensions.
Assume the 2-D unit cell of the periodic array of the posts has dimensionless size $a$ and $b$ in $x_1$- and $x_2$-directions respectively.
We denote the unit cell to be $Y = \left[-\frac{a}{2}, \frac{a}{2}\right]\times \left[-\frac{b}{2}, \frac{b}{2}\right]$.
See Figure \ref{figure: geometry of the domain}.
The following discussion also applies to unit cells in other shapes with minor modification.
For example, for a staggered periodic array, hexagonal unit cells can be more convenient choices than rectangular ones.

In the sequel, for convenience, we will interchangeably use the notion of a post and its cross-section.
Let $O$ denote the domain occupied by the post inside $Y$ and let $\omega = Y\backslash O$ be the domain filled with fluid, in which the rods can swim freely.
The interior and exterior boundaries of $\omega$ are denoted by $\gamma = \partial O$ and $\partial Y$ respectively; see Figure \ref{figure: unit cell}.

To study the statistical behavior of the swimming rods, we look at the spatial and orientational distribution of an ensemble of rods in the bulk and on the boundary.
We assume that the rod concentration is so small that their interactions are negligible.
We also treat the rods as points with orientation but no size.

%
%

\begin{figure}
\centering
\includegraphics[width = 0.85\textwidth]{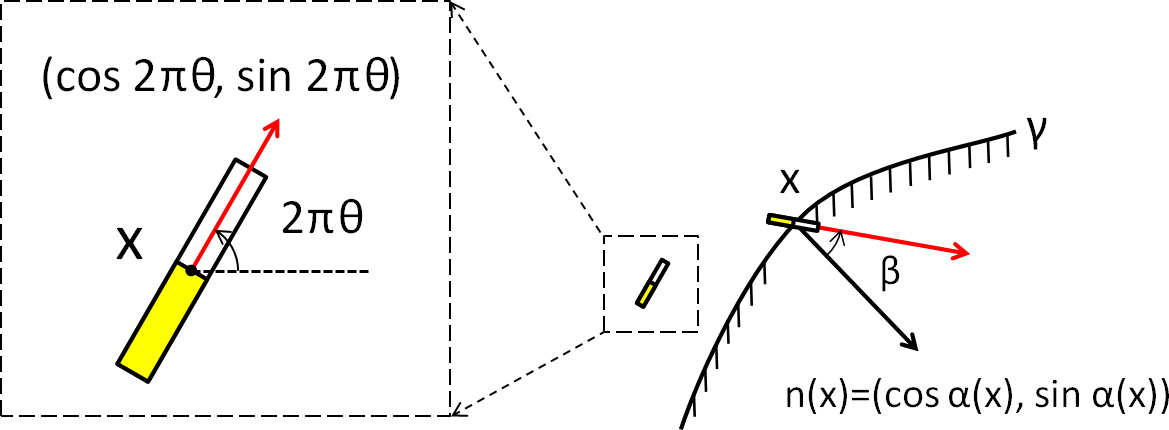}
\caption{We use $x\in \omega$ and $2\pi\theta$, with $\theta \in [0,1)$, to represent position and orientational angle of a swimming Au-Pt rod, respectively. The enlarged picture in the big dashed box shows the state $(x,\theta)$ of the rod in the small dashed box, which is swimming away from the post. The red arrow represents its orientation, given by the direction of its Pt end (white block); the dashed line is the positive $x_1$-axis. Note that rods are modeled as points with orientations but no size, although we have drawn a white-yellow rod in the picture for the sake of clarity.
In the right half of the figure, with abuse of notations, a rod hitting the post boundary $\gamma$ at $x$ with relative angle $\beta$ is shown. The shaded side of $\gamma$ is occupied by the post. The black arrow is the normal vector of $\gamma$ at $x$, denoted by $n(x) = (\cos\alpha(x),\sin\alpha(x))$, while the red arrow again represents the rod orientation. The (signed) angle between them is the relative angle $\beta$.}
\label{figure: rod state and boundary geometry}
\end{figure}

The motion of the rods is modeled as swimming in the bulk $\omega$, and swimming along the boundary $\gamma$.
For a single rod swimming in the bulk, we use $X_t\in \omega$ and $2\pi\Theta_t$ to denote its position and orientation, respectively.
The latter is the angle between the swimming direction of the rod and the positive $x_1$-axis; see Figure \ref{figure: rod state and boundary geometry}.
Here $X_t$ is defined up to a natural periodicity on the exterior boundary of $\omega$, while $\Theta_t\in[0,1)$ defined in the modulus of $1$.
We assume that the rod deterministically swim in its axial direction with velocity $v_0$, while both its position and orientation are subject to random fluctuations.
The stochastic dynamics of $(X_t,\Theta_t)$ is then written as follows
\begin{equation}
\begin{split}
&\;dX_t = v_0(\cos(2\pi\Theta_t),\sin(2\pi\Theta_t))^T+\sqrt{2D_t} dW^{(2)}_t,\\
&\;d\Theta_t = \sqrt{2D_r} dW^{(1)}_t.
\end{split}
\label{equation: SDE for single rod motion}
\end{equation}
Here $D_t$ and $D_r$ are scalar (dimensionless) translational and rotational diffusion coefficients, respectively; they are assumed to be constant throughout the state space $\Omega \triangleq \omega\times [0,1]$.
$W_t^{(1)}$ and $W_t^{(2)}$ are the standard Brownian motions in one and two dimensions, respectively; they are independent with each other.
Let $p(x,\theta,t)\geq 0$ be the distribution of rods in $\Omega$, where $x\in \omega$ and $2\pi\theta\in[0,2\pi]$.
The evolution of $p$ is then governed by the following Fokker-Planck equation associated with \eqref{equation: SDE for single rod motion}
\begin{equation}
\partial_t p(x,\theta, t) = D_t\Delta_x p(x,\theta, t)+D_r\Delta_\theta p(x,\theta, t) - v_0(\cos 2\pi\theta, \sin 2\pi\theta)^T\cdot\nabla_{x} p(x,\theta,t),\quad (x,\theta)\in \Omega.
\label{equation: equation in the bulk non steady state with drift}
\end{equation}
Here $\Delta_x = \partial_{x_1x_1}+\partial_{x_2x_2}$, and $\Delta_\theta = \partial_{\theta\theta}$; $\nabla_x = (\partial_{x_1},\partial_{x_2})^T$ is the gradient operator in spatial components only.
The term $-v_0(\cos 2\pi\theta, \sin 2\pi\theta)^T\cdot\nabla_{x} p$ represents convection in $\Omega$ due to the directed swimming in the axial direction. 

For rods moving along the boundary $\gamma$, we assume they always swim tangentially to $\gamma$; thus only the rod position along $\gamma$ needs to be considered.
We also assume that rods swim at the constant speed $v_0$ \cite{takagi2014hydrodynamic} and can never switch swimming direction before it leaves the boundary.
This assumption is suitable for bimetallic swimming rods, but might not be true for some biological swimmers \cite{cisneros2006reversal}. 
For $x\in \gamma$, let $p_B^+(x,t),p_B^-(x,t)\geq 0$ be the boundary distributions of rods that swim counterclockwise and clockwise, respectively.
The evolution of $p_B^\pm$'s is given by
\begin{equation}
\partial_t p_B^\pm(x,t) = D_t\Delta_\gamma p_B^\pm(x,t) \mp v_0 \partial_\gamma p_B^\pm(x,t)+ F_\mathrm{in}^\pm(x,t) - F_\mathrm{out}^\pm(x,t),\quad x\in\gamma,
\label{equation: equation on the boundary non steady state with drift}
\end{equation}
where $\Delta_\gamma$ and $\partial_\gamma$ are the Laplace operator and the derivative along $\gamma$ with respect to its arclength.
Here $\gamma$ is parameterized counterclockwise.
On the right hand side of \eqref{equation: equation on the boundary non steady state with drift}, $D_t\Delta_\gamma p_B^\pm$ is the spatial diffusion along $\gamma$; for simplicity, we assume the same diffusion coefficient $D_t$ as in the free space.
The term $\mp v_0 \partial_\gamma p_B^\pm(x,t)$ comes from the deterministic swimming along $\gamma$.
$F_\mathrm{in}^\pm(x,t)$ and $F_\mathrm{out}^\pm(x,t)$ are rod absorption and desorption fluxes at $x$ on and off the boundary $\gamma$, respectively.
They 
depend on the local geometry of $\gamma$ and how rods hit or leave the boundary.

\begin{figure}
\centering
\subfigure[]{\includegraphics[scale = 0.65]{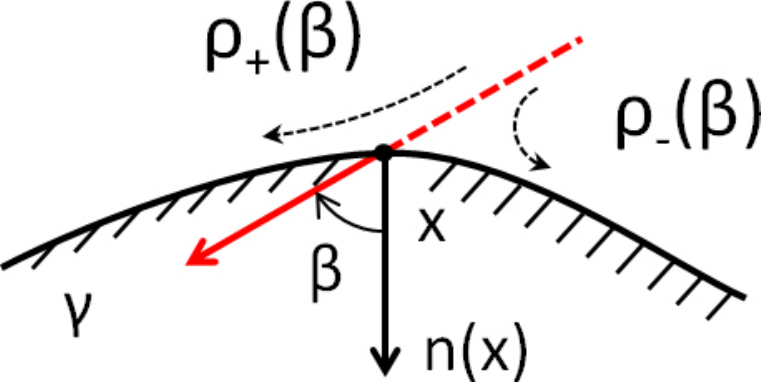}\label{figure: illustration of rho_pm}}\hspace{0.1\textwidth}
\subfigure[]{\includegraphics[scale = 0.65]{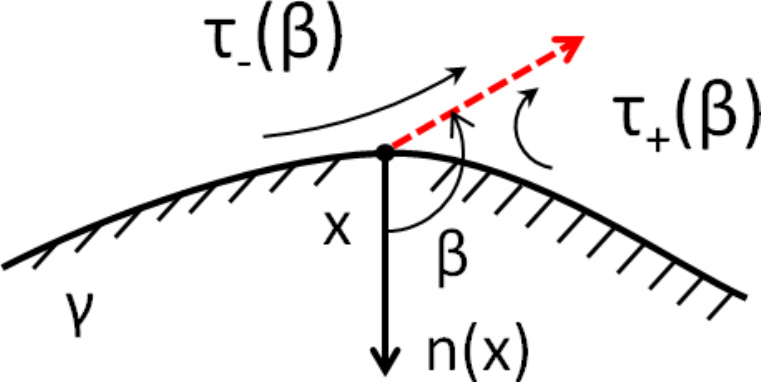}\label{figure: illustration of tau_pm}}
\caption{$\rho_\pm(\beta)$ and $\tau_{\pm}(\beta)$. The black curve represents $\gamma$ and the shaded side is occupied by the post. The black arrow is the normal vector $n(x)$ of $\gamma$ at $x$, while the red arrow represents the direction in which the rod hits $\gamma$, or the potential direction in which the rod is going to leave $\gamma$. (a) When the rod hits $\gamma$ with relative angle $\beta$, it will have probabilities $\rho_+(\beta)$ and $\rho_-(\beta)$, respectively, of swimming counter-clockwise and clockwise afterwards along $\gamma$. (b) When a rod is going to leave $\gamma$, its relative angle $\beta$ with respect to $\gamma$ is determined via the distributions $\tau_\pm(\beta)$. The subscripts indicate its original swimming direction before leaving $\gamma$.}
\label{figure: illustration of rho and tau}
\end{figure}

To model this pair of fluxes, we need some notation.
Let $\kappa(x)$ be the curvature of $\gamma$ at $x$, and $\alpha(x)$ be the orientation of the outer normal of $\gamma$ at $x$ with respect to $\omega$, i.e., the normal vector is given by $(\cos\alpha(x), \sin\alpha(x))^T$.
When a rod appears at a boundary point $x\in \gamma$ with orientational angle $2\pi\theta$, we define its relative angle with respect to $\gamma$ to be $\beta = 2\pi \theta - \alpha(x)\,(\mathrm{mod}\,2\pi)$; see Figure \ref{figure: rod state and boundary geometry}.
To this end, we introduce empirical rate functions $r_{\mathrm{in}}(\kappa)$ and $r_{\mathrm{in}}(\kappa)$, and angular functions $\rho_\pm(\beta)$ and $\tau_\pm(\beta)$, to be explained later, and write
\begin{align}
F_\mathrm{in}^\pm(x,t)=&\;r_\mathrm{in}(\kappa(x))\int_0^1 p(x,\theta,t)\rho_\pm(2\pi\theta-\alpha(x))\,d\theta,\quad x\in \gamma,\label{eqn: absorption flux at the boundary} \\
F_\mathrm{out}^\pm(x,t)=&\;r_\mathrm{out}(\kappa(x))p_B^\pm(x,t)\int_0^1 2\pi\tau_\pm(2\pi\theta-\alpha(x))\,d\theta,\quad x\in \gamma.\label{eqn: desorption flux at the boundary}
\end{align}
Here $r_{\mathrm{in}}(\kappa)$ and $r_{\mathrm{out}}(\kappa)$ are called absorption and desorption (Poisson) rates, respectively, 
which are assumed to be functions of curvature only.
In fact, experiments have shown that at least the desorption rate also depends on the rod speed \cite{wykes2017guiding}.
However, our assumption is valid since the rod speed is fixed to be $v_0$.
The functions $r_{\mathrm{in}}(\kappa(x))$ and $r_\mathrm{out}(\kappa(x))$ then characterize how fast the boundary $\gamma$ can absorb and desorb rods at $x\in\gamma$, respectively.
Their precise characterization will be clear after we choose the characteristic scales and do non-dimensionalization in Section \ref{section: model choices}.
The functions $\rho_\pm(\beta)$ account for the orientation dependence in the absorption.
We assume that rods hitting $\gamma$ with relative angle $\beta$ will have probabilities $\rho_+(\beta)$ and $\rho_-(\beta)$ of subsequently swimming counter-clockwise and clockwise along $\gamma$, respectively; see Figure \ref{figure: illustration of rho_pm}.
To make sense of \eqref{eqn: absorption flux at the boundary}, we note that rods hitting $x\in \gamma$ can come from the bulk in all directions;
the function $r_\mathrm{in}(\kappa(x))p(x,\theta,t)\rho_+(2\pi\theta-\alpha(x))$ is the amount of rods getting absorbed at $x$ with angle $2\pi\theta$, and sliding counter-closewise along $\gamma$ afterwards.
The function $r_\mathrm{in}(\kappa(x))p(x,\theta,t)\rho_-(2\pi\theta-\alpha(x))$ can be interpreted similarly.
If we take an integral over all possible orientations of the incoming rods, we obtain the absorption fluxes.

Similarly, for desorption, rods that are leaving $\gamma$ are assumed to leave at a random angle, with its probability distribution characterized by $\tau_\pm(\beta)$, where $\beta$ is the relative angle defined before.
The subscripts $\pm$ indicate the rods originally move counter-clockwise or clockwise before leaving $\gamma$; see Figure \ref{figure: illustration of tau_pm}.
We argue as before to obtain \eqref{eqn: desorption flux at the boundary}.
Via $\beta$ and the tangent direction of $\gamma$, the initial orientational angle of the rod when it returns to the bulk can be determined.
This will be useful in deriving the boundary condition of $p$ below.

To summarize, the assumptions on $r_\mathrm{in}$, $r_\mathrm{out}$, $\rho_\pm$ and $\tau_\pm$ are as follows:
\begin{enumerate}
  \item $r_\mathrm{in}(\kappa),\,r_\mathrm{out}(\kappa) \geq 0$;
  \item $\rho_\pm(\beta)\geq 0$ and $\rho_+ (\beta)+\rho_- (\beta)\leq 1$;
  \item $\tau_\pm(\beta)\geq 0$ and $\int_{-\pi}^{\pi} \tau_\pm (\beta)\,\mathrm{d}\beta = 1$;
  \item $\rho_+(\beta) = \rho_-(-\beta)$ and $\tau_+(\beta) = \tau_-(-\beta)$, by symmetry.
\end{enumerate}
Since $\tau_\pm$'s are normalized, 
Eq.\;\eqref{eqn: desorption flux at the boundary} reduces to $F_\mathrm{out}^\pm= r_\mathrm{out} p_B^\pm$.

Lastly, the boundary condition of $p$ on $\Gamma\triangleq \gamma\times [0,1]$, the inner curved part of $\partial\Omega$, is derived from the conservation law, which gives
\begin{equation}
\begin{split}
D_t \frac{\partial p}{\partial n_\Gamma}(x,\theta,t) - &\;v_0p(x,\theta,t)\cdot(\cos2\pi\theta,\sin 2\pi\theta,0)\cdot n_\Gamma(x)\\ =&\;r_\mathrm{out}(\kappa(x))[p_B^+(x,t)\cdot 2\pi\tau_+(2\pi\theta-\alpha(x))+p_B^-(x,t)\cdot 2\pi\tau_-(2\pi\theta-\alpha(x))]\\
&\;-r_\mathrm{in}(\kappa(x))p(x,\theta,t)[\rho_+(2\pi\theta-\alpha(x))+\rho_-(2\pi\theta-\alpha(x))],\quad (x,\theta)\in\Gamma.
\end{split}
\label{equation: boundary condition non steady state with drift}
\end{equation}
Here $n_{\Gamma}(x) = (\cos \alpha(x), \sin\alpha(x),0)^T$ is the unit outer normal vector of $\Gamma$ with respect to $\Omega$.
The left hand side of \eqref{equation: boundary condition non steady state with drift} represents the boundary flux at $(x,\theta)\in \Gamma$ generated by the spatial diffusion and the swimming; it is balanced by the flux coming into the bulk due to desorption and absorption on the right hand side.
Using the formula for $n_\Gamma$, \eqref{equation: boundary condition non steady state with drift} is simplified to be
\begin{equation*}
\begin{split}
D_t \frac{\partial p}{\partial n_{\Gamma}}(x,\theta,t) - v_0p(x,\theta,t)\cos\beta =&\;2\pi r_\mathrm{out}(\kappa(x))[p_B^+(x,t)\tau_+(\beta)+p_B^-(x,t)\tau_-(\beta)]\\
&\;-r_\mathrm{in}(\kappa(x))p(x,\theta,t)[\rho_+(\beta)+\rho_-(\beta)],\quad (x,\theta)\in \Gamma,
\end{split}
\end{equation*}
with $\beta = 2\pi\theta -\alpha(x)$.
For the outer flat surfaces of $\partial\Omega$, i.e.\;$\partial\Omega\backslash\Gamma$, we assign periodic boundary conditions for $p$.


In this paper, we only consider the steady-state solution.
Omitting the $t$-dependence in \eqref{equation: equation in the bulk non steady state with drift} and \eqref{equation: equation on the boundary non steady state with drift}, the equations become
\begin{align}
&\;D_t\Delta_x p(x,\theta)+D_r\Delta_\theta p(x,\theta) - v_0(\cos 2\pi\theta, \sin 2\pi\theta)^T\cdot\nabla_{x} p(x,\theta) = 0 \quad(x,\theta)\in \Omega,\label{equation: model in the steady state with drift equation in the bulk}\\
&\;- D_t\Delta_\gamma p_B^\pm(x) \pm v_0 \partial_\gamma p_B^\pm(x) = \int_0^1 f_\pm(x,\theta)\,\mathrm{d}\theta,\quad x\in \gamma,\label{equation: model in the steady state with drift equation on the boundary}\\
&\;D_t \frac{\partial p}{\partial n_{\Gamma}}(x,\theta) - v_0p(x,\theta)\cos\beta+ f_+(x,\theta)+f_-(x,\theta)=0,\quad(x,\theta)\in\Gamma,\label{equation: model in the steady state with drift boundary condition}\\
&\; p\mbox{ satisfies periodic boundary condition on }\partial \Omega\backslash \Gamma,\label{eqn: periodic boundary condition on p}
\end{align}
where $\beta = 2\pi\theta-\alpha(x)$ and
\begin{equation}
f_\pm(x,\theta) = r_\mathrm{in}(x)\rho_\pm(\beta)p(x,\theta) -r_\mathrm{out}(x)\cdot2\pi\tau_\pm(\beta)p_B^\pm(x),\quad(x,\theta)\in\Gamma.\label{equation: model in the steady state with drift definition of boundary fluxes}
\end{equation}
As we are describing the probability distribution of rods, the following normalization condition is needed
\begin{equation}
N(\Omega) \triangleq \int_{\Omega} p(x,\theta)\,\mathrm{d}x\mathrm{d}\theta+\int_\gamma [p_B^+(x)+p_B^-(x)]\,\mathrm{d}\gamma = 1. \label{equation: normalization condition}
\end{equation}
Note that if \eqref{equation: normalization condition} is not assumed, that $(p,p_B^+,p_B^-)$ is a solution of the equations \eqref{equation: model in the steady state with drift equation in the bulk}-\eqref{equation: model in the steady state with drift definition of boundary fluxes} implies that $(\lambda p,\lambda p_B^+,\lambda p_B^-)$ is also a solution for $\forall\, \lambda >0$.

Suppose we have obtained a nontrivial solution $(p,p_B^+,p_B^-)$ to \eqref{equation: model in the steady state with drift equation in the bulk}-\eqref{equation: model in the steady state with drift definition of boundary fluxes} (obviously $(0,0,0)$ is a trivial solution which is not interesting), without necessarily satisfying \eqref{equation: normalization condition}.
We wish to characterize the intensity of the spontaneous directed migration of the rods induced by the post in some particular direction, say the positive $x_2$-direction.
In our model, it is exactly the probability flux crossing the part of $\partial\Omega$ where $x_2 = b/2$.
The unnormalized net flux is defined to be
\begin{equation}
F(\Omega) = \int_{\partial\Omega \cap \{x_2 = b/2\}} -D_t\frac{\partial p}{\partial x_2} + v_0p\sin2\pi\theta\,dA.
\label{eqn: definition of F_Omega}
\end{equation}
The first term comes from the spatial diffusion of rods, while the second is due to the directed swimming, where the rod orientation plays a role.
Since this flux is generated by an amount of rods given by $N(\Omega)$, the normalized net flux is thus defined to be $E(\Omega)\triangleq F(\Omega)/N(\Omega)$.
$E(\Omega)$ will be the key quantity in the rest of the paper.

The numerical method for solving the coupled system \eqref{equation: model in the steady state with drift equation in the bulk}-\eqref{equation: model in the steady state with drift definition of boundary fluxes} and computing $E(\Omega)$ will be given in the Supplementary Materials.

\subsection{Seeking stronger directed migration of the rods}\label{section: shape optimization}





It is clear that the intensity of the directed migration of the rods is governed by many geometric features of the array, such as the sizes of gaps between neighboring posts, and the shapes of posts.
An interesting and practical question to ask is how to make the directed migration stronger by cleverly designing the array and post shapes.
In our model, this could be formulated as the following optimization problem: given the functions $r_{\mathrm{in}}(\kappa)$, $r_{\mathrm{out}}(\kappa)$, $\rho_\pm(\beta)$ and $\tau_\pm(\beta)$, find $Y = \left[-\frac{a}{2},\frac{a}{2}\right]\times \left[-\frac{b}{2},\frac{b}{2}\right]$ and $\omega\subset Y$, such that $E(\Omega)$ is maximized.

This is an infinite dimensional optimization problem and it would be very hard to find the actual maximizer.
We are only able to pose it formally.
Our strategy here is as follows.
Firstly, we shall study how the spacings of the posts affect $E(\Omega)$, simply by fixing the shape of the post and tuning $a$ and $b$ to see how $E(\Omega)$ changes.
Secondly, we would like to look for a better shape of the posts.
Using the theory of shape optimization, we develop a numerical method that evolves $\gamma$ in an iterative manner, so that the corresponding $E(\Omega)$ keeps increasing with each iteration.
The mathematical formulation of the shape optimization problem is exceedingly long.
We shall leave an overview of the theory of shape optimization, a complete derivation of the equations needed in the shape optimization, and an introduction of the associated numerical method to the Supplementary Materials.

\section{Model Choices}\label{section: model choices}
Before presenting numerical results, we specify the dimensionless parameters and rates used in the simulations.



We take the characteristic length scale in the model to be the typical size of the unit cell in the experiment \cite{wykes2017guiding}, which is $L = 45\,\mu m$.
We use the angular diffusion of the rods to determine the characteristic time scale.
In the experiments, the typical angular diffusion scale is $0.5\,rad/s$.
Hence, we take the characteristic time scale to be $T = 4\pi\,s$.
In this way, the dimensionless angular diffusion coefficient is $D_r=1$.
Note that it is $2\pi\theta$ instead of $\theta$ that represents the rod orientation.

Under the above choice, we take the dimensionless parameter in the model as follows: $D_t = 0.002$, $D_r = 1$, and $v_0 = 1$.
Indeed, this corresponds to the case where the spatial diffusion of rods is approximately $0.002\cdot [45\,\mu m]^2/[4\pi\,s]\approx 0.322 \, \mu m^2 /s$; the angular diffusion is $0.5\, rad/s$; and rods swim with approximate speed $1\cdot [45\,\mu m]/[4\pi \,s]\approx 3.58\,\mu m/s$.
All of these agree qualitatively with the measurements in the typical experiments \cite{wykes2017guiding}.

\begin{figure}
\centering
\includegraphics[width = \textwidth]{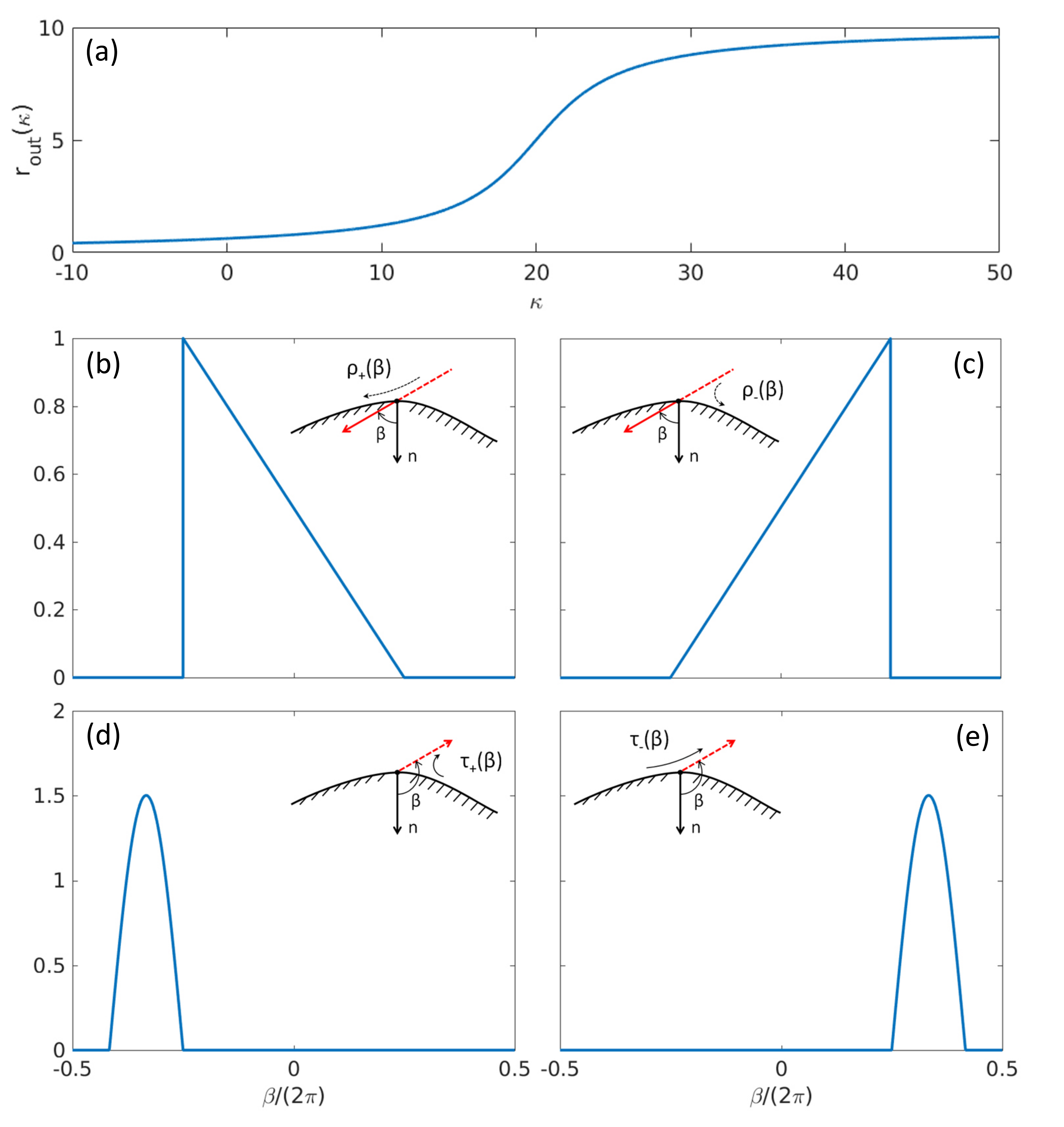}
\caption{Functions $r_\mathrm{out}(\kappa)$, $\rho_\pm(\beta)$ and $\tau_\pm(\beta)$ used in the simulations. (a) The graph of $r_\mathrm{out}(\kappa)= \frac{10}{\pi}\arctan\left(\frac{\kappa-20}{4}\right)+5$. It is positive, increasing in $\kappa$ and bounded as $\kappa \rightarrow +\infty$. (b-e)
Graphs of functions $\rho_\pm(\beta)$ and $\tau_\pm(\beta)$ used in the simulations. The insets illustrate the situations where these functions come into play; see \eqref{eqn: formula for rho}, \eqref{eqn: formula for tau} and Figure \ref{figure: illustration of rho and tau} for more details. Note that $\tau_\pm$ are normalized in the sense that $\int_{-\pi}^\pi \tau_\pm(\beta)\,d\beta = 1$.} 
\label{figure: choice of r_out rho and tau}
\end{figure}

There has been little systematic measurement of $r_\mathrm{in}(\kappa)$, $r_\mathrm{out}(\kappa)$, $\rho_\pm(\beta)$ and $\tau_\pm(\beta)$.
However, some experiments indicate that for swimming Au-Pt rods \cite{wykes2017guiding} or swimming Janus particles \cite{simmchen2016topographical}, $r_\mathrm{out}(\kappa)$ is an increasing function of the dimensionless curvature $\kappa$.
In our study of the hydrodynamic capture of bacteria to a solid surface, a threshold $\pi/9$ of the take-off angle is used in the stochastic simulation of a lubrication theory model to fit an exponential distribution to the trapping times \cite{takagi2014hydrodynamic}.
With this in mind, we use functions above such that they qualitatively agree with the existing experimental observations and physical intuition.
We take $r_\mathrm{in}(\kappa) = 1$, which implies that each segment of the post boundary with equal length has equal efficacy in absorbing rods, regardless of its curvature.
More precisely, given the unit density of rods (one rod per unit cell on average, i.e.\;$\frac{1}{45^2}\,\mathrm{rod}/\mu m^2$) in the vicinity of the post boundary, there is on average one absorbing event occurring over a boundary section with arclength $45\,\mu m$ every $4\pi\, s \approx 12.6\,s$, which is on the right scale.
The function $r_\mathrm{out}(\kappa)$ is defined as the inverse of the expected trapping time of rods sliding along a circular boundary with constant dimensionless curvature $\kappa$.
We take
\begin{equation}
r_\mathrm{out}(\kappa) = \frac{10}{\pi}\arctan\left(\frac{\kappa-20}{4}\right)+5
\label{eqn: formula for r_out}
\end{equation}
as a convenient choice; see Figure \ref{figure: choice of r_out rho and tau}(a).
Indeed, we only need the property that $r_{\mathrm{out}}(\kappa)$ is increasing in $\kappa$, which implies that the efficacy of the boundary in releasing rods is higher where its curvature is larger (more convex).
That $r_{\mathrm{out}}(\kappa)$ saturates as $\kappa\rightarrow +\infty$ is a convenient feature that makes the numerical simulations more tractable.
We shall discuss its effect in Section \ref{section: optimization of the shape of posts}.
To give a sense of the choice of $r_\mathrm{out}(\kappa)$, we have $r_\mathrm{out}(20) = 5$, which implies that when a rod is sliding along the boundary of a circular post with radius $2.25\,\mu m = [45\,\mu m]/20$, the expected sliding time is $\frac{4\pi}{5}\,s\approx 2.51\,s$. In other words, the Poisson leaving rate of a rod on a circular boundary with radius $2.25\,\mu m$ is approximately $0.398\,s^{-1}$, which agrees qualitatively with the experimental data \cite{wykes2017guiding}.

The functions $\rho_\pm(\beta)$ are taken to be
\begin{equation}
\rho_\pm(\beta) = \mp\frac{1}{\pi}\beta+\frac{1}{2},\quad \beta\in[-\pi/2,\pi/2],
\label{eqn: formula for rho}
\end{equation}
and $\rho_\pm(\beta)\equiv 0$ for $\beta \in[-\pi,\pi/2)\cup (\pi/2,\pi)$.
They are plotted in Figure \ref{figure: choice of r_out rho and tau}(b) and \ref{figure: choice of r_out rho and tau}(c).
They imply that when a rod hits the boundary perpendicularly ($\beta = 0$), it has equal probability of going in either direction along the boundary.
When it approaches in the tangent directions ($\beta = \pm \pi/2$), it will go forward in that direction with probability $1$.
When a rod reaches the boundary with its orientation pointing away from the boundary ($\beta \in[-\pi,\pi/2)\cup (\pi/2,\pi)$), it will not get absorbed.
Indeed, the last case can arise when spatial diffusion pushes the rod to the boundary even though its orientation points away.

The functions $\tau_\pm(\beta)$ are defined as follows.
Let
\begin{equation}
\tau_+(\beta) =
\begin{cases}
\frac{3}{2}\cos(3\beta)&\mbox{ for }\beta\in[-5\pi/6,-\pi/2],\\
0&\mbox{ for }\beta\in [-\pi,-5\pi/6)\cup (-\pi/2, \pi).
\end{cases}
\label{eqn: formula for tau}
\end{equation}
We then define $\tau_-(\beta) = \tau_+(-\beta)$.
These are plotted in Figure \ref{figure: choice of r_out rho and tau}(d) and \ref{figure: choice of r_out rho and tau}(e).
In this setting, rods are going to leave the boundary with take-off angles, defined to be the (unsigned) angle between the boundary and the rods, ranging from $0$ to $\pi/3$ ($\beta \in [-5\pi/6, -\pi/2]$ for rods originally sliding counter-clockwise on $\gamma$, or $\beta \in [\pi/2, 5\pi/6]$ for rods originally sliding clockwise).
The most likely take-off angle is $\pi/6$ ($\beta = -2\pi/3$ for rods originally sliding counter-clockwise, or $\beta = 2\pi/3$ for rods originally sliding clockwise).

\section{Numerical Results}\label{section: numerical results}

With implementation details now behind us, in the section, we shall present numerical results concerning how post spacings and post shapes can affect the induced normalized net flux $E(\Omega)$.

\subsection{The effect of post spacings}\label{section: optimization with respect to the post spacings}
Our first numerical simulation is devoted to investigating how post spacings affect the normalized net flux $E(\Omega)$.
To be more precise, we fix the shape of the post specified later, and change size of the unit cell specified by $a$ and $b$, to see how $E(\Omega)$ changes correspondingly.
Note that the set $\Omega$ implicitly depends on both $a$ and $b$.

The post used in this simulation has a teardrop-shaped cross-section, defined by two circular arcs smoothly connected by two straight lines; see Figure \ref{figure: unit cell}.
Under the non-dimensionalization, the radii of the larger and the smaller circular arcs are $0.192$ and $0.0154$ respectively.
Their corresponding dimensionless curvatures $\kappa$ are $5.21$ at the large end and $65.1$ at the small tip, respectively.
The angle between the two straight sides is $48$ degrees.
The tip of the post points to the positive $x_2$-direction.
The dimensionless size of the post is approximately $0.642$ 
in the vertical direction and $0.384$ 
in the horizontal direction.
Its perimeter is approximately $1.59$. 

\begin{figure}
\centering
\subfigure[]{\includegraphics[scale = 0.5]{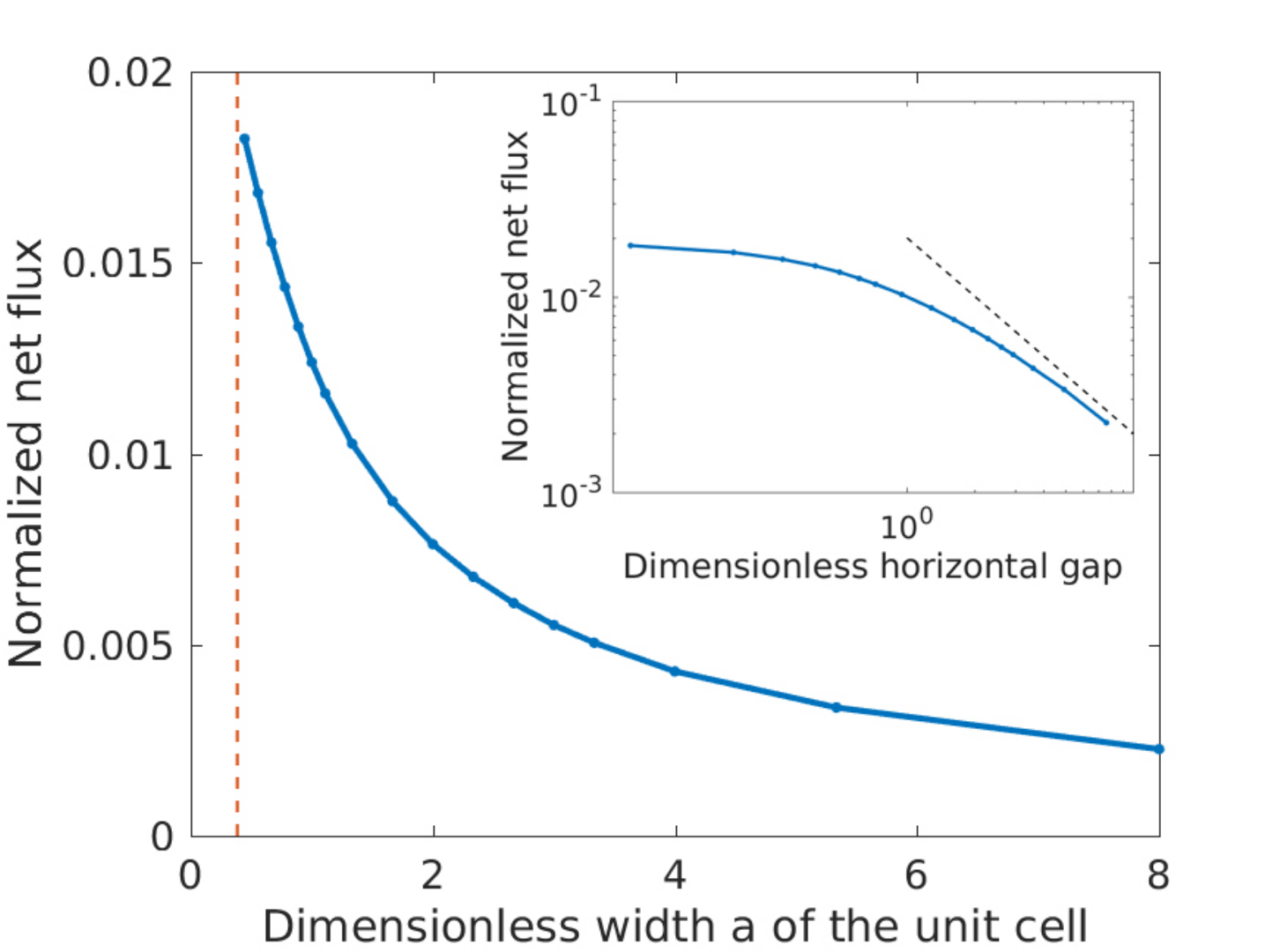}\label{figure: flux vs width}}
\subfigure[]{\includegraphics[scale = 0.5]{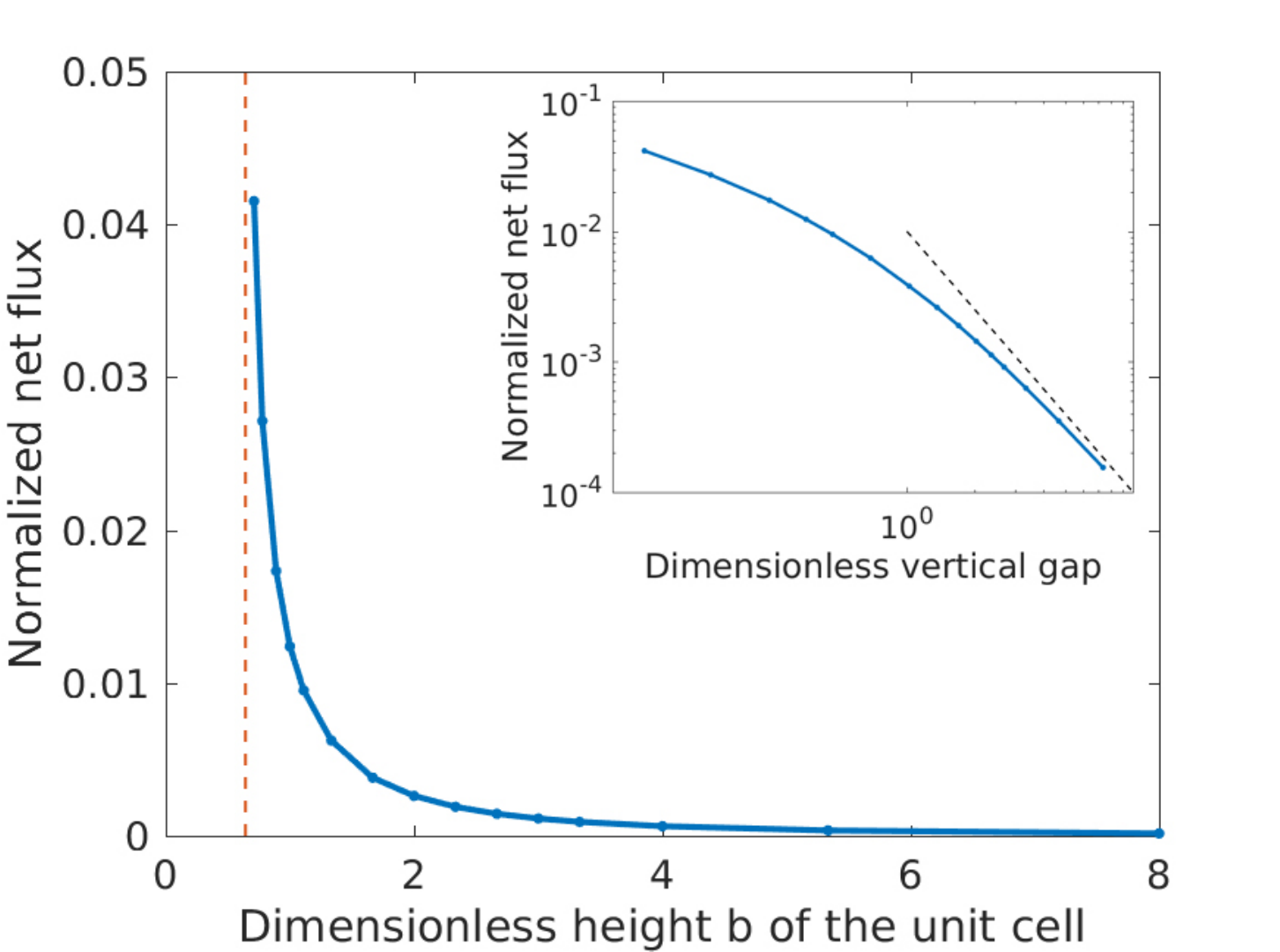}\label{figure: flux vs height}}
\caption{The normalized net flux $E(\Omega)$ depending on the dimensionless width $a$ and the height $b$ of the unit cell. (a) With $b = 1$, $E(\Omega)$ decreases as $a$ increases. The red dashed line indicates the minimum possible width $a_\mathrm{min} = 0.384$. The inset shows a log-log plot of $E(\Omega)$ depending on the dimensionless horizontal gap, $a - a_\mathrm{min}$. The black dashed line in the inset has slope $-1$. (b) With $a = 1$, $E(\Omega)$ decreases as $b$ increases. The red dashed line indicates the minimum possible height $b_\mathrm{min} = 0.642$. The inset is a log-log plot of $E(\Omega)$ depending on the dimensionless vertical gap, $b - b_\mathrm{min}$. The black dashed line in the inset has slope $-2$.}
\label{flux vs width and height}
\end{figure}

In Figure \ref{flux vs width and height}, we show how $E(\Omega)$ changes as we vary the dimensionless width $a$ and height $b$ of the unit cell.
These results are obtained by solving the model \eqref{equation: model in the steady state with drift equation in the bulk}-\eqref{equation: model in the steady state with drift definition of boundary fluxes} for each pair of $(a,b)$ and calculating the corresponding $E(\Omega)$.
In Figure \ref{figure: flux vs width}, we fix $b = 1$ and vary $a$.
It shows that $E(\Omega)$ decreases as $a$ increases.
The red dashed line indicates the minimum $a$ ($a_\mathrm{min} = 0.384$) we can possibly achieve, in which case the neighboring teardrop-shaped posts will touch each other.
As $a$ approaches $a_\mathrm{min}$, $E(\Omega)$ converges a finite value.
The inset of Figure \ref{figure: flux vs width} shows a log-log plot of $E(\Omega)$ vs.\;$a - a_\mathrm{min}$.
The black dashed line in the inset has slope $-1$, which implies that when $a$ is large, $E(\Omega)$ decays like $(a-a_{\mathrm{min}})^{-1}$, or equivalently, as $a^{-1}$.
In Figure \ref{figure: flux vs height}, we take $a = 1$ and vary $b$.
$E(\Omega)$ decreases as $b$ increases.
The red dashed line again indicates the minimum $b$ ($b_\mathrm{min} = 0.642$) we can achieve, when the teardrop-shaped posts in two neighboring rows will touch each other.
When $b$ approaches $b_\mathrm{min}$, $E(\Omega)$ apparently diverges to infinity.
The inset of Figure \ref{figure: flux vs height} shows a log-log plot of $E(\Omega)$ vs.\;$b - b_\mathrm{min}$.
The black dashed line in the inset has slope $-2$.
In other words, when $b$ is large, $E(\Omega)$ decays like $(b-b_{\mathrm{min}})^{-2}$, or equivalently $b^{-2}$.

The above results agree with the intuition that the open-space swimming away from the boundary smears out the anisotropy or bias \cite{wykes2017guiding} in the swimming direction induced by the boundary shape, thus making $E(\Omega)$ weaker.
As a result, $E(\Omega)$ should get boosted if we compress the open space by shrinking the gaps between neighboring posts in both directions.
Indeed, shrinking $b$ might be particularly effective due to the unboundedness of the graph in Figure \ref{figure: flux vs height} when $b\rightarrow b_\mathrm{min}$.
However, we should remark that making the gaps too narrow is not always favored in practice.
Swimming rods can hit into each other or even cause traffic jams in very narrow gaps, which impairs their mobility and makes the directed migration weaker.
Narrow gaps may significantly change the swimming behavior of the self-propelled rods, which is not considered in our model.
For example, experiments show that active rods can increase speed by up to five times in confining channels with ceiling \cite{PhysRevLett.117.198001}.
Further, the size of the rods, which is ignored in our model, becomes important when they swim in confined spaces like narrow gaps.

With this in mind, we study the net flux under a different normalization
by normalizing the number of rods per unit area instead of one unit cell.
This is more useful in practice, since we may use suspensions of swimming rods with some particular concentration to do experiments, and so the number of rods in the unit area is approximately given.
Besides, we may want to compute the net flux per unit width of array, since that characterizes the efficacy of the whole device in transporting rods within horizontal cross sections of unit length.
The net flux under this new normalization is thus given by $\tilde{E}(\Omega) = abE(\Omega)/a = bE(\Omega)$, which is the net flux over unit horizontal cross section with the rod concentration is normalized.

From the discussion above, it is known that when $a$ or $b$ is large, $\tilde{E}(\Omega)$ should decay like $a^{-1}$ or $b^{-1}$ respectively.
This can be justified by the following formal analysis.
\begin{enumerate}
\item When $b$ is fixed and $a$ is large, posts in different vertical columns function almost independently as they are so far away. 
That $a$ gets doubled is almost equivalent to removing half of the posts in one row, which will naturally decrease $\tilde{E}(\Omega)$ by one half.
This implies that $\tilde{E}(\Omega)$ will decay like $a^{-1}$ when $a$ is large.

\item When $a$ is fixed and $b$ is large, posts in different rows are now far away from each other.
If a rod leaves a post and swims into the open space towards posts in another neighboring row, then midway its orientation is already randomized.
An important length scale here is $v_0/D_r = 1$, which is the characteristic distance a rod can travel away from the post before it forgets its initial direction when leaving the boundary.
Therefore, $p(x,\theta)$ should have little $\theta$-dependence when the distance between $x$ and any posts in the array is significantly larger than $1$.
As a result, when $b$ is large, the motion of the rods midway between two rows can be characterized by an enhanced isotropic diffusion with the effective diffusion coefficient $D_{\mathrm{eff}} = D_t+\frac{v_0^2}{4D_r}$ \cite{howse2007self}.
Instead of \eqref{equation: model in the steady state with drift equation in the bulk}, one can solve $D_{\mathrm{eff}} \Delta \tilde{p}(x) = 0$ for the spatial distribution of rods, while the orientational distribution of rods there should be almost uniform in all directions.
From the far-field point of view, the effect of one post can be modeled as a dipole in the positive $x_2$-direction, since it effectively sucks rods from one end (the larger end) and releases them from the other (the smaller tip).
The magnitude of the dipole is insensitive to $b$, since we have normalized the rod concentration instead of the number of rods per unit cell.
Hence, the far-field spatial density of rods $\tilde{p}(x)$ induced by one single dipole at the origin, up to an additive constant, should be approximately of the form
\begin{equation*}
\tilde{p}(x) \sim \frac{C x_2}{x_1^2+x_2^2},
\end{equation*}
where $C$ is a constant depending on $D_{\mathrm{eff}}$ and the dipole magnitude.

We wish to align such dipole in an array with horizontal spacing $a$ and vertical spacing $b$, and calculate the net flux crossing the segment $\{(x_1,x_2):\; x_1\in[-a/2, a/2],\;x_2 = b/2\}$.
We start from one row of posts all centered at $x_2 = 0$.
The spatial density in a neighborhood of the line $x_2 = b/2$ contributed by this row should be well approximated, up to an additive constant, by $\tilde{p}_0(x) = \sum_{k\in\mathds{Z}} \tilde{p}(x_1+ka, x_2)$.
According to \eqref{eqn: definition of F_Omega}, we calculate the net flux contributed by this row as
\begin{equation*}
\tilde{E}_0 = \int_{\partial\Omega \cap \{x_2 = b/2\}} -D_t\frac{\partial \tilde{p}_0}{\partial x_2} + v_0\tilde{p}_0\sin2\pi\theta\,dA.
\end{equation*}
The subscripts $0$ implies this portion of the net flux comes from the row of posts centered at $x_2 = 0$ (i.e.\;$\tilde{p}_0$).
The second term in the integral above should vanish, since there is no $\theta$-dependence in $\tilde{p}_0(x)$ and the $\theta$-integral of $\sin 2\pi\theta$ is zero.
Hence,
\begin{equation*}
\begin{split}
\tilde{E}_0 =&\; -D_t\left[\int_{-a/2}^{a/2} \frac{\partial \tilde{p}_0}{\partial x_2}\,dx_1\right]_{x_2 = b/2} = -D_t\sum_{k\in \mathds{Z}}\left[\int_{-a/2}^{a/2} \frac{\partial \tilde{p}(x_1+ka, x_2)}{\partial x_2}\,dx_1\right]_{x_2 = b/2}\\
= &\;-D_t\sum_{k\in \mathds{Z}}\left[\int_{ka-a/2}^{ka+a/2} \frac{\partial \tilde{p}(x_1, x_2)}{\partial x_2}\,dx_1\right]_{x_2 = b/2}= -D_t\left[\int_{\mathds{R}} \frac{\partial \tilde{p}}{\partial x_2}\,dx_1\right]_{x_2 = b/2}\\
= &\;-D_t\left[\int_\mathds{R} \partial_{x_2}\left(\frac{C x_2}{x_1^2+x_2^2}\right) \,dx_1\right]_{x_2 = b/2} = \frac{C_0}{b},
\end{split}
\end{equation*}
where $C_0$ is a positive constant depending on $D_t$, $D_{\mathrm{eff}}$ and the dipole magnitude.
By symmetry, the net flux contributed by the row of posts centered at $x_2 = b$ should be $\tilde{E}_1 = \tilde{E}_0$.
For the rows of posts farther away from the line $x_2 = b/2$, their contributions to the net flux are suppressed by the screening effect of the rows that are closer to $x_2 = b/2$.
Indeed, the probability of a rod leaving from a post centered at $x_2 = kb$ and reaching the horizontal line $x_2 = b/2$ without being captured any other posts one the way should decay exponentially as $|k-1/2|\rightarrow \infty$;
in other words, the contributions to the net flux from these farther rows cannot be fully seen by the line $x_2 = b/2$ due to the existence of closer rows.
Assume the screening factor to be $\alpha_k$ for the row of posts centered at $x_2 = kb$, with $\sum_{k\in \mathds{Z}} \alpha_k <\infty$.
We write the contribution of the row centered at $x_2 = kb$ to the net flux crossing the segment $\{(x_1,x_2):\; x_1\in[-a/2, a/2],\;x_2 = b/2\}$ to be $\tilde{E}_k = \alpha_k \tilde{E}_0$.
Presumably, $\alpha_k$'s should be independent of $b$, but only depend on the horizontal spacing of neighboring posts in one row, which is fixed here.
The total net flux induced by the whole array then becomes
\begin{equation*}
\tilde{E}(\Omega) = \sum_{k\in\mathds{Z}} \alpha_k \tilde{E}_0 = \frac{\tilde{C}}{b},
\end{equation*}
where $\tilde{C}$ is a constant depending on $\alpha_k$'s, $D_t$, $D_\mathrm{eff}$ and the dipole magnitude, but independent of $b$.
This justifies the $b^{-1}$-decay of $\tilde{E}(\Omega)$ when $b$ is large.
\end{enumerate}

\subsection{Optimization of post shape}\label{section: optimization of the shape of posts}
In the second family of simulations, we fix the size of the unit cell by setting $a = b = 1$, and apply the shape optimization method mentioned in Section \ref{section: shape optimization} to find posts that induce larger normalized net flux $E(\Omega)$.
See Section \ref{section: derivation of shape derivative p' and pB'}-\ref{section: linear dependence of p' p_B' on Vn} for the complete mathematical derivations of the equations involved, and see Section \ref{section: numerical method for shape optimization} for the numerical methods.
The parameters and rates have been chosen in Section \ref{section: model choices}.

\begin{figure}
\centering
\subfigure[]{\includegraphics[height = 0.17\textheight]{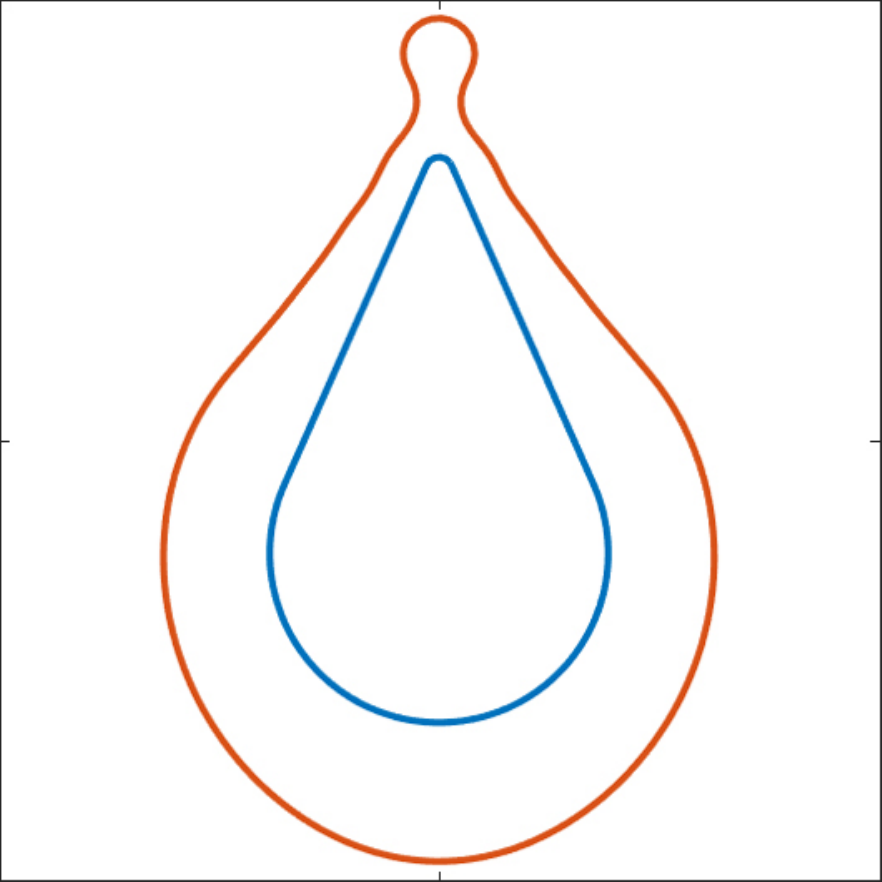}\label{figure: teardrop shape optimization initial and final}}\hspace{0.03\textwidth}
\subfigure[]{\includegraphics[height = 0.17\textheight]{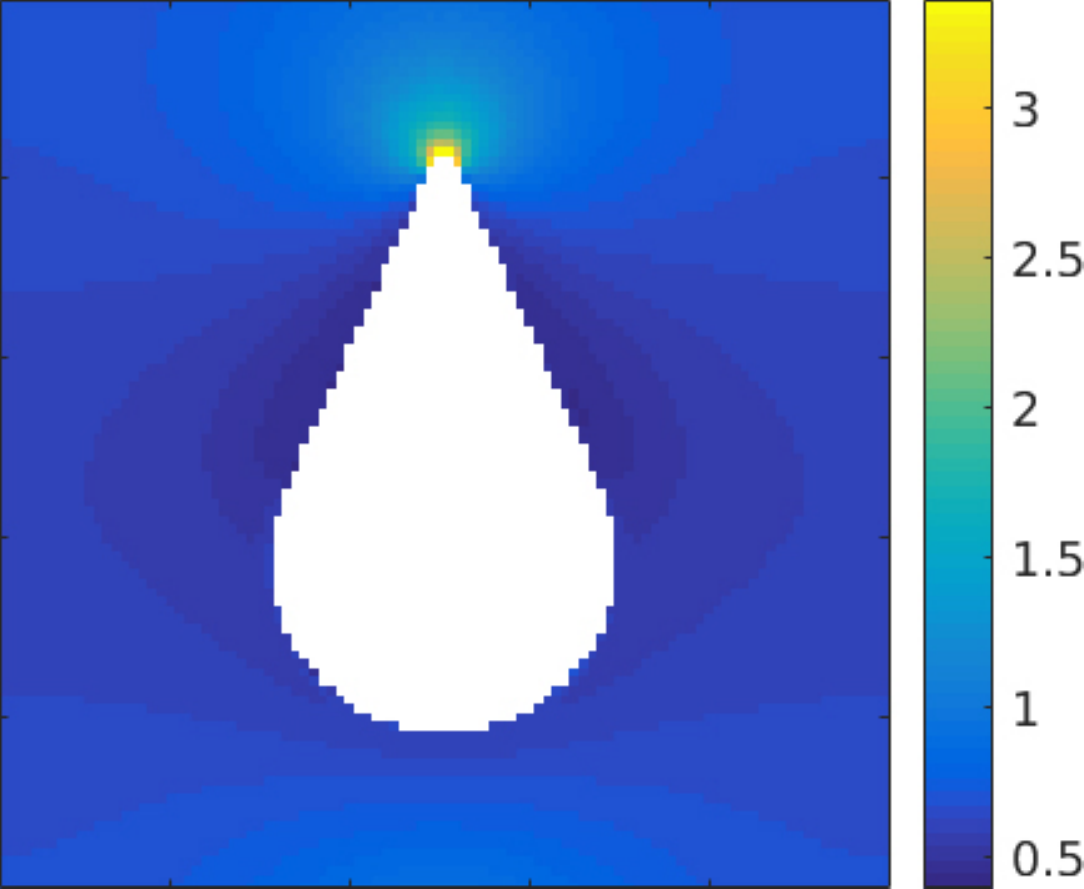}\label{figure: spatial density teardrop initial}}\hspace{0.03\textwidth}
\subfigure[]{\includegraphics[height = 0.17\textheight]{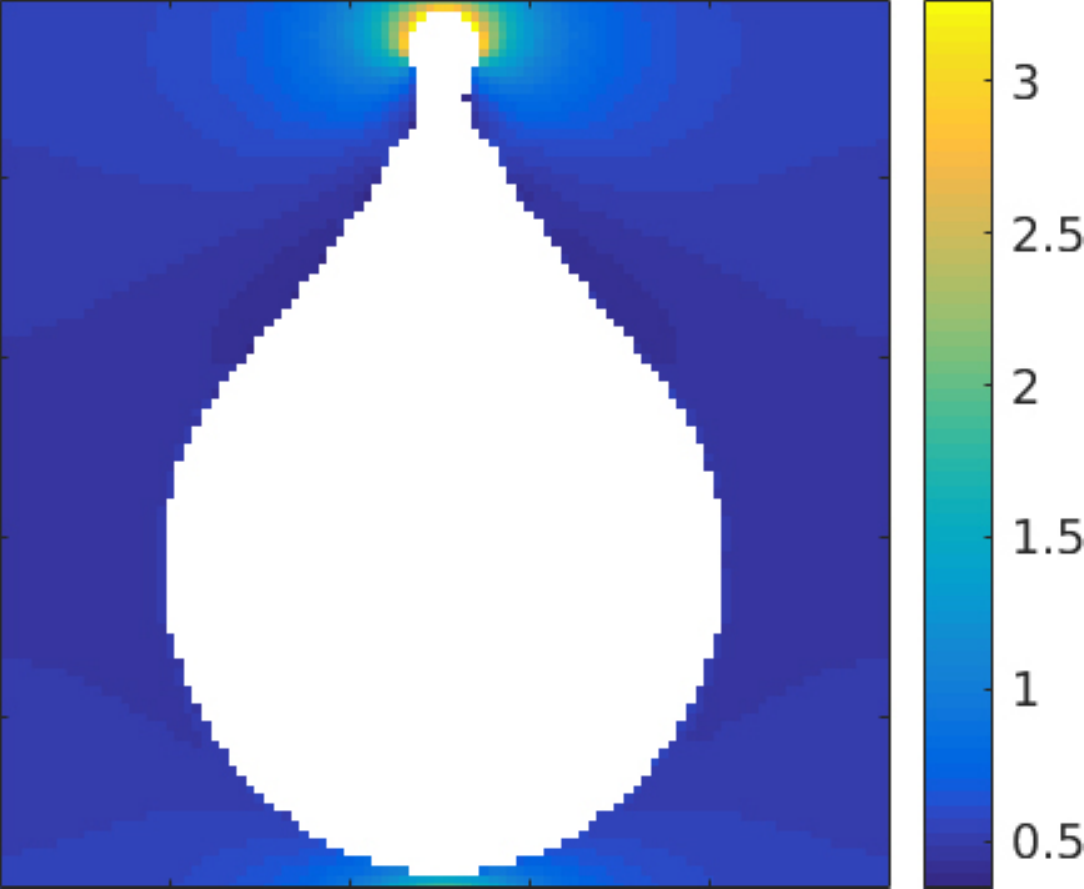}\label{figure: spatial density teardrop terminal}}
\caption{Performing shape optimization of a teardrop-shaped post. (a) The initial and the optimized shapes are represented by blue and red curves, respectively. In the course of the shape optimization, the post swells, with the sharp tip at the top evolving into a small round head. The lower half of the post remains a circular arc, while the overall post shape becomes non-convex. $E(\Omega)$ increases from $1.24\times 10^{-2}$ to $8.44\times 10^{-2}$. (b-c) The spatial concentrations of rods $c(x)$ in the presence of the posts with the initial and the optimized shapes in (a) are plotted in (b) and (c), respectively. The fraction of rods captured on the post boundary is not included in these figures.}
\label{figure: shape optimization on teardrop}
\end{figure}

We start from the post in the convex teardrop shape introduced in Section \ref{section: optimization with respect to the post spacings}.
Figure \ref{figure: teardrop shape optimization initial and final} shows the comparison before and after we apply the shape optimization to the teardrop-shaped candidate.
The blue curve represents the initial shape while the red curve is the optimized one (the final shape).
Note that the iterative optimization gets terminated when the post gets too close (dimensionless distance $\leq 0.02$) to the border of the unit cell.
We observe that with shape optimization, the post swells significantly and becomes non-convex.
A small round head forms at the top, connected by a thin neck with the larger belly.
The dimensionless curvature around the head is in a narrow range $[24.5, 25.5]$; recall that it is $65.1$ at the tip of the initial shape.
The side parts of the post, initially flat, become slightly wavy.
This can be artificial since only low-frequency modes are used in describing the boundary evolution.
Indeed, we use $240$ equally-spaced points to represent the curve, but only the first $41$ Fourier modes are used in the boundary evolution.
See Section \ref{section: numerical method for shape optimization} for more details about the numerical method.
The lower half of the post largely remains a circular arc.
In the course of the shape optimization, $E(\Omega)$ increases almost seven-fold from $1.24\times 10^{-2}$ to $8.44\times 10^{-2}$.

\begin{figure}
\centering
\subfigure[]{\includegraphics[height = 0.17\textheight]{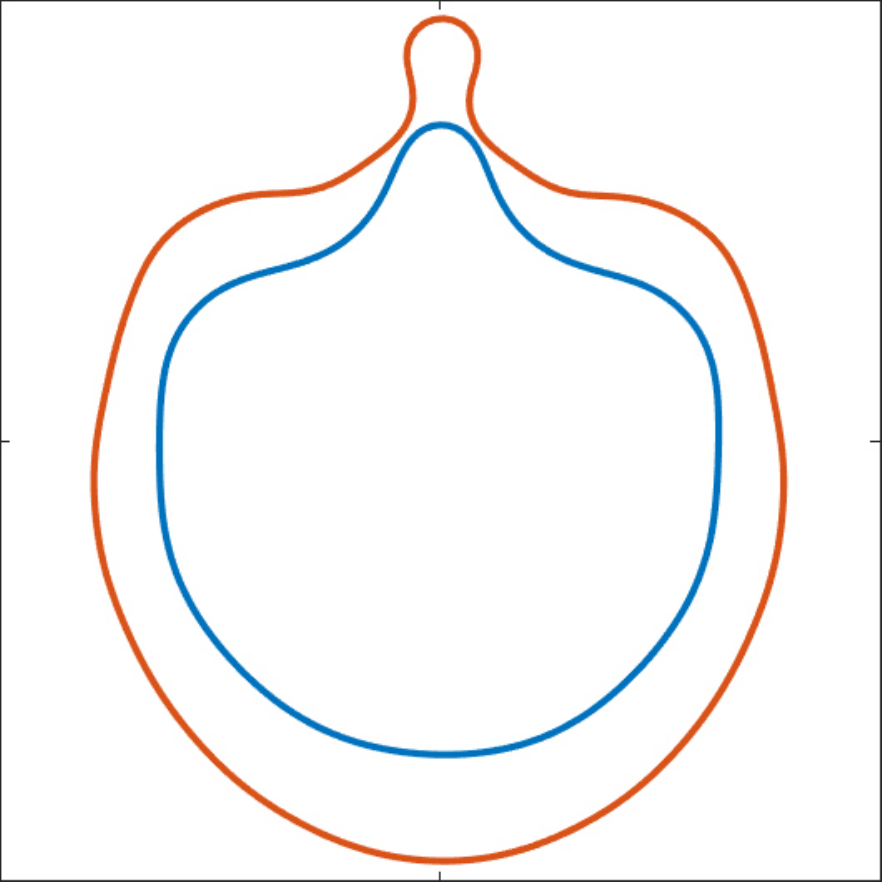}\label{figure: nut shape optimization initial and final}}\hspace{0.03\textwidth}
\subfigure[]{\includegraphics[height = 0.17\textheight]{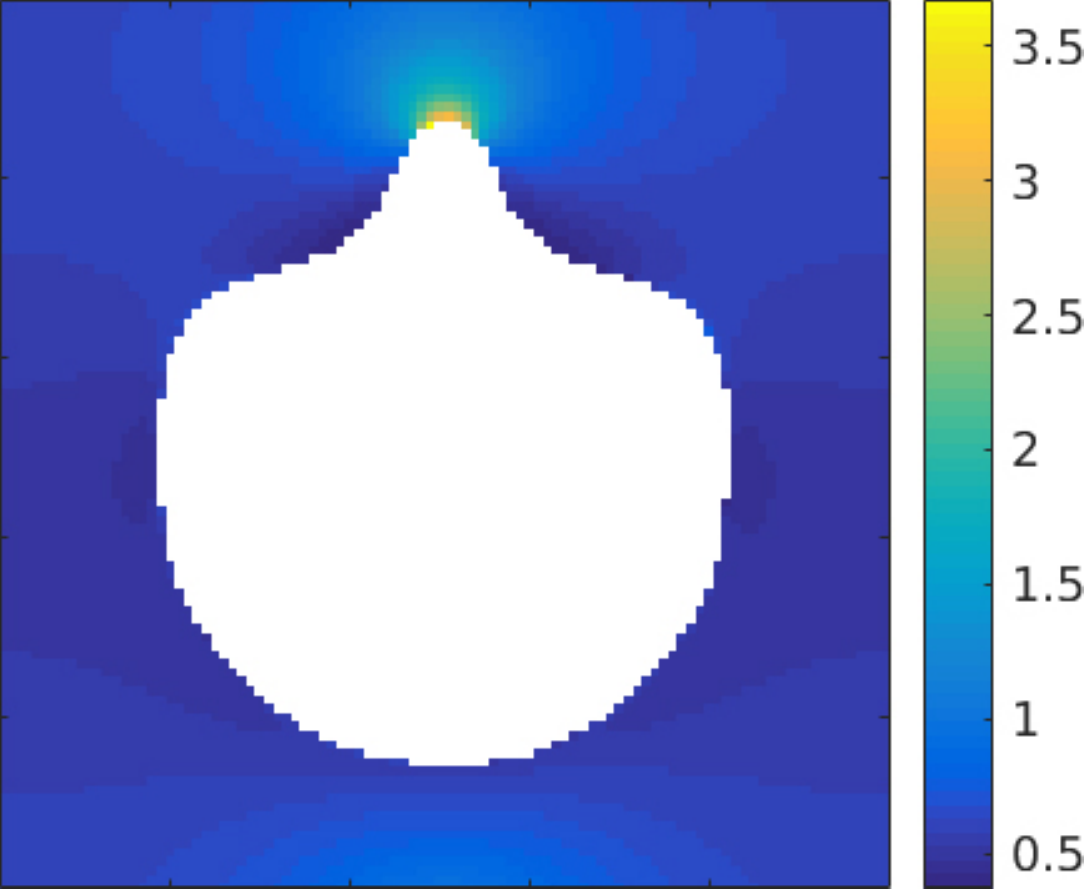}\label{figure: spatial density nut initial}}\hspace{0.03\textwidth}
\subfigure[]{\includegraphics[height = 0.17\textheight]{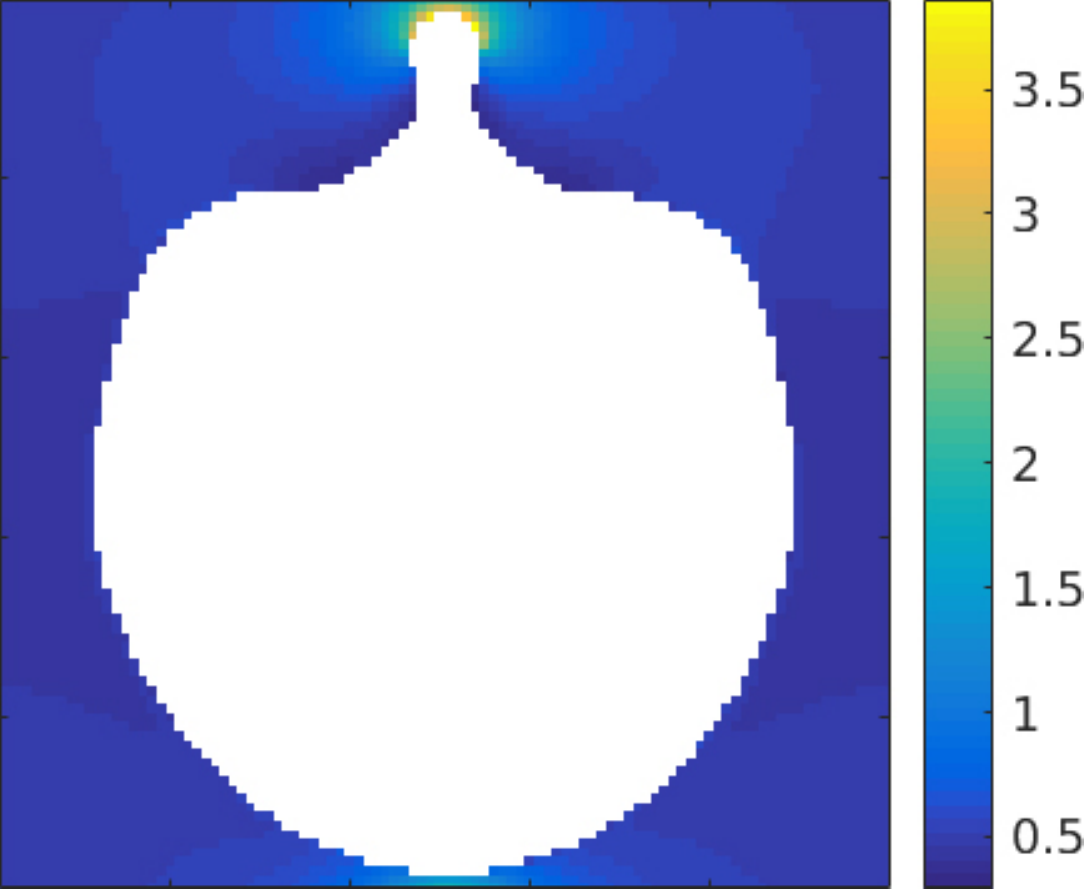}\label{figure: spatial density nut terminal}}
\caption{Performing shape optimization of a nut-shaped post. (a) The initial and the optimized shapes are represented by blue and red curves, respectively. In the course of the shape optimization, the post swells and a small round head forms at the top, while the lower half of the post remains a circular arc. $E(\Omega)$ increases from $1.81\times 10^{-2}$ to $7.82\times 10^{-2}$. (b-c) The spatial concentrations of rods $c(x)$ in the presence of the posts with the initial and the optimized shapes in (a) are plotted in (b) and (c), respectively. Again, the fraction of rods captured on the post boundary is not included.}
\label{figure: shape optimization on nut}
\end{figure}

To investigate how rods are spatially distributed in the presence of the posts with the initial or the optimized shapes in Figure \ref{figure: teardrop shape optimization initial and final}, we plot in Figure \ref{figure: spatial density teardrop initial} and \ref{figure: spatial density teardrop terminal} their corresponding spatial concentrations of rods $c(x) \triangleq \int_0^1 p(x,\theta)\,d\theta$ for $x\in \omega$.
Note that the fraction of rods captured on the boundary is not included in $c(x)$.
Figure \ref{figure: spatial density teardrop initial} and \ref{figure: spatial density teardrop terminal} share some common features:
\begin{enumerate}
\item The rod concentrations in both cases have their peaks near the top of the posts, implying that the top parts are the sites where strong net desorption occurs.
\item There are regions of depletion near the flat sides of the posts in both cases.
\item The rod concentrations around the bottoms are also relatively high, due to the high concentrations near the tops of the posts in the next unit cells right below.
\item The rod concentrations in both cases have a negative normal derivative near the bottom of the posts (i.e.\;$c(x)$ decreases when we approach the bottom from some distance away), which implies that the bottoms are effectively absorbing rods.
\end{enumerate}
The main difference between Figure \ref{figure: spatial density teardrop initial} and \ref{figure: spatial density teardrop terminal} is that, the area of the strong desorption site at the top in Figure \ref{figure: spatial density teardrop terminal} is much larger than that in Figure \ref{figure: spatial density teardrop initial}, although the curvature there ($\kappa\in [24.5,25.5]$) is much smaller than that ($\kappa =65.1$) in the initial shape.
This can partially explain why the optimized shape can induce a much larger net flux.

Next, we apply the shape optimization to a new nut-shaped post.
The motivation of choosing this as the new initial shape is that the head-forming process can potentially increase $E(\Omega)$.
It might accelerate this process if we start from an initial shape which already has a head.
In Figure \ref{figure: nut shape optimization initial and final}, we plot the shape of post before and after the shape optimization.
Again, the blue curve represents the initial shape while the red curve is the final one.
The normalized net flux increases from $1.81\times 10^{-2}$ to $7.82\times 10^{-2}$, which is a big improvement but not as good as the previous case.
We see that once again, the shape swells to fill almost the whole height of the unit cell; a round head and a neck forms at the top, while the lower half of the post largely remains a circular arc.
The dimensionless curvature at the head of the final shape ranges in $[24,25.5]$.
Again we plot the spatial concentrations of rods corresponding to posts with the initial and optimized shapes in Figure \ref{figure: spatial density nut initial} and \ref{figure: spatial density nut terminal} respectively.
It is clear that the top parts of both posts are the strong desorption sites, while it is larger in the optimized shape than in the initial shape.
Besides, there are noticeably two more sites near the post boundary with relatively high rod concentration and positive normal derivatives of $c(x)$.
They are the ``shoulders" of both the initial and the optimized shapes, which refer to the curved parts on both sides of the posts; they are also efficient in releasing rods.
The bottoms of the posts are the main absorption sites as before.

The effect of the ``shoulders" on the net flux is unclear though.
The net flux can benefit from larger desorbing sites.
On the other hand, however, the ``shoulders" are not as efficient desorbing sites as the heads at the top, since the ``shoulders" are far away from the top border of the unit cell and the rods released there may fail to reach the top border.
In this sense, the presence of the ``shoulders" can impair the capability of the head releasing rods at the top, and thus reduce $E(\Omega)$.
This may explain why the optimized shape in this case does not have as high $E(\Omega)$ as the one in the Figure \ref{figure: teardrop shape optimization initial and final}.

\begin{figure}
\centering
\subfigure[]{\includegraphics[height = 0.23\textheight]{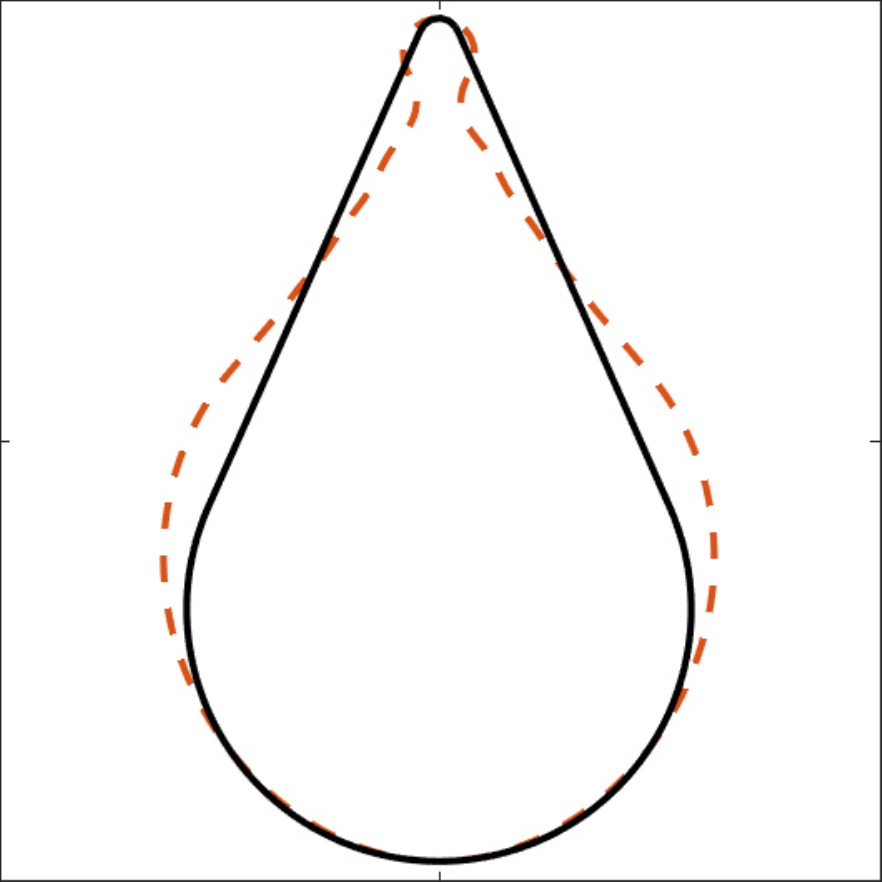}\label{figure: comparison between optimized teardrop and an enlarged teardrop}}\hspace{0.1\textwidth}
\subfigure[]{\includegraphics[height = 0.23\textheight]{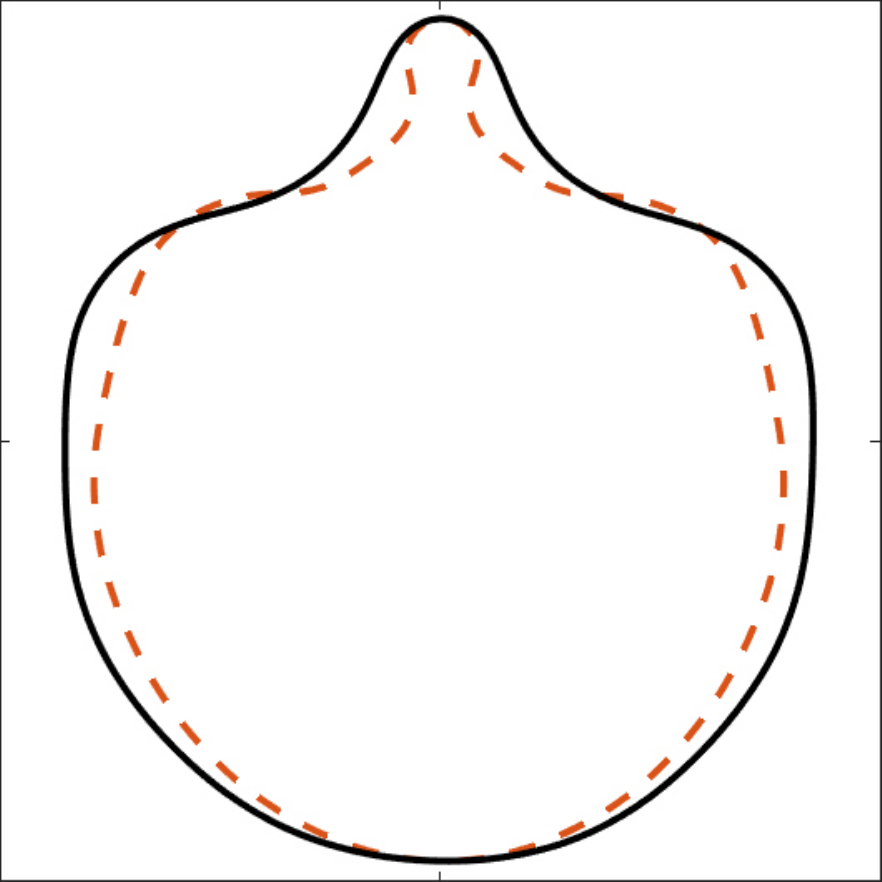}\label{figure: comparison between optimized nut and an enlarged nut}}
\caption{Enlarged posts with the initial shapes (solid curves) and posts with the optimized shapes (dashed curves). The heights of the enlarged posts are set to be the same as their corresponding optimized posts.}
\label{figure: comparison between optimized shape and an enlarged shape}
\end{figure}

It is shown in Section \ref{section: optimization with respect to the post spacings} that shrinking gaps between neighboring posts, especially the vertical gap, can effectively increase $E(\Omega)$.
One may question whether the increase in $E(\Omega)$ in these two cases above is purely due to the enlargement of the post (and thus shrinking of the vertical gap) or it does benefit from the changes in shape.
To rule out the first possibility, we make up enlarged copies of the initial teardrop-shaped and nut-shaped posts, such that they have the same height (and thus the same vertical gaps between neighboring posts) as their corresponding optimized shapes.
We plot the boundaries of enlarged posts in Figure \ref{figure: comparison between optimized shape and an enlarged shape} using solid curves, together with the corresponding optimized shapes that are plotted as dashed curves.
We calculate $E(\Omega)$ for the enlarged posts.
The enlarged teardrop-shaped post induces a normalized net flux $5.00\times 10^{-2}$, and the enlarged nut-shaped post gives $2.58\times 10^{-2}$.
In both cases, the enlarged posts can generate much larger net fluxes than the original ones, but still cannot compete with the optimized shapes.
In this way, we justify that the shape optimization does help us find better designs of the post.

To summarize, in addition to the overall swelling of the post in the shape optimization, we empirically find different evolutions of the upper and lower halves of the post boundary.
Round heads tend to form at the top of the posts with curvature there being in a narrow range $\kappa \in[24.5, 26]$; a convex shape can evolve to become non-convex.
By contrast, the lower half of the post always prefers to be a circular arc.

\section{An Approximate, but Informative, Analysis}\label{section: approximate simplified model}

\begin{figure}
\centering
\includegraphics[height = 0.23\textheight]{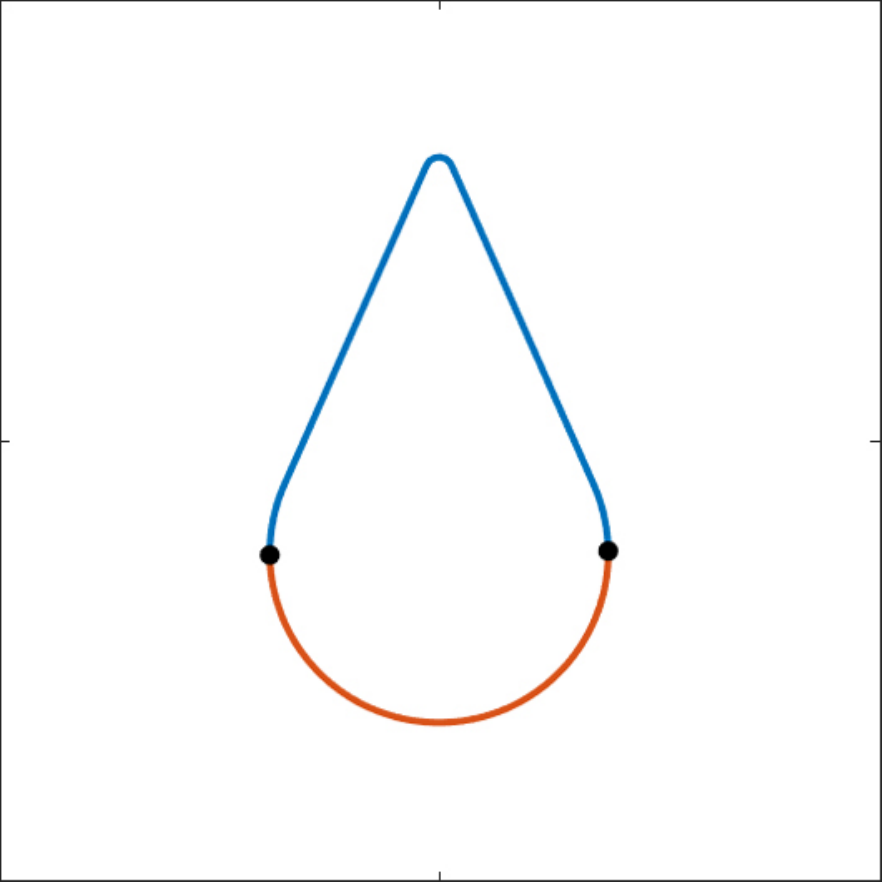}
\caption{Dividing the post boundary $\gamma$ into its upper (blue) and lower (red) halves in the case of a teardrop-shaped post. The left- and right-most points on $\gamma$ are marked as black dots.}
\label{figure: illustration of upper and lower halves}
\end{figure}

To better understand our findings and explore the possibility of designing posts with yet better shapes, we consider a simplified model.
We divide the post boundary curve $\gamma$ into its upper and lower halves by cutting it at the left-most and the right-most points on $\gamma$.
If there are several such points, pick the lowest one whenever necessary.
See Figure \ref{figure: illustration of upper and lower halves} for an illustration.
We denote the upper and lower halves of $\gamma$ by $\gamma_U$ and $\gamma_L$ respectively.
Since $r_\mathrm{in} \equiv 1$ on $\gamma$, when there is no a priori information about the spatial distribution of rods, then
\begin{equation*}
F_U \triangleq \int_{\gamma_U} r_\mathrm{out}(\kappa(s))\,ds
\end{equation*}
becomes a good characterization of the capability of $\gamma_U$ releasing rods.
Here $s$ is the arclength parameter of $\gamma$.
Similarly,
\begin{equation*}
F_L \triangleq \int_{\gamma_L} r_\mathrm{out}(\kappa(s))\,ds
\end{equation*}
is the corresponding quantity for $\gamma_L$.
We naively assume that rods leaving from $\gamma_U$ are more likely to reach the top border of the unit cell than the bottom border, while rods released from $\gamma_L$ are more likely to cross the bottom border.
By this assumption, the former type of rods contributes positively to $E(\Omega)$ since they cross the border in the positive $x_2$-direction, while the latter type contributes negatively.
Hence, in a simplified manner, we use $F_U-F_L$ to characterize the overall capability of $\gamma$ inducing spontaneous migration in the positive $x_2$-direction.
Here we ignore the variance in position and orientation of rods when they cross either part of the boundary, and neither do we incorporate any information about the spatial distribution of rods.
Therefore, to find a post with a good shape, we formally consider the following optimization problem,
\begin{equation}
\max_\gamma (F_U - F_L) = \max_\gamma\left[\int_{\gamma_U} r_\mathrm{out}(\kappa(s))\,ds - \int_{\gamma_L} r_\mathrm{out}(\kappa(s))\,ds\right].
\label{eqn: maximize the over-simplified flux}
\end{equation}
Here $r_\mathrm{out}(\kappa)$ is given by \eqref{eqn: formula for r_out} and we take maximum over all admissible curves $\gamma$.
By admissible, we mean $\gamma$ that is sufficiently smooth and that does not intersect with itself.
If needed, we may also impose a constraint that the curvature of $\gamma$ is bounded above and below by some constants.

To formally solve the maximization problem \eqref{eqn: maximize the over-simplified flux}, we first note that
\begin{equation}
\max_\gamma (F_U - F_L) \leq  \max_\gamma\int_{\gamma_U} r_\mathrm{out}(\kappa(s))\,ds - \min_{\gamma'}\int_{\gamma'_L} r_\mathrm{out}(\kappa(s))\,ds.
\label{eqn: over-simplified optimziation optimize upper and lower halves separately}
\end{equation}
On the right hand side, the $\gamma$ that attains the maximum in the first term and the $\gamma'$ that attains the minimum in the second term do not have to be the same one.
The famous Gauss-Bonnet Theorem \cite{do1976differential} will be useful in the following discussion, which states that for any admissible $\gamma$ in our context,
\begin{equation}
\int_{\gamma_U} \kappa(s)\,ds = \int_{\gamma_L} \kappa(s)\,ds = \pi,
\label{eqn: Gauss-Bonnet theorem}
\end{equation}
where $s$ is the arclength parameterizing $\gamma$ in a counter-clockwise orientation.

\begin{figure}
\centering
\includegraphics[scale = 0.6]{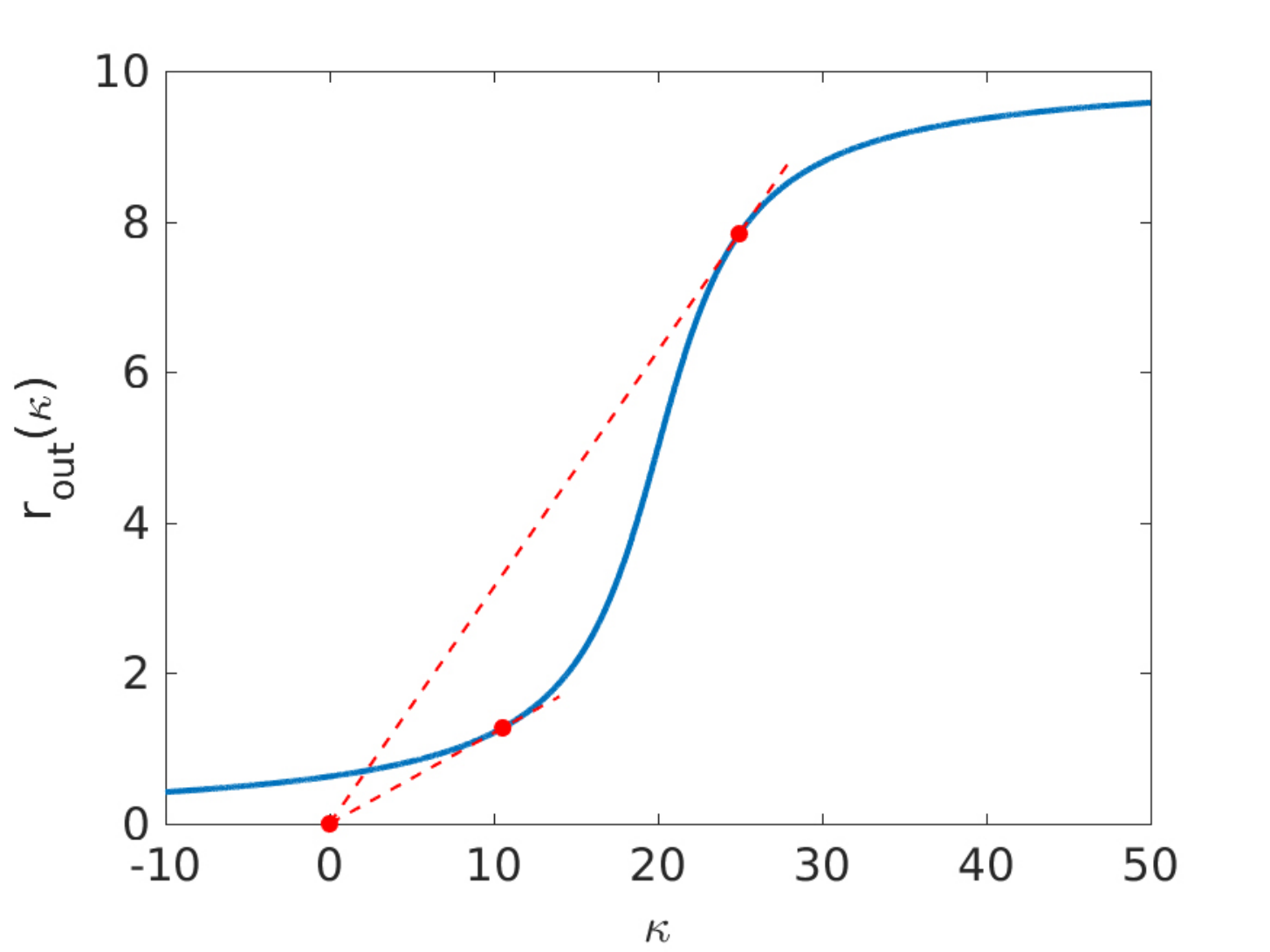}
\caption{$r_\mathrm{out}(\kappa)/\kappa$ reaches a local maximum at $\kappa_\mathrm{max} \approx 25$ and a local minimum at $\kappa_\mathrm{min} \approx 10.5$, marked as red dots on the curve. Note that $r_\mathrm{out}(\kappa)/\kappa$ is the slope of straight line connecting the point $(\kappa, r_\mathrm{out}(\kappa))$ with the origin.}
\label{figure: analysis of r_out}
\end{figure}

We first consider the maximizing problem in \eqref{eqn: over-simplified optimziation optimize upper and lower halves separately}.
We rewrite
\begin{equation*}
\max_{\gamma}\int_{\gamma_U} r_\mathrm{out}(\kappa(s))\,ds = \max_{\gamma}\int_{\gamma_U} \frac{r_\mathrm{out}(\kappa(s))}{\kappa(s)}\kappa(s)\,ds.
\end{equation*}
Here $r_\mathrm{out}(\kappa)/\kappa$ can be understood as the efficiency of utilizing the curvature to generate a desorption flux, given that \eqref{eqn: Gauss-Bonnet theorem} implies a fixed ``budget" of the curvature on both $\gamma_U$ and $\gamma_L$.
In order that the integral is maximized, $r_\mathrm{out}(\kappa)/\kappa$ needs to be as large as possible when $\kappa>0$ and as small as possible when $\kappa<0$.
It is known by \eqref{eqn: formula for r_out} that $r_\mathrm{out}(\kappa)/\kappa \rightarrow \pm\infty$ as $\kappa\rightarrow 0^\pm$.
Moreover, it is easy to show that $r_\mathrm{out}(\kappa)/\kappa$ reaches a local maximum at $\kappa_{\mathrm{max}} \approx 25$.
See Figure \ref{figure: analysis of r_out}.
This implies that in the upper half of the post, the curvature tends to be close to $0$ or $\kappa_{\mathrm{max}}$.
This explains why the curvature at the heads in both examples above lies in a narrow range close to $25$ instead of even larger values, such as $\kappa = 65.1$ at the tip of the initial teardrop-shaped post.
In addition, it does not harm to have negative curvature in $\gamma_U$ as it potentially increases the arclength where the $\kappa_{\mathrm{max}}$ could be attained, thus improving the overall capability of $\gamma_U$ releasing rods.
This explains why convex shapes can evolve into non-convex ones in the shape optimization.

Now we turn to the minimization problem involving $\gamma'_L$ in \eqref{eqn: over-simplified optimziation optimize upper and lower halves separately}.
We note that $r_\mathrm{out}(\kappa)/\kappa$ has a local minimum at $\kappa_{\mathrm{min}} \approx 10.5$.
See Figure \ref{figure: analysis of r_out}.
If we rule out the case when $\kappa$ can be very large along $\gamma'_L$, say assuming $\kappa < 80$, we will find
\begin{equation*}
r_\mathrm{out}(\kappa) \geq \frac{r_\mathrm{out}(\kappa_{\mathrm{min}})}{\kappa_\mathrm{min}} \kappa, \quad \forall\,\kappa < 80.
\end{equation*}
Hence,
\begin{equation*}
\min_{\gamma'}\int_{\gamma'_L} r_\mathrm{out}(\kappa(s))\,ds \geq \frac{r_\mathrm{out}(\kappa_{\mathrm{min}})}{\kappa_\mathrm{min}} \int_{\gamma'_L} \kappa(s)\,ds = \frac{\pi r_\mathrm{out}(\kappa_{\mathrm{min}})}{\kappa_\mathrm{min}}.
\end{equation*}
Therefore, the minimum is achieved if $\gamma'_L$ is a semicircle with dimensionless curvature $\kappa_\mathrm{min}$.
This might not be obtained in general, because for example, the arclength between the left-most and the right-most points on $\gamma'$ may not match the arclength of the semicircle with curvature $\kappa_\mathrm{min}$.
In such case, $\gamma'_L$ still has to be largely a semicircle.
We can prove this under the assumption that $\kappa < 20$ on $\gamma'_L$, simply by noticing that $r_\mathrm{out}(\kappa)$ is a convex function for $\kappa<20$ and then applying Jensen's inequality.

\section{Explorations of Other Designs}\label{section: explorations of other designs}
\begin{figure}[htbp]
\centering
\subfigure[]{\includegraphics[height = 0.23\textheight]{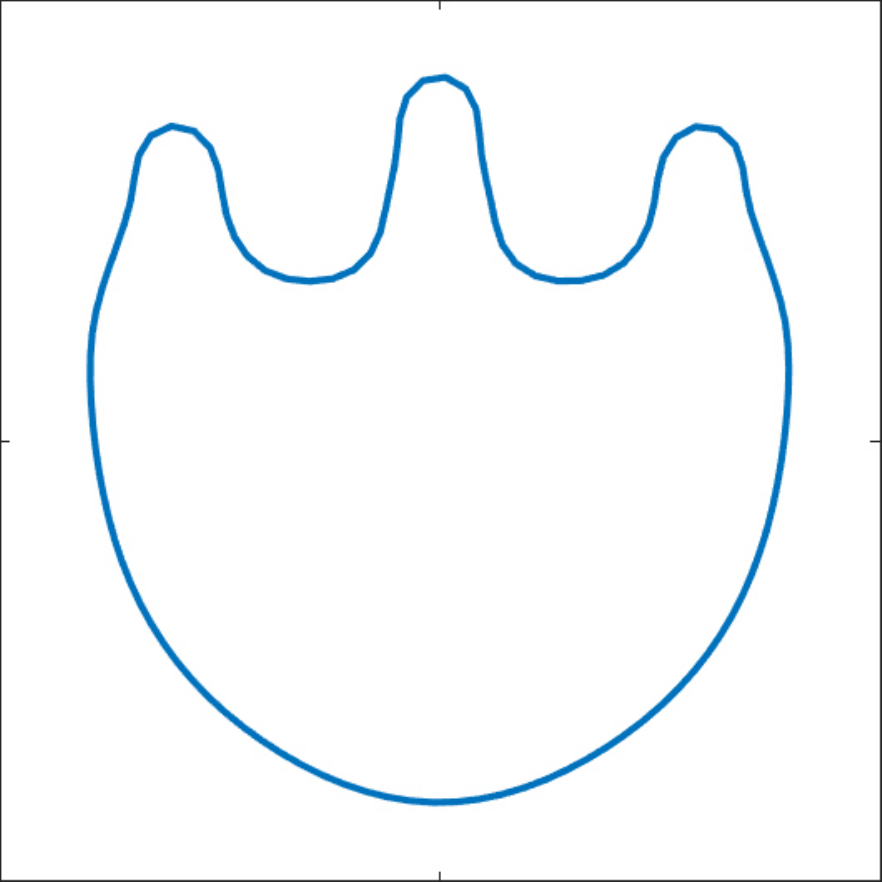}\label{figure: three finger post}}\hspace{0.1\textwidth}
\subfigure[]{\includegraphics[height = 0.23\textheight]{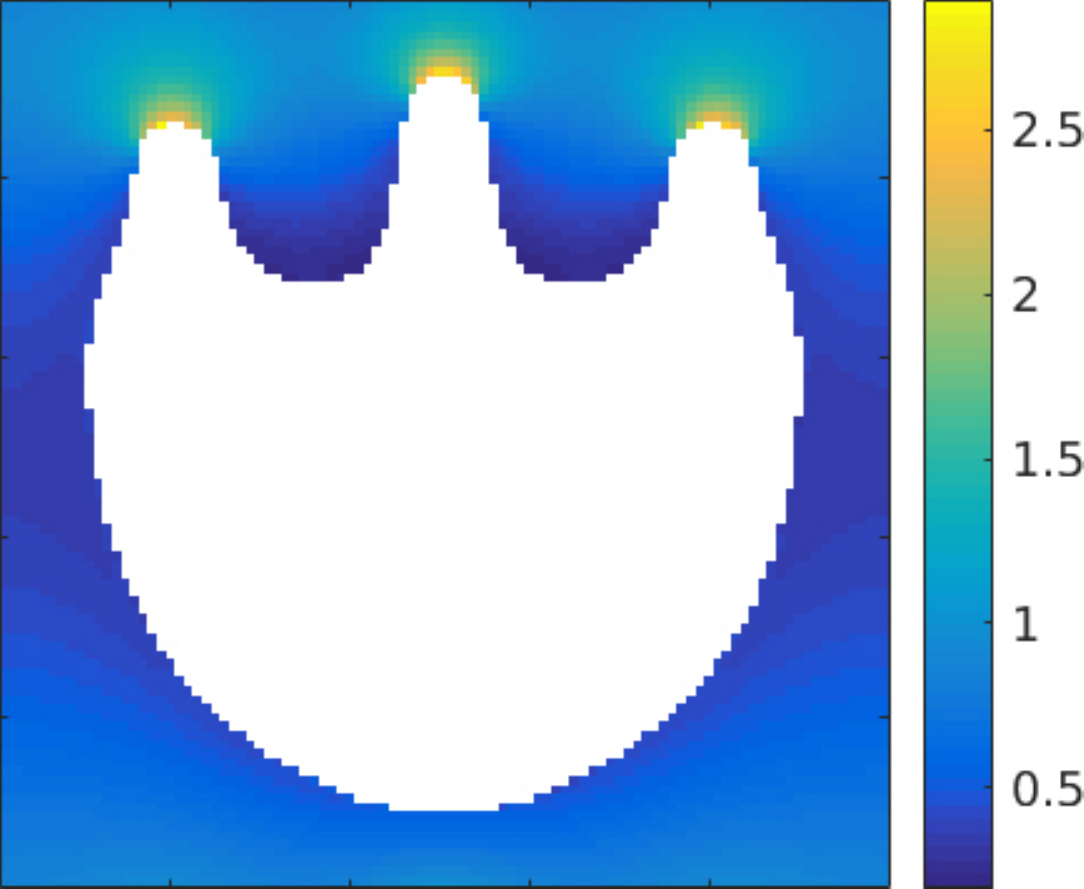}\label{figure: three finger post spatial distribution}}\\
\subfigure[]{\includegraphics[height = 0.23\textheight]{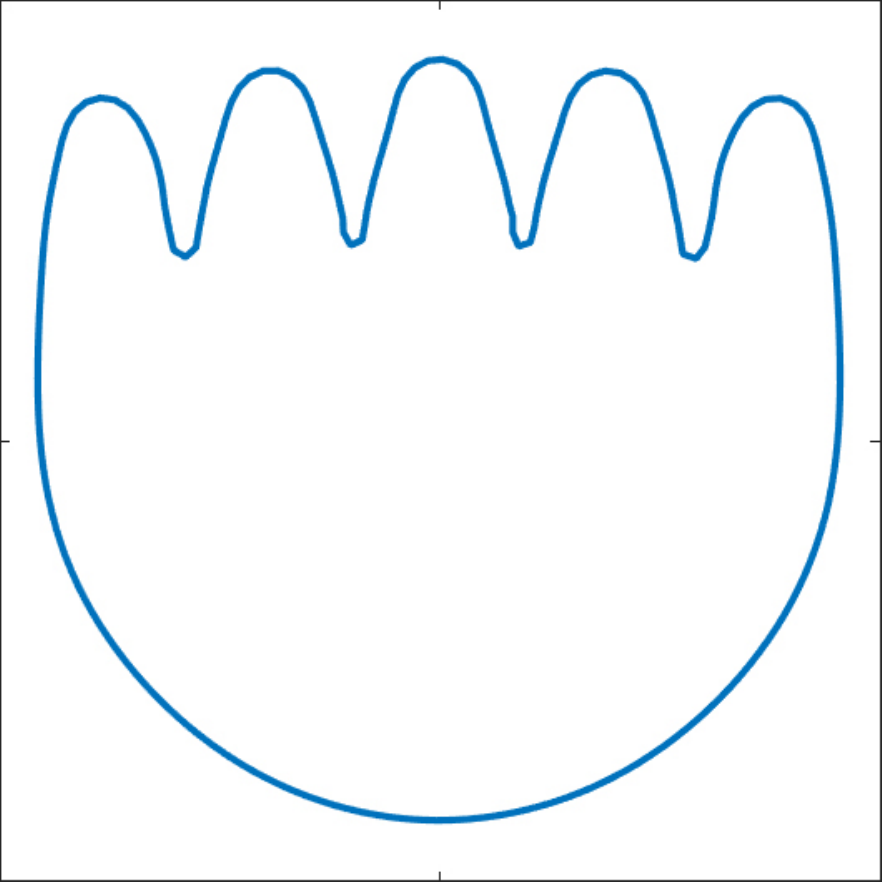}\label{figure: five finger post}}\hspace{0.1\textwidth}
\subfigure[]{\includegraphics[height = 0.23\textheight]{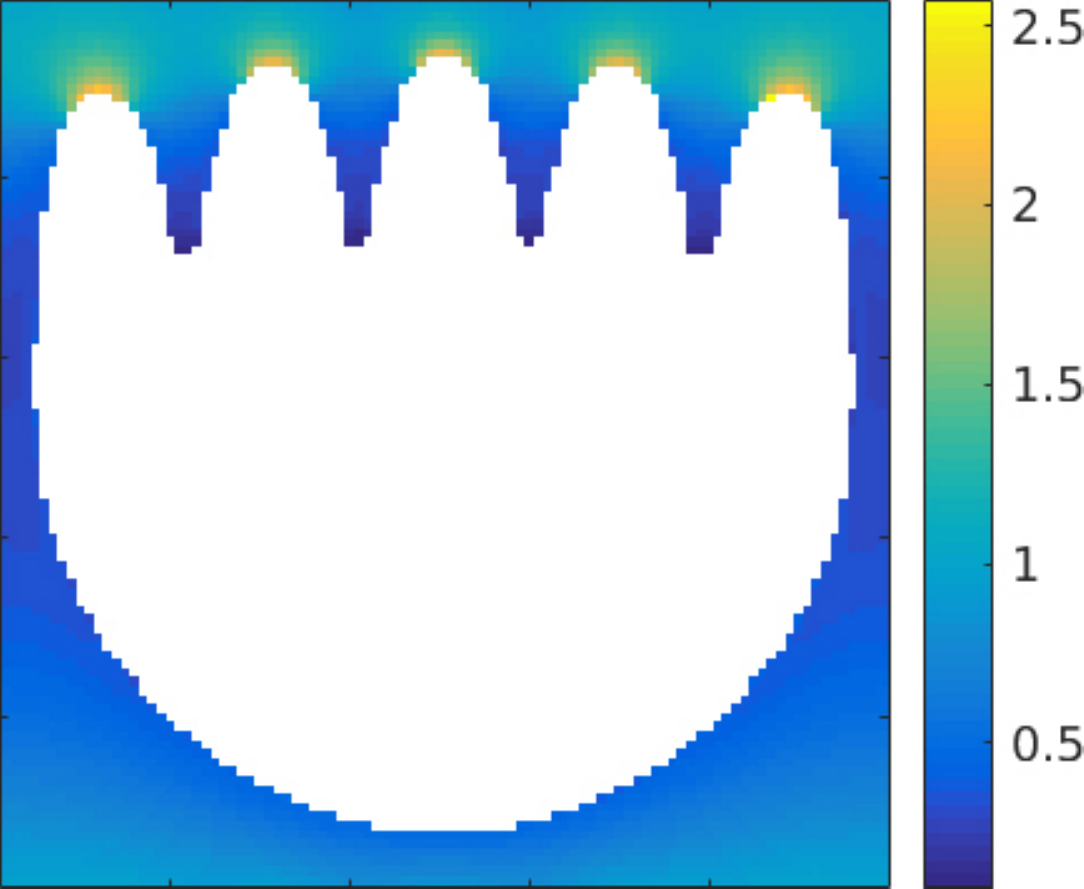}\label{figure: five finger post spatial distribution}}
\caption{Two new candidate post shapes with multiple fingers at the top as strong desorption sites. All fingertips have curvature $\kappa \in [23,26]$ at the top, while the lower halves of the posts are designed to be largely circular arcs. (a-b) The three-finger post and its induced spatial rod concentration. In this case, $E(\Omega)=4.78\times 10^{-2}$. (c-d) The five-finger post and its induced spatial rod concentration. In this case, $E(\Omega) = 6.18\times 10^{-2}$.}
\label{figure: multi-finger candidates}
\end{figure}

In what follows, we shall design better posts based on the above simulations and analysis.
We have seen that the round heads formed in Figure \ref{figure: teardrop shape optimization initial and final} and \ref{figure: nut shape optimization initial and final} act as strong desorption sites, which contribute a lot to increase $E(\Omega)$.
It is natural to believe that $E(\Omega)$ can benefit from putting more strong desorption sites close to the top border of the unit cell.
This inspires us to consider the new posts plotted in Figure \ref{figure: three finger post} and \ref{figure: five finger post}, with multiple fingers at the top.
Indeed, we choose these two shapes, such that the curvature at all fingertips satisfies $\kappa \in [23,26]$, presumably making them into strong desorption sites.
Besides, we take the lower halves of these two shapes to be largely circular arcs.
We compute $E(\Omega)$ for these two posts, without performing shape optimization.
For the three-finger post, $E(\Omega)=4.78\times 10^{-2}$; for the five-finger post, $E(\Omega) = 6.18\times 10^{-2}$, already close to our optimized single-head case.
The spatial concentrations of the rods corresponding to these two candidate posts are also plotted in Figure \ref{figure: three finger post spatial distribution} and Figure \ref{figure: five finger post spatial distribution} respectively.
It is clear that all the finger tips in both shapes are indeed strong desorption sites that are close to the top border of the unit cell.
Efforts are being made to experimentally study the directed migration of swimmers induced by such posts with complex shapes.

Although these multi-finger posts already give strong net flux even without shape optimization, they still cannot compete the final optimized shape we obtained in Figure \ref{figure: teardrop shape optimization initial and final}.
One of the reasons is that there are still lots of open spaces between the post and the border of the unit cell.
In particular, further reducing the vertical gap by enlarging the posts can hopefully lead to stronger fluxes.
Another reason is the sub-optimality of their shapes.
For example, the curvature of the lower half of these two posts are not close to $\kappa_\mathrm{min}$ found before.
We also note that if we change the fingertips into small round heads with curvature close to $\kappa_\mathrm{max}$, the area of the strong desorption site can increase considerably and thus $E(\Omega)$ may increase as well.

\begin{figure}[htbp]
\centering
\includegraphics[height = 0.4\textheight]{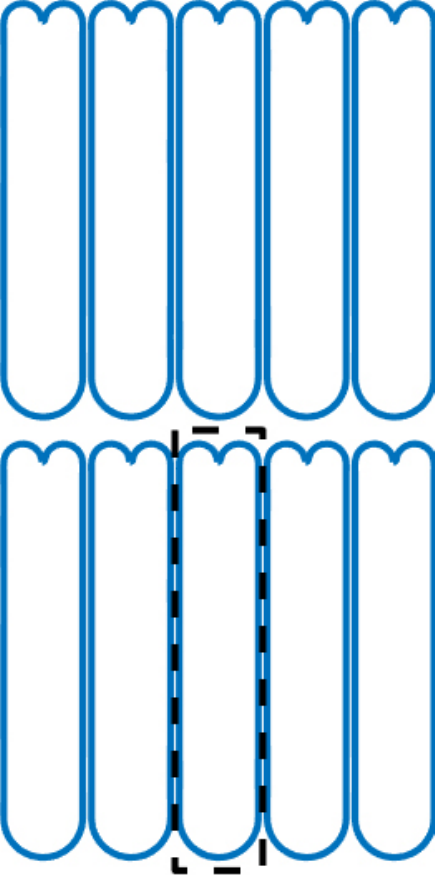}
\caption{A $2\times 5$-array of a slim post. Its unit cell has $a = 0.2$ and $b = 1$. The post has flat sides and curved top and bottom parts. The top part consists of two semi-circles with $\kappa \approx 23$, while the bottom part is a semi-circle with $\kappa \approx 11$. There are only narrow gaps between neighboring posts in the same row. Such an array induces a normalized net flux $E(\Omega) = 1.15\times 10^{-1}$.}
\label{figure: bone-shaped candidate}
\end{figure}

Now we propose a shape that is almost the optimal in the sense that: it leaves very little open space for the free swimming of the rods; and it almost solves the simplified optimization problem on the right hand side of \eqref{eqn: over-simplified optimziation optimize upper and lower halves separately} by putting the right curvatures in the upper and lower halves of $\gamma$.
We plot it in Figure \ref{figure: bone-shaped candidate} using blue curves in a $2\times 5$-array.
The dashed box represents one unit cell of the array, with $a = 0.2$ and $b = 1$.
The post in it has a slim shape, whose height and width are $0.940$ and $0.180$ respectively.
It has flat sides and curved top and bottom parts.
The top part consists of two semi-circular arcs with $\kappa \approx 23$, which is close to $\kappa_\mathrm{max}\approx 25$; the bottom part is a semi-circular arc, with $\kappa \approx 11$, which is close to $\kappa_\mathrm{min}\approx 10.5$.
There are only narrow gaps between neighboring posts.
We compute its normalized net flux to obtain $E(\Omega) = 1.15\times 10^{-1}$, which surpasses any other posts considered in this paper!
Note that we do not rule out the existence of even more judicious designs of posts.
In sum, we showed that using what we have learned in the numerical shape optimization and the formal analysis above, we can design better shapes of posts effectively.

\section{Conclusion and Discussion}\label{section: discussion}
In this paper, we propose a kinetic-type model to study Au-Pt rods swimming and directionally migrating in a periodic array of posts with non-circular cross-sections.
Both position and orientation of the rods are taken into account.
The absorption and desorption of the swimming rods on the post boundaries are modeled via empirically defined rate functions of the boundary curvature, and angular distributions.
Within this model, we define and compute the normalized net flux induced by a periodic rectangular array of posts, which characterizes the intensity of the spontaneous directed migration of rods in the array due to the asymmetry in geometry.
We study how to design the array judiciously so that it can induce stronger directed migration.
It is shown that the net flux increases if the horizontal and vertical spacings between neighboring posts shrink.
On the other hand, we apply the numerical shape optimization to find better shapes of posts that induce yet larger flux.
Inspired by the numerical results on two candidate posts --- a teardrop-shaped post and a nut-shaped post, we propose a simplified model to show the key geometric features a good post should have.
Based on that, we come up with three new candidate shapes that generate large fluxes.
In this way, we show the shape optimization technique can help design good posts effectively.

Our results crucially rely on the choices of (hopefully reasonable) universal rate functions $r_{\mathrm{in}}(\kappa)$ and $r_{\mathrm{out}}(\kappa)$, and the angular distributions $\rho_\pm(\beta)$ and $\tau_\pm(\beta)$, for which little experimental measurement exists.
In this paper, we choose these functions so that they qualitatively agree with the existing experimental observations and physical intuition.
We note that our choices of $\rho_\pm(\beta)$ and $\tau_\pm(\beta)$ are natural and reasonable.
The numerical experience is that, even if we alter the choice of these angular distributions, the numerical results are qualitatively unchanged.
By contrast, $r_{\mathrm{in}}(\kappa)$ and $r_{\mathrm{out}}(\kappa)$ could have bigger impacts on the numerical results.
It has been shown in the simplified model in Section \ref{section: numerical results} that given $r_\mathrm{in}\equiv 1$, the local maximum and minimum of $r_{\mathrm{out}}(\kappa)/\kappa$ can be very crucial quantities that determine geometric features of a good post.
For example, if we alternatively take $r_{\mathrm{out}}(\kappa)$ to be $e^{\kappa}$, one would expect a good post to have very sharp tips at its top instead of small round heads or circular arcs, since $r_{\mathrm{out}}(\kappa)/\kappa$ formally achieves maximum at $\kappa =+\infty$.
If we further choose $r_{\mathrm{in}}(\kappa)$ to be a function depending on $\kappa$, the situation would be more complicated and it can hardly be characterized by our simplified model.

In the current model, the absorption and desorption of the swimming rods on the post boundary are handled in a phenomenological manner, which is the reason we need the functions above.
It would be ideal to build up a hydrodynamic model to fully characterize the interaction between the swimming bimetallic rods and the curved solid boundary, from which we can hopefully derive the rate functions and angular distributions that are needed.
Previous work by Takagi et al.\;\cite{takagi2014hydrodynamic} applying lubrication theory between a swimming rod and a flat solid boundary cannot be immediately generalized to the case of the curved boundary, especially when the radius of curvature of the boundary is on the same scale as or even below the rod length, which is the case when a rod comes to the sharp tip of a teardrop-shaped post.
Spagnolie et al.\;\cite{spagnolie2015geometric} models hydrodynamic capture and escape of microswimmers on an obstacle, by assuming the swimmers to be force dipoles.
However, this analysis assumes that the swimmer can preserve its orientation when it hits the boundary, which is not true in our case.
More delicate modeling is thus needed to understand the hydrodynamic interaction between swimmers and complex boundaries in fluid environments.

\section*{Acknowledgement}
We want to thank our colleagues in Courant Institute and Department of Chemistry of New York University, Dr.\;Megan S.\;Davies Wykes, Dr.\;Xiao Zhong, Prof.\;Leif Ristroph, Prof.\;Jun Zhang, Prof.\;Michael D.\;Ward, Prof.\;Yanpeng Liu, Jinzi Mac Huang, Yang Wu and Dr.\;Abtin Rahimian, for many inspiring discussions and providing useful experimental data. The project is partially supported by NSF under Grant DMS-1463962.

\newpage
\appendix
\section*{Supplementary Materials}

The Supplementary Materials contain all the numerical methods and mathematical derivations omitted in the main body of the paper.
It is organized as follows.
In Section \ref{subsection: numerical method to solve the original coupled system}, we discuss the numerical methods of solving the model \eqref{equation: model in the steady state with drift equation in the bulk}-\eqref{equation: model in the steady state with drift definition of boundary fluxes} for $p$ and $p_B^\pm$ under the normalization condition \eqref{equation: normalization condition} and calculating the corresponding flux $E(\Omega)$.
Section \ref{section: a formal overview of the theory of shape optimization} is a quick overview of the theory of shape optimization, which is the mathematical foundation of the derivation in Section \ref{section: derivation of shape derivative p' and pB'}, \ref{section: shape functional and auxiliary functions} and \ref{section: linear dependence of p' p_B' on Vn}.
In Section \ref{section: derivation of shape derivative p' and pB'}, we derive the equations for the shape derivatives $p'$ and $p_B^\pm{}'$, whose meanings will be clear in Section \ref{section: a formal overview of the theory of shape optimization}.
In Section \ref{section: shape functional and auxiliary functions}, we rewrite the Eulerian derivative of the shape functional $E(\Omega)$ with respect to the perturbation of the shape of the post into an integral on the boundary that is being perturbed, before which we define several auxiliary functions.
Section \ref{section: linear dependence of p' p_B' on Vn} is used to justify the linear dependence of the Eularian derivative on the normal component of the perturbation vector field.
The numerical method to evolve and optimize the boundary curve of the post is discussed in Section \ref{section: numerical method for shape optimization}.

\section{Numerical methods of solving the model \eqref{equation: model in the steady state with drift equation in the bulk}-\eqref{equation: model in the steady state with drift definition of boundary fluxes} and calculating $E(\Omega)$ in the static geometry}\label{subsection: numerical method to solve the original coupled system}

We start with the following formal analysis based on the model \eqref{equation: model in the steady state with drift equation in the bulk}-\eqref{equation: model in the steady state with drift definition of boundary fluxes}.
Given $p_B^\pm(x)$ as functions on $\gamma$, define
\begin{equation}
g(x,\theta) = r_\mathrm{out}(\kappa(x))[2\pi\tau_+(\beta)p_B^+(x)+2\pi\tau_-(\beta)p_B^-(x)],\quad (x,\theta)\in\Gamma.
\label{equation: formula for Robin boundary data in subproblem}
\end{equation}
We denote $g(x,\theta) = g[p_B^+, p_B^-]$, where $g[\cdot]$ is obviously linear in $(p_B^+,p_B^-)$.
Given this $g(x,\theta)$ on $\Gamma$, we solve the following sub-problem adapted from \eqref{equation: model in the steady state with drift equation in the bulk} and \eqref{equation: model in the steady state with drift boundary condition}-\eqref{equation: model in the steady state with drift definition of boundary fluxes},
\begin{align}
&\;D_t\Delta_x \tilde{p}(x,\theta)+D_r\Delta_\theta \tilde{p}(x,\theta) - v_0(\cos 2\pi\theta, \sin 2\pi\theta)^T\cdot\nabla_{x} \tilde{p}(x,\theta) = 0, \quad(x,\theta)\in\Omega,\label{equation: subproblem in solving the whole system equation in the bulk}\\
&\;D_t \frac{\partial \tilde{p}}{\partial n_{\Gamma}}(x,\theta) - v_0\tilde{p}(x,\theta) \cos\beta+ r_\mathrm{in}(\kappa(x))(\rho_+(\beta)+\rho_-(\beta))\tilde{p}(x,\theta) = g(x,\theta),\quad(x,\theta)\in\Gamma,\label{equation: subproblem in solving the whole system boundary condition}\\
&\; \tilde{p}\mbox{ satisfies periodic boundary condition on }\partial \Omega\backslash \Gamma,\label{eqn: periodic boundary condition on the subproblem}
\end{align}
where $\beta = 2\pi\theta-\alpha(x)$.
Suppose the equation is solvable for any $g(x,\theta)$. The solution is denoted by $\tilde{p}(x,\theta)$. The linear map from $g$ to $\tilde{p}|_\Gamma$ is denoted to be $T$, i.e.\;$\tilde{p}|_\Gamma \triangleq Tg$.

With the $\tilde{p}|_\Gamma$ in hand, we solve the following equation for $\tilde{p}_B^\pm$ on $\gamma$ adapted from \eqref{equation: model in the steady state with drift equation on the boundary} and \eqref{equation: model in the steady state with drift definition of boundary fluxes},
\begin{equation}
- D_t\Delta_\gamma \tilde{p}_B^\pm(x) \pm v_0 \partial_\gamma \tilde{p}_B^\pm(x) + r_\mathrm{out}(\kappa(x))\tilde{p}_B^\pm(x)= \int_0^1 r_\mathrm{in}(\kappa(x))\rho_\pm(\beta)\tilde{p}|_\Gamma(x,\theta)\,\mathrm{d}\theta,\quad x\in \gamma.
\label{eqn: the step of solving for p_B's in the numerical method of solving model with fixed geometry}
\end{equation}
Here we used the assumption that $\tau_\pm$ are normalized.
This equation for $p_B^\pm$ is always solvable as long as $r_{\mathrm{out}}(\kappa(x))$ is positive and sufficiently smooth.
We denote that $\tilde{p}_B^\pm = Q^\pm Tg$.
Here $Q^\pm$ are also linear maps.

To this end, we find that
\begin{equation}
(\tilde{p}_B^+, \tilde{p}_B^-) = (Q^+ T g[p_B^+, p_B^-],~ Q^- T g[p_B^+, p_B^-]) \triangleq K[p_B^+, p_B^-],
\label{eqn: defintion of operator K}
\end{equation}
where $K$ maps $[p_B^+, p_B^-]$ linearly to $[\tilde{p}_B^+, \tilde{p}_B^-]$.
Now solving the model \eqref{equation: model in the steady state with drift equation in the bulk}-\eqref{equation: model in the steady state with drift definition of boundary fluxes} reduces to solving the following fixed point problem
\begin{equation}
K[p_B^+, p_B^-] = (p_B^+, p_B^-),
\label{equation: the linear system leading to the solution}
\end{equation}
i.e.\;the desired solution $(p_B^+, p_B^-)$ should be in the kernel of the linear operator $(K-Id)$.
Once such $(p_B^+, p_B^-)$ is found out, $p$ could be solved by first using \eqref{equation: formula for Robin boundary data in subproblem} to calculate $g$, and then solving the subproblem \eqref{equation: subproblem in solving the whole system equation in the bulk}-\eqref{eqn: periodic boundary condition on the subproblem}.
We may have to normalize $p$ and $p_B^\pm$ so that \eqref{equation: normalization condition} is satisfied.
The normalized net flux $E(\Omega)$ is then calculated by \eqref{eqn: definition of F_Omega}.

\begin{remark}\label{rmk: existence and uniqueness of positive solution of the coupled system}
There exists a unique positive smooth solution $(p,p_B^+, p_B^-)$ of the model \eqref{equation: model in the steady state with drift equation in the bulk}-\eqref{equation: model in the steady state with drift definition of boundary fluxes} satisfying \eqref{equation: normalization condition}, provided that
\begin{enumerate}
\item $\gamma$ is sufficiently smooth, and thus $\alpha(x)$ and $\kappa(x)$ are sufficiently smooth in $x\in\gamma$;
\item $r_\mathrm{in}(\kappa)> 0$, $r_\mathrm{out}(\kappa)> 0$, $\rho_\pm(\beta)$ and $\tau_\pm(\beta)$, where $\kappa = \kappa(x)$ and $\beta = 2\pi\theta-\alpha(x)$, are sufficiently smooth in $(x,\theta)\in\Gamma$;
\item and in addition, $r_\mathrm{in}(\kappa(x))(\rho_+(\beta)+\rho_-(\beta))-v_0\cos\beta \geq 0$ but not identically zero.
\end{enumerate}
This could be proved by applying elliptic regularity theory to study the operator $K$ and using the Schauder fixed point theorem \cite{gilbarg2015elliptic}. We shall omit the proof.
In what follows, we always assume $(K-Id)$ has a one-dimensional kernel.
\end{remark}

\begin{remark}\label{rmk: r_out r_out is in the orthogonal compliment of K-Id}
It could be shown that $(r_\mathrm{out}(x),r_\mathrm{out}(x))$ is orthogonal to the image of $(K-Id)$ in the inner product on $L^2(\gamma)\times L^2(\gamma)$.
Indeed, by \eqref{equation: formula for Robin boundary data in subproblem}-\eqref{eqn: the step of solving for p_B's in the numerical method of solving model with fixed geometry},
\begin{equation}
\begin{split}
\int_\gamma r_\mathrm{out} (\tilde{p}_B^+ + \tilde{p}_B^-) =&\; \int_\Gamma D_t\Delta_\gamma (\tilde{p}_B^++\tilde{p}_B^-) -v_0 \partial_\gamma (\tilde{p}_B^+-\tilde{p}_B^-)+r_\mathrm{in}(\rho_+ + \rho_-)\tilde{p}\\
 =&\; \int_\Gamma r_\mathrm{in}(\rho_+ + \rho_-)\tilde{p}=\int_\Gamma g(x,\theta) - D_t\frac{\partial \tilde{p}}{\partial n_\Gamma} + v_0 \tilde{p}\cos\beta\\
=&\;\int_\Gamma g(x,\theta) - \int_\Omega D_t\Delta_x \tilde{p}+D_r\Delta_\theta \tilde{p} - v_0(\cos 2\pi\theta, \sin 2\pi\theta)^T\cdot\nabla_{x} \tilde{p}\\
=&\;\int_\Gamma r_{\mathrm{out}}(2\pi\tau_+p_B^+ +2\pi\tau_-p_B^-)=\int_\gamma r_{\mathrm{out}}(p_B^+ +p_B^-).
\end{split}
\label{eqn: proof of r_out r_out is orthogonal to the image of K-Id}
\end{equation}
Therefore, we find for $\forall\,(p_B^+, p_B^-)$,
\begin{equation*}
\langle(r_\mathrm{out}, r_\mathrm{out}), (K-Id)[p_B^+, p_B^-]\rangle = 0,
\end{equation*}
where $\langle\cdot,\cdot\rangle$ denotes the inner product on $L^2(\gamma)\times L^2(\gamma)$.
This property is preserved, up to small numerical error, when \eqref{equation: formula for Robin boundary data in subproblem}-\eqref{eqn: the step of solving for p_B's in the numerical method of solving model with fixed geometry} are discretized and solved numerically.
This indicates that the matrix representation of $(K-Id)$ in the discretized case should have one (and exactly one due to the assumption in Remark \ref{rmk: existence and uniqueness of positive solution of the coupled system}) zero singular value, up to numerical error.
\end{remark}

Based on the above analysis, the numerical method to solve the model should go as follows.
It aims at forming a discrete representation of the operator $(K-Id)$, finding out its null space, and calculating $E(\Omega)$.
\begin{enumerate}
  \item With abuse of notations $x_1$ and $x_2$, we represent $\gamma$ by $N$ points $\{x_1,\cdots, x_{N}\}\triangleq X\subset\gamma$, which are equally aligned on $\gamma$ counterclockwise with spacing $\Delta s$, i.e.\;the arclength of $\gamma$ between two neighboring points are assumed to be $\Delta s$.
  Here $N$ is assumed to be even.
  We use central difference scheme to evaluate $\alpha(x_j)$, $n(x_j)$ and $\kappa(x_j)$ for $j = 1,2,\cdots,N$, i.e.\;take $\alpha(x_j)$ such that
  \begin{equation}
  (\cos\alpha(x_j),\sin\alpha(x_j)) = \frac{x_{j+1}-x_{j-1}}{|x_{j+1}-x_{j-1}|}.
  \label{eqn: central difference scheme to calculate alpha}
  \end{equation}
  where the subscripts are understood in a cyclic fashion. Then
  \begin{align}
  n(x_j) = &\;(-\sin\alpha(x_j), \cos\alpha(x_j)),  \label{eqn: central difference scheme to calculate n}\\
  \kappa(x_j) = &\;\frac{1}{2\Delta s}[(\cos\alpha(x_{j+1}),\sin\alpha(x_{j+1}))-(\cos\alpha(x_{j-1}),\sin\alpha(x_{j-1}))]\cdot n(x_j).  \label{eqn: central difference scheme to calculate kappa}
  \end{align}
  In the $\theta$-direction, $[0,1]$ is discretized evenly using $M$ points $\theta_m = \frac{2m-1}{2M}$ with $m = 1,2,\cdots,M$.
  Hence, the grid on $\Gamma$ is given by
  \begin{equation}
  X_\Gamma \triangleq X\times\{\theta_1,\theta_2,\cdots,\theta_M\} = \{(x_j,\theta_m):\;j = 1,2,\cdots,N,\;m = 1,2,\cdots,M\}.
  \label{eqn: defintion of X_Gamma}
  \end{equation}
  \item\label{numerical scheme step: set the initial values of p_Bs}
  Let $\{\mathbf{u}_1, \cdots,\mathbf{u}_{2N}\}$ be an orthogonal basis of $\mathds{R}^{2N}$; $\mathbf{u}_i$'s are understood as column vectors.
  Denote $U = (\mathbf{u}_1, \cdots,\mathbf{u}_{2N})$.
  We shall use Fourier modes to generate $\mathbf{u}_i$'s.
  To be more precise, let $U_0$ be an $N\times N$-matrix, whose $(i,j)$-entry is given by
  \begin{align}
  U_{0,ij} = &\;\sin(2\pi ij/N),\quad i\in\{1,\cdots, N/2-1\},j\in \{1,\cdots, N\},\label{eqn: definition of U_0 sin modes}\\
  U_{0,ij} = &\;\cos(2\pi (i-N/2)j/N),\quad i\in\{N/2,\cdots, N\},j\in \{1,\cdots, N\}.\label{eqn: definition of U_0 cos modes}
  \end{align}
  Note that the rows of $U_0$ form an orthgonal basis of $\mathds{R}^N$.
  Then we define
  \begin{equation*}
  U = \left(
  \begin{array}{cc}
  U_0^T&0\\
  0&U_0^T
  \end{array}
  \right)_{2N\times 2N}= (\mathbf{u}_1, \cdots,\mathbf{u}_{2N}).
  \end{equation*}
  $\{\mathbf{u}_1, \cdots,\mathbf{u}_{2N}\}$ form a basis of all possible vectorial representations of $(p_B^+,p_B^-)^T$ on the grid points $x_n$'s.
  \item\label{numerical scheme step: solve the equations for p}
  Put
  \begin{equation}
  (p_{B,i}^+(x_1),\cdots, p_{B,i}^+(x_{N}), p_{B,i}^-(x_1),\cdots, p_{B,i}^-(x_{N}))^T = \mathbf{u}_i.
  \label{eqn: assign values of p_B when solving the model}
  \end{equation}
  For $j = 1,2,\cdots,N$ and $m = 1,2,\cdots,M$, calculate 
  \begin{equation*}
  g_i(x_j,\theta_m) = r_\mathrm{out}(\kappa(x_j))[2\pi\tau_+(\beta(x_j,\theta_m))p_{B,i}^+(x_j)+2\pi\tau_-(\beta(x_j,\theta_m))p_{B,i}^-(x_j)],
  \end{equation*}
  where $\beta(x_j,\theta_m)= \alpha(x_j)-2\pi\theta_m$.
%
  With $g_i$ well-defined on $X_\Gamma$, we use the commercial finite element package COMSOL Multiphysics \cite{multiphysics4livelink} to solve the following subproblem in dimension three.
    \begin{align*}
    &\;D_t\Delta_x \tilde{p}_i(x,\theta)+D_r\Delta_\theta \tilde{p}_i(x,\theta) - v_0(\cos 2\pi\theta, \sin 2\pi\theta)^T\cdot\nabla_{x} \tilde{p}_i(x,\theta) = 0, \quad(x,\theta)\in\Omega,\\
    &\;D_t \frac{\partial \tilde{p}_i}{\partial n_{\Gamma}}(x,\theta) - v_0\tilde{p}_i(x,\theta) \cos\beta+ r_\mathrm{in}(\kappa(x))(\rho_+(\beta)+\rho_-(\beta))\tilde{p}_i(x,\theta) = g_i(x,\theta),\quad(x,\theta)\in\Gamma,\\
    &\; \tilde{p}_i\mbox{ satisfies periodic boundary condition on }\partial \Omega\backslash \Gamma.
    \end{align*}
  We obtain the Dirichlet boundary data of $\tilde{p}_i$ on $X_\Gamma$.

  \item\label{numerical scheme step: solve the equations for p_B's}
  Now consider the equation for $\tilde{p}_{B,i}^\pm$
  \begin{equation}
  - D_t\Delta_\gamma \tilde{p}_{B,i}^\pm(x) \pm v_0 \partial_\gamma \tilde{p}_{B,i}^\pm(x) + r_\mathrm{out}(\kappa(x))\tilde{p}_{B,i}^\pm(x)= \int_0^1 r_\mathrm{in}(\kappa(x))\rho_\pm(\beta)\tilde{p}_i|_\Gamma(x,\theta)\,\mathrm{d}\theta,\quad x\in \gamma.
  \label{eqn: p_Bi equations in solving the coupled system}
  \end{equation}
  Here the integral is evaluated using the trapezoidal rule:
  \begin{equation}
  \left.\left[\int_0^1 r_\mathrm{in}(\kappa(x))\rho_\pm(\beta)\tilde{p}_i|_\Gamma(x,\theta)\,\mathrm{d}\theta\right]\right|_{x=x_j} = \frac{1}{M}\sum_{m = 1}^M r_\mathrm{in}(\kappa(x_j))\rho_\pm(\beta(x_j,\theta_m))\tilde{p}_i(x_j,\theta_m).
  \label{eqn: trapezoidal rule}
  \end{equation}
  By virtue of the periodic boundary condition in the $\theta$-direction, the trapezoidal rule can achieve high accuracy even when $M$ is not very large.
  Then we solve \eqref{eqn: p_Bi equations in solving the coupled system} for $\tilde{p}_{B,i}^\pm$ using the usual finite difference scheme.
  Denote
  \begin{equation*}
  (\tilde{p}_{B,i}^+(x_1),\cdots, \tilde{p}_{B,i}^+(x_{N}), \tilde{p}_{B,i}^-(x_1),\cdots, \tilde{p}_{B,i}^-(x_{N}))^T \triangleq  \mathbf{w}_i\in\mathds{R}^{2N}.
  \end{equation*}
   In this way, we obtain $\mathbf{w}_i = K\mathbf{u}_i$, where $K$, with abuse of notations, is a discrete representation of the operator $K$ defined in \eqref{eqn: defintion of operator K}. It is not explicitly defined at this moment.
  \item Going through Step \ref{numerical scheme step: solve the equations for p} and \ref{numerical scheme step: solve the equations for p_B's} for $i = 1,2,\cdots, 2N$, we obtain a $2N\times 2N$ matrix
      \begin{equation*}
      W\triangleq(\mathbf{w}_1,\cdots,\mathbf{w}_{2N}) = KU.
      \end{equation*}
      Now it suffices to find out an eigenvector of $K-I_{2N}$ corresponding to zero eigenvalue.
      Assume it can be represented by $U\xi$ for some $\xi\in\mathds{R}^{2N}$.
      Hence, $(W-U)\xi = (K-I_{2N})U\xi = 0$.
      By singular value decomposition, we can find out the zero eigenvector of $W-U$, or equivalently the right singular vector corresponding to zero singular value, denoted by $\xi_*$.
      Then values of the unnormalized solution $p_B^\pm$ on $X$ are given by
      \begin{equation}
      (p_B^+(x_1),\cdots, p_B^+(x_{N}), p_B^-(x_1),\cdots, p_B^-(x_{N}))^T = U\xi_*.
      \label{eqn: unnormalized solution p_B's}
      \end{equation}
  \item We go through Step \ref{numerical scheme step: solve the equations for p} again with \eqref{eqn: assign values of p_B when solving the model} replaced by \eqref{eqn: unnormalized solution p_B's} and $(\tilde{p}_i,p_{B,i}^\pm)$ replaced by $(p,p_B^\pm)$ to find out the (unnormalized) solution $p$ in $\Omega$.
      Normalize $p$ and $p_B^\pm$ such that \eqref{equation: normalization condition} is satisfied.
      Finally, we calculate $E(\Omega)$ by \eqref{eqn: definition of F_Omega} using trapezoidal rule.
      Note that we can obtain all the necessary boundary data from COMSOL and COMSOL can calculate $\int_\Omega p\,dxd\theta$ automatically.
\end{enumerate}

\begin{remark}
In practice, the matrix $W-U$ does not necessarily have a zero singular value due to numerical errors. 
In typical numerical experiments, we always find that one of the singular values of $W-U$ is very close to $0$, while the rest are all of order $1$.
Hence, we always take $\xi_*$ to be the right singular vector corresponding to the smallest singular value, i.e.
\begin{equation*}
\xi_* = \argmin_{\xi\in\mathds{R}^{2N}, |\xi| = 1}|(W-U)\xi|.
\end{equation*}
\end{remark}

\section{A formal overview of the theory of shape optimization}\label{section: a formal overview of the theory of shape optimization}
In this section, we give a formal overview of the theory of shape optimization, which is the technique used in Section \ref{section: shape optimization} to search for a better shape of post.
We use the notations in Section \ref{section: modeling}.

Let $V(t,x)$ be a two-dimensional vector field defined for $(t,x)\in[0,\varepsilon)\times \mathds{R}^2$ for some $\varepsilon>0$;
it is assumed that $V(t,x)$ is smooth in all coordinates and is compactly supported in space in a neighborhood of $\gamma$ in the interior of $Y$.
Then $\tilde{V}(t,x,\theta)^T = (V(t,x)^T,0)$ is its natural three-dimensional extension, defined for $(t,x,\theta)\in[0,\varepsilon)\times \mathds{R}^2\times\mathds{R}$. 
$V$ and $\tilde{V}$ are referred to be perturbation vector fields, under which the domain is going to be changed.
Let $T_t$ be the transformation associated with $V(t,x)$.
To be more precise, for $\forall\, x\in \mathds{R}^2$, we solve
\begin{equation*}
\frac{d T(t,x)}{dt} = V(t,T(t,x)),\quad T(0,x) = x,
\end{equation*}
and define $T_t: x\mapsto T(t,x)$.
With abuse of the notation $\varepsilon$, it is known that $\{T_t\}_{t\in[0,\varepsilon)}$ is a family of smooth diffeomorphisms of $\mathds{R}^2$ for some $\varepsilon>0$ \cite[\S2.9]{sokolowski1992introduction}.
In particular, for $t\in[0,\varepsilon)$, $T_t$ is a one-to-one transformation from $Y$ to itself with its inverse well-defined and smooth, which is called a \emph{flow map}.
For an arbitrary domain $\omega\subset \mathds{R}^2$ with smooth boundary, let $\omega_t$ be the image of $\omega$ under $T_t$.
By assumption, $\omega_0 = \omega$ and $\omega_t \subset Y$.
Also let $\gamma_t = T_t(\gamma)$.
Let $\tilde{T}_t$ is the flow map in 3-D associated with $\tilde{V}$.
It is easy to show that $\Omega_t \triangleq \omega_t\times[0,1]= \tilde{T}_t(\Omega)$.
Similarly, define $\Gamma_t = \tilde{T}_t(\Gamma)$.
In what follows, we shall see how integral functionals depending on $\Omega_t$ and/or $\Gamma_t$ vary as $t$ changes.

Suppose $\{u(\Omega_t)\}_{t\in[0,\varepsilon)}$ is a family of functions defined on $\Omega_t$ and determined by $\Omega_t$.
The limit
\begin{equation*}
\dot{u}(\Omega,\tilde{V})(x,\theta) \triangleq \lim_{t\rightarrow 0^+}\frac{u(\Omega_t)(\tilde{T}_t(x,\theta))-u(\Omega)(x,\theta)}{t},\quad (x,\theta)\in \Omega,
\label{eqn: formal defintion of material derivative in the bulk}
\end{equation*}
if well-defined, is called the \emph{material derivative} of $u$ in $\Omega$ with respect to $\tilde{V}$ {\cite[\S2.25]{sokolowski1992introduction}}.
The \emph{shape derivative} of $u$ in $\Omega$ with respect to $\tilde{V}$ is defined to be {\cite[\S2.30]{sokolowski1992introduction}}
\begin{equation*}
u'(\Omega,\tilde{V})(x,\theta) \triangleq \dot{u}(\Omega,\tilde{V})(x,\theta) - \nabla( u(\Omega))(x,\theta)\cdot \tilde{V}(0,x,\theta).
\label{eqn: formal defintion of shape derivative in the bulk}
\end{equation*}
In other words, the material derivative characterizes the rate of change of $u(\Omega_t)$ on a particle drifted by $\tilde{V}$, while the shape derivative is the rate of change of $u(\Omega_t)$ by canceling the convective effect of $\tilde{T}_t$.
In the sequel, we shall always write material and shape derivatives of $u$ as $\dot{u}$ and $u'$ respectively.
The usual derivatives will be denoted in a more explicit way, e.g.\;$\frac{d}{dt}$.

Let $\{z(\Gamma_t)\}_{t\in[0,\varepsilon)}$ be a family of functions defined on $\Gamma_t$ and determined by $\Gamma_t$.
Note that $\Gamma$ is a submanifold in $\mathds{R}^3$ with co-dimension $1$.
The material derivative of $z$ with respect to $\tilde{V}$ is defined to be
\begin{equation*}
\dot{z}(\Gamma,\tilde{V})(x,\theta) \triangleq \lim_{t\rightarrow 0^+}\frac{z(\Gamma_t)(\tilde{T}_t(x,\theta))-z(\Gamma)(x,\theta)}{t},\quad (x,\theta)\in \Gamma,
\label{eqn: formal defintion of material derivative on the surface}
\end{equation*}
and the shape derivative is defined to be
\begin{equation*}
z'(\Gamma,\tilde{V})(x,\theta) \triangleq \dot{z}(\Gamma,\tilde{V})(x,\theta) - \nabla_\Gamma z(\Gamma)(x,\theta)\cdot \tilde{V}(0,x,\theta).
\label{eqn: formal defintion of shape derivative on the surface}
\end{equation*}
Here \cite[\S2.19]{sokolowski1992introduction}
\begin{equation*}
\nabla_\Gamma z = \nabla \tilde{z}|_{\Gamma} - \frac{\partial \tilde{z}}{\partial n_{\Gamma}}n_{\Gamma},
\end{equation*}
where $\tilde{z}$ is an arbitrary extension of $z$ from $\Gamma$ to its ambient space, and $n_{\Gamma}$ is the outer unit normal vector on $\Gamma$ with respect to $\Omega$ defined as before.
For conciseness, in what follows, we shall always denote both $\tilde{V}(0,x,\theta)$ and its trace on $\Gamma$ by $\tilde{V}(0)$.
\begin{remark}
For any family of functions $u(\Omega_t)$ defined in $\Omega_t$ and determined by $\Omega_t$, if its material and shape derivatives are well-defined, then $(u|_\Gamma)\dot{} = \dot{u}|_\Gamma$.
Here $(u|_\Gamma)\dot{}$ is the material derivative of $\{u(\Omega_t)|_{\Gamma_t}\}_{t\in[0,\varepsilon)}$, treated as a family of functions defined on $\Gamma_t$ and determined by $\Gamma_t$, while $\dot{u}|_\Gamma$ is the restriction of $\dot{u}$ on $\Gamma$.
However,
\begin{equation*}
(u|_\Gamma)' = u'|_\Gamma + \frac{\partial u}{\partial n_\Gamma} (\tilde{V}(0)\cdot n_\Gamma).
\end{equation*}
Here $(u|_\Gamma)'$ is the shape derivative of $\{u(\Omega_t)|_{\Gamma_t}\}_{t\in[0,\varepsilon)}$ as a family of functions defined on $\Gamma_t$ and determined by $\Gamma_t$, while $u'|_\Gamma$ is the restriction of $u'$ on $\Gamma$, also written as $u'$ for simplicity.
We shall always use these notations to distinguish the two different quantities.
\end{remark}

The following two formulae are useful in calculating Eularian (or directional) derivatives of integral functionals with respect to $\tilde{V}$.
\begin{enumerate}
\item Suppose $u(\Omega_t)$ is a function on $\Omega_t$ and determined by $\Omega_t$. Then the derivative of the domain integral
\begin{equation*}
J_1(\Omega_t) = \int_{\Omega_t} u (\Omega_t)\, dxd\theta
\end{equation*}
with respect to the vector field $\tilde{V}$ is that \cite[(2.168)]{sokolowski1992introduction}
\begin{equation}
dJ_1(\Omega; \tilde{V}) \triangleq \left.\frac{dJ_1(\Omega_t)}{dt}\right|_{t = 0} = \int_{\Omega} u'(\Omega;\tilde{V})\,dxd\theta +\int_{\Gamma} u(\Omega) (\tilde{V}(0)\cdot n_{\Gamma})\,d\Gamma.
\label{eqn: derivative of domain integral}
\end{equation}
Here we used the fact that $\tilde{V}(t,\cdot,\theta)$ is supported in the interior of $Y$ and has zero $\theta$-component everywhere.
Note that only $\Gamma$ comes into the second integral but no other parts of $\partial\Omega$.

\item Suppose $z = z(\Gamma_t)$ is a function defined on $\Gamma_t$ and determined by $\Gamma_t$.
Then the derivative of the boundary integral
\begin{equation*}
J_2(\Gamma_t) = \int_{\Gamma_t} z(\Gamma_t)\,d\Gamma_t
\end{equation*}
with respect to the vector field $\tilde{V}$ is that \cite[(2.173) and (2.174)]{sokolowski1992introduction}
\begin{equation}
dJ_2(\Gamma; \tilde{V})\triangleq \left.\frac{dJ_2(\Gamma_t)}{dt}\right|_{t = 0}  = \int_\Gamma z'(\Gamma;\tilde{V})-\kappa z(\Gamma)(\tilde{V}(0)\cdot n_{\Gamma})\,d\Gamma,
\label{eqn: derivative of boundary integral pre-form}
\end{equation}
where $\kappa$ is the sum of principal curvatures of $\Gamma$, or equivalently, the curvature of $\gamma$ in our context.
In particular, if $z(\Gamma_t)=u(\Omega_t)|_{\Gamma_t}$, where $u(\Omega_t)$ is some function defined on $\Omega_t$ and determined by $\Omega_t$,
\begin{equation}
z'(\Gamma;\tilde{V}) = u'(\Omega;\tilde{V})|_\Gamma+\frac{\partial u}{\partial n_\Gamma}(\Omega) (\tilde{V}(0)\cdot n_{\Gamma}).
\label{eqn: derivative of boundary integral}
\end{equation}
Hence,
\begin{equation}
dJ_2(\Omega; \tilde{V}) = \int_\Gamma u'(\Omega;\tilde{V})|_\Gamma+\left(\frac{\partial u}{\partial n_\Gamma}(\Omega)-\kappa u(\Gamma)\right)\tilde{V}(0)\cdot n_{\Gamma}\,d\Gamma.
\label{eqn: Eulerian derivative of an integral shape functional on a submanifold}
\end{equation}
\end{enumerate}

The following theorem states that the Eulerian derivative with respect to vector field $\tilde{V}$ of a functional in a smooth domain $\Omega$ can always have an integral representation involving only the normal component of $\tilde{V}(0)$ on $\partial\Omega$.
\begin{theorem}[The Hadamard Formula {\cite[Theorem 2.27]{sokolowski1992introduction}}]\label{thm: Hadamard formula}
Let $J$ be a real-valued shape functional defined on a smooth domain $\Omega$, which is shape differentiable.
There exists a scalar distribution $g_J(\partial\Omega)\in (C^\infty(\partial\Omega))'$ on $\partial\Omega$, such that for $\forall\,\tilde{V}\in C^\infty([0,\varepsilon)\times\Omega)$,
\begin{equation*}
dJ(\Omega;\tilde{V}) = \langle g_J(\partial\Omega), \tilde{V}(0)|_{\partial\Omega}\cdot n_{\partial\Omega}\rangle_{\partial\Omega},
\label{eqn: structure theorem integral representation}
\end{equation*}
where $\tilde{V}(0)|_{\partial\Omega}$ is the trace of $\tilde{V}(0)$ on $\partial\Omega$; $\langle\cdot,\cdot\rangle_{\partial\Omega}$ is the pairing of distribution and smooth functions on $\partial\Omega$; and $n_{\partial\Omega}$ is the unit outer normal vector to $\Omega$. If $g(\partial\Omega)\in L^1(\partial\Omega)$,
\begin{equation}
dJ(\Omega;\tilde{V}) = \int_{\partial\Omega} g_J(\partial\Omega) (\tilde{V}(0)|_{\partial\Omega}\cdot n_{\partial\Omega})\,dA.
\label{eqn: structure theorem integral representation if g is smooth}
\end{equation}
\end{theorem}

\begin{remark}\label{rmk: shape gradient as a measure on gamma instead of on Gamma}
By assumption, $\tilde{V}$ has zero $\theta$-component everywhere and is only supported in a neighborhood of $\Gamma$ in the interior of $Y\times \mathds{R}$.
Hence, \eqref{eqn: structure theorem integral representation if g is smooth} could be rewritten as
\begin{equation*}
dJ(\Omega;\tilde{V}) = \int_{\Gamma} g_J(\partial\Omega) (\tilde{V}(0)\cdot n_{\Gamma})\,d\Gamma = \int_{\gamma} d\gamma\,(V(0)\cdot n_{\gamma}) \int_0^1 d\theta\,g_J(\partial\Omega)|_\Gamma,
\label{eqn: structure theorem integral representation only integral on gamma}
\end{equation*}
where $V(0) = V(0,\cdot)|_{\gamma}$.
Here $g(\partial\Omega)|_\Gamma$ is understood as a function of $(x,\theta) \in \gamma\times [0,1]$, and thus $\int_0^1 d\theta\,g_J(\partial\Omega)|_\Gamma$ is a function in $x\in\gamma$.
\end{remark}

Formally, the shape functional $J$ increases fastest if $\gamma$ evolves in the direction given by $V(0) = n_\gamma\cdot\int_0^1 d\theta\,g_J(\partial\Omega)|_\Gamma$.
In the context of multi-variable calculus, this corresponds exactly to the steepest ascent method in maximizing a smooth multi-variable function $f$.
Hence, in what follows, we shall find out the representation of $\int_0^1 d\theta\,g_E(\partial\Omega)|_\Gamma$, associated to the shape functional $E(\Omega)$, in terms of the shape derivatives $p'$ and $p_B^\pm{}'$, the given functions $r_{\mathrm{in}}(\kappa)$, $r_{\mathrm{out}}(\kappa)$, $\rho_\pm(\beta)$ and $\tau_\pm(\beta)$, and other auxiliary functions.
By letting $\gamma$ evolve by the steepest ascent method, we can find a better shape of $\omega$ (or equivalently $\Omega = \omega\times[0,1]$) that induces larger $E(\Omega)$.

\section{Derivation of equations for the shape derivatives $p'$ and $p_B^\pm{}'$}\label{section: derivation of shape derivative p' and pB'}
Let $\omega_t$, $\Omega_t$, $\gamma_t$ and $\Gamma_t$ be defined as before.
Denote $(p^t,p_B^{+,t},p_B^{-,t})$ to be the normalized solution of the equations \eqref{equation: model in the steady state with drift equation in the bulk}-\eqref{equation: normalization condition} with $\Omega$ and $\gamma$ replaced by $\Omega_t$ and $\gamma_t$ respectively.

\subsection{Equations for $p'$}
We start from the equation for $p^t$ in $\Omega^t$.
\begin{equation}
 \begin{split}
   &\; \mathrm{div}(D\nabla p^t(x,\theta)) - v_0(\cos 2\pi\theta, \sin 2\pi\theta,0)\cdot\nabla p^t(x,\theta) = 0,\quad(x,\theta)\in\Omega_t,\\
   &\; D_t\frac{\partial p^t}{\partial n_{\Gamma_t}}(x,\theta) - v_0 p^t(x,\theta)\cos\beta^t+ r_{\mathrm{in}}(\kappa^t(x))p^t(x,\theta)[\rho_+(\beta^t)+\rho_-(\beta^t)] = g^t(x,\theta),\quad(x,\theta)\in\Gamma_t,\\
   &\; g^t(x,\theta) = r_{\mathrm{out}} (\kappa^t(x))[2\pi\tau_+(\beta^t)p_B^{+,t}(x) +2\pi\tau_-(\beta^t)p_B^{-,t}(x)],\\
   &\;\beta^t = 2\pi\theta - \alpha^t(x),\\
   &\; p^t\mbox{ satisfies periodic boundary condition on }\partial \Omega_t\backslash \Gamma_t.
 \end{split}
 \label{eqn: equation for p}
\end{equation}
where $D = \mathrm{diag}(D_t, D_t, D_r)$, and $\beta^t = 2\pi\theta -\alpha^t(x)$. $\kappa^t(x)$ and $\alpha^t(x)$ are the curvature and the angle of the unit outer normal vector of $\gamma_t$, respectively.

\begin{definition}
$$
C^\infty_\#(Y\times[0,1]) = \{u|_{Y\times[0,1]}:\;u\in C^\infty(\mathds{R}^3),\;u\mbox{ is }Y\times[0,1]\mbox{-periodic}\}.
$$
Recall that $Y = \left[-\frac{a}{2},\frac{a}{2}\right]\times \left[-\frac{b}{2},\frac{b}{2}\right]$.
\end{definition}

The weak formulation of \eqref{eqn: equation for p} is that, for $\forall\,\varphi \in C^\infty_\#(Y\times[0,1])$,
\begin{equation}
\begin{split}
\int_{\Omega_t} (D\nabla p^t)\cdot \nabla \varphi &\;- v_0 p^t(\cos 2\pi\theta, \sin 2\pi\theta, 0)\cdot\nabla\varphi \\
&\;= \int_{\Gamma_t}\varphi\left(g^t(x,\theta)-r_{\mathrm{in}}(\kappa^t(x))p^t(x,\theta)[\rho_+(\beta^t)+\rho_-(\beta^t)]\right).
\end{split}
\label{eqn: weak form of p equation}
\end{equation}
Take $t$-derivative on both sides at $t = 0$.
By \eqref{eqn: derivative of domain integral} (see also \cite[pp.~120]{sokolowski1992introduction}) and the fact that $\varphi' = 0$, we find that the Eulerian derivative of the left hand side of \eqref{eqn: weak form of p equation} is
\begin{equation}
\begin{split}
&\;\int_{\Omega}  (D\nabla p')\cdot\nabla \varphi - v_0 p' (\cos 2\pi\theta, \sin 2\pi\theta,0)\cdot\nabla \varphi\\
&\;+\int_\Gamma [ (D\nabla p)\cdot \nabla \varphi - v_0 p(\cos 2\pi\theta, \sin 2\pi\theta, 0)\cdot\nabla\varphi] (\tilde{V}(0)\cdot n_\Gamma).
\end{split}
\label{eqn: left hand side of the weak p' equation}
\end{equation}
Recall that $\tilde{V}(t,\cdot,\theta)$ is compactly supported in the interior of $Y$ and has zero $\theta$-components everywhere.
Hence the second term above only involves an integral on $\Gamma$ instead of on the entire $\partial\Omega$.
Similarly, by \eqref{eqn: derivative of boundary integral pre-form} and \eqref{eqn: derivative of boundary integral}, the Eulerian derivative of the right hand side of \eqref{eqn: weak form of p equation} is
\begin{equation}
\begin{split}
&\;\int_\Gamma (\varphi g - \varphi r_\mathrm{in} p (\rho_+ +\rho_-))'-\int_\Gamma\kappa(\varphi g-\varphi r_\mathrm{in} p (\rho_+ +\rho_-))(\tilde{V}(0)\cdot n_{\Gamma})\\
= &\;\int_\Gamma \varphi g'-\varphi (r_\mathrm{in}(\rho_++\rho_-))' p-\varphi r_\mathrm{in} (\rho_++\rho_-)\left[p'|_\Gamma +\frac{\partial p}{\partial n_\Gamma}\cdot(\tilde{V}(0)\cdot n_{\Gamma})\right]\\
&\;+\int_\Gamma (\varphi|_\Gamma)'(g-r_\mathrm{in}p(\rho_+ +\rho_-))-\int_\Gamma \kappa\varphi(\tilde{V}(0)\cdot n_{\Gamma})(g-r_\mathrm{in}p(\rho_++\rho_-)),
\end{split}
\label{eqn: expand RHS of p' equation}
\end{equation}
where
\begin{equation}
 \begin{split}
  g'(x,\theta) = &\;r_\mathrm{out}'(\kappa(x))[2\pi\tau_+(\beta)p_B^+(x)+2\pi\tau_-(\beta)p_B^-(x)]\\
  &\;+r_\mathrm{out}(\kappa(x))\cdot 2\pi[\tau_+(\beta) p_B^+(x) +\tau_-(\beta) p_B^-(x)]'\\
  =&\;\frac{dr_\mathrm{out}}{d\kappa}(\kappa(x))\kappa'(x)[2\pi\tau_+(\beta)p_B^+(x)+2\pi\tau_-(\beta)p_B^-(x)]\\
&\;+r_\mathrm{out}(\kappa(x))\cdot 2\pi(-\alpha'(x))\cdot\left[\frac{d\tau_+}{d\beta}(\beta) p_B^+(x) +\frac{d\tau_-}{d\beta}(\beta) p_B^-(x)\right]\\
&\;+r_\mathrm{out}(\kappa(x))\cdot 2\pi\left[\tau_+(\beta) p_B^+{}'(x)+\tau_-(\beta)p_B^-{}'(x)\right],
 \end{split}
 \label{eqn: g'}
\end{equation}
\begin{equation}
[r_\mathrm{in}(\rho_++\rho_-)]' = \frac{dr_\mathrm{in}}{d\kappa}(\kappa(x))\kappa'(x)(\rho_++\rho_-)+r_\mathrm{in}(\kappa(x))(-\alpha'(x))\cdot\left[\frac{d\rho_+}{d\beta}(\beta)+\frac{d\rho_-}{d\beta}(\beta)\right],
\label{eqn: (r_in(rho+ + rho-))'}
\end{equation}
and \cite[pp.~118]{sokolowski1992introduction}
\begin{equation}
 (\varphi|_\Gamma)'(x,\theta) = \frac{\partial \varphi}{\partial n_\Gamma}(x,\theta)(\tilde{V}(0)\cdot n_\Gamma),\quad\forall\,\varphi\in C^\infty_\#(Y\times[0,1]).
 \label{eqn: phi'}
\end{equation}
To this end, if we take $\varphi\in C_{0,\#}^\infty(\Omega)\subset C^\infty_\#(Y\times[0,1])$, all the integrals on $\Gamma$ vanish in \eqref{eqn: left hand side of the weak p' equation} and \eqref{eqn: expand RHS of p' equation}.
By equating \eqref{eqn: left hand side of the weak p' equation} and \eqref{eqn: expand RHS of p' equation}, we find that
\begin{equation*}
\int_{\Omega}(D\nabla p')\cdot  \nabla \varphi - v_0 p' (\cos 2\pi\theta, \sin 2\pi\theta,0)\cdot\nabla \varphi = 0,
\end{equation*}
which gives the equation for $p'$
\begin{equation}
 \mathrm{div}(D\nabla p') = v_0(\cos2\pi\theta, \sin 2\pi\theta,0)\cdot\nabla p',\quad(x,\theta)\in\Omega.
\label{eqn: equation for p'}
\end{equation}
Then we alternatively take $\varphi\in C_\#^\infty(Y\times[0,1])$ in \eqref{eqn: left hand side of the weak p' equation} and \eqref{eqn: expand RHS of p' equation} such that $\frac{\partial \varphi}{\partial n_\Gamma} = 0$ on $\Gamma$.
This gives
\begin{equation}
\begin{split}
\eqref{eqn: left hand side of the weak p' equation} = &\;\int_\Gamma[(D\nabla p')\cdot n_\Gamma] \varphi - v_0p'|_\Gamma\varphi(\cos2\pi\theta, \sin 2\pi\theta, 0)\cdot n_\Gamma\\
&\;+\int_\Gamma [(D\nabla p)\cdot \nabla_\Gamma \varphi - v_0 p(\cos2\pi\theta, \sin 2\pi\theta, 0)\cdot\nabla_\Gamma\varphi](\tilde{V}(0)\cdot n_\Gamma).
\end{split}
\label{eqn: weak form of p boundary condition}
\end{equation}
To obtain the first line in \eqref{eqn: weak form of p boundary condition}, we applied \eqref{eqn: equation for p'} and integration by parts.
Note that $\nabla \varphi = \nabla_\Gamma\varphi$ by the assumption $\frac{\partial \varphi}{\partial n_\Gamma} = 0$.
By integration by parts on $\Gamma$, \eqref{eqn: weak form of p boundary condition} can be rewritten as
\begin{equation}
\begin{split}
\eqref{eqn: left hand side of the weak p' equation} = &\;\int_\Gamma D_t\frac{\partial p'}{\partial n_\Gamma}\varphi - v_0p'\varphi\cos\beta \\
&\;-\int_\Gamma \varphi\mathrm{div}_\Gamma[(\tilde{V}(0)\cdot n_\Gamma) D\nabla_\Gamma p - v_0 (\tilde{V}(0)\cdot n_\Gamma) p \mathds{P}_\Gamma(\cos2\pi\theta, \sin 2\pi\theta, 0)]\\
\end{split}
\label{eqn: final form of LHS of p'}
\end{equation}
where $\mathds{P}_\Gamma$ is the projection to the tangent space of $\Gamma$.
Here we used the fact that $D\nabla_\Gamma p$ is tangent to $\Gamma$, and $(\cos2\pi\theta, \sin 2\pi\theta, 0)^T\cdot n_\Gamma = \cos(2\pi\theta -\alpha(x)) = \cos\beta$.

Combining \eqref{eqn: expand RHS of p' equation}, \eqref{eqn: g'}, \eqref{eqn: (r_in(rho+ + rho-))'}, \eqref{eqn: phi'} and \eqref{eqn: final form of LHS of p'}, we obtain the boundary condition for $p'$
\begin{equation}
 \begin{split}
  D_t\frac{\partial p'}{\partial n_\Gamma} - &\;v_0p'\cos\beta +r_\mathrm{in} (\rho_++\rho_-)p'\\
    =&\;r_\mathrm{out}(\kappa(x))\cdot 2\pi[\tau_+(\beta) p_B^+{}'(x)+\tau_-(\beta)p_B^-{}'(x)]\\
    &\;+\frac{dr_\mathrm{out}}{d\kappa}(\kappa(x))\kappa'(x)[2\pi\tau_+(\beta)p_B^+(x)+2\pi\tau_-(\beta)p_B^-(x)]\\
    &\;+r_\mathrm{out}(\kappa(x))\cdot 2\pi(-\alpha'(x))\cdot\left[\frac{d\tau_+}{d\beta}(\beta) p_B^+(x) +\frac{d\tau_-}{d\beta}(\beta) p_B^-(x)\right]\\
    &\;-\frac{dr_\mathrm{in}}{d\kappa}(\kappa(x))\kappa'(x)(\rho_+(\beta)+\rho_-(\beta))p\\
    &\;-r_\mathrm{in}(\kappa(x))(-\alpha'(x))\cdot\left[\frac{d\rho_+}{d\beta}(\beta)+\frac{d\rho_-}{d\beta}(\beta)\right]p\\
    &\;-r_\mathrm{in} (\rho_++\rho_-)\frac{\partial p}{\partial n_\Gamma}(\tilde{V}(0)\cdot n_\Gamma) -\kappa(\tilde{V}(0)\cdot n_\Gamma)(g-r_\mathrm{in}p(\rho_++\rho_-))\\
    &\;+\mathrm{div}_\Gamma[(\tilde{V}(0)\cdot n_\Gamma) D\nabla_\Gamma p - v_0 (\tilde{V}(0)\cdot n_\Gamma) p \mathds{P}_\Gamma(\cos2\pi\theta, \sin 2\pi\theta, 0)],
 \end{split}
 \label{eqn: boundary condition of p'}
\end{equation}
where $g$ is given by \eqref{equation: formula for Robin boundary data in subproblem}.
Here
\begin{align}
\alpha'(s) = &\;\partial_s[V(0)\cdot n(s)],\label{eqn: alpha' in the main body}\\
\kappa'(s) = &\;\partial_{ss}(V(0)\cdot n(s))+\kappa^2(s) (V(0)\cdot n(s)),\label{eqn: kappa' in the main body}
\end{align}
which will be shown in Section \ref{section: derivation of k' and alpha'}.

\subsection{Equations for $p_B^\pm{}'$}\label{section: equations for p_B pm}
Let $s_t$ be the arclength parameter of $\gamma_t$.
The equations for $p_B^{\pm,t}$ on $\gamma_t$ are
\begin{equation*}
-D_t\Delta_{\gamma_t} p_B^{\pm,t}(x) \pm v_0\partial_{s_t} p^{\pm,t}_b(x) + r_\mathrm{out}(\kappa^t(x)) p_B^{\pm,t}(x) = r_\mathrm{in}(\kappa^t(x))\int_0^1\rho_\pm(\beta^t)p^t(x,\theta)\,d\theta.
\label{eqn: equation for p_B}
\end{equation*}
Their weak formulations are that, for $\forall\,\psi\in \mathcal{D}_\#(Y)$,
\begin{equation}
\int_{\gamma_t} - D_t p_B^{\pm,t} \Delta_{\gamma_t} \psi|_{\gamma_t} \mp v_0 \partial_{s_t}\psi|_{\gamma_t} p_B^{\pm,t} +r_\mathrm{out}(\kappa^t) p_B^{\pm,t} \psi|_{\gamma_t}= \int_{\gamma_t}
\psi|_{\gamma_t}r_\mathrm{in}(\kappa^t) \int_0^1\,d\theta \rho_\pm(\beta^t)p^t(x,\theta).
\label{eqn: weak form of p_B equations}
\end{equation}


Take $t$-derivative at $t = 0$ on both sides of \eqref{eqn: weak form of p_B equations}.
By \eqref{eqn: Eulerian derivative of an integral shape functional on a submanifold}, we find the Eulerian derivative of the right hand side of \eqref{eqn: weak form of p_B equations} is
\begin{equation}
\begin{split}
&\;\int_\Gamma [\psi|_\gamma r_\mathrm{in}(\kappa(x))\rho_\pm(\beta)p(x,\theta)]'-\int_\Gamma \kappa(V(0)\cdot n)\psi r_\mathrm{in}(\kappa(x))\rho_\pm(\beta)p(x,\theta)\\
=
&\;\int_\Gamma \frac{\partial \psi}{\partial n}(V(0)\cdot n) r_\mathrm{in}(\kappa(x))\rho_\pm(\beta) p(x,\theta) + \int_\Gamma \psi \frac{dr_\mathrm{in}}{d\kappa}(\kappa(x)) \kappa'(x)\rho_\pm(\beta) p(x,\theta)\\
&\;+\int_\Gamma \psi r_\mathrm{in}(\kappa(x))\frac{d\rho_\pm}{d\beta}(\beta)(-\alpha'(x)) p(x,\theta)\\
&\;+\int_\Gamma \psi r_\mathrm{in}(\kappa(x))\rho_\pm \left[p'(x,\theta)|_\Gamma+\frac{\partial p}{\partial n_\Gamma}(x,\theta)(V(0)\cdot n)\right]-\int_\Gamma \kappa(V(0)\cdot n)\psi r_\mathrm{in}\rho_\pm p(x,\theta).
\end{split}
\label{eqn: RHS of weak p_B' equation}
\end{equation}
Here the functions originally defined for $x\in\gamma$ are interpreted in a natural way as functions defined for $(x,\theta)\in\Gamma = \gamma\times [0,1]$.
On the other hand, the Eulerian derivative of the left hand side of \eqref{eqn: weak form of p_B equations} is that
\begin{equation}
\begin{split}
&\; \int_\gamma [-D_t p_B^\pm\Delta_\gamma \psi|_\gamma \mp v_0\partial_s\psi|_\gamma p_B^\pm+r_\mathrm{out}(\kappa(x)) p_B^\pm\psi|_\gamma]'\\
&\;-\int_\gamma\kappa(V(0)\cdot n) (-D_t p_B^\pm \Delta_\gamma\psi\mp v_0 \partial_s \psi p_B^\pm+r_\mathrm{out}(\kappa(x))p_B^\pm\psi)\\
=&\; \int_\gamma -D_t p_B^\pm{}'\Delta_\gamma \psi \mp v_0\partial_s\psi p_B^\pm{}'+r_\mathrm{out}(\kappa(x)) p_B^\pm{}'\psi + \frac{dr_\mathrm{out}}{d\kappa}(\kappa(x))\kappa'(x) p_B^\pm\psi\\
&\; +\int_\gamma -D_t p_B^\pm(\Delta_\gamma \psi|_\gamma)' \mp v_0 p_B^\pm (\partial_s \psi|_\gamma)' +r_\mathrm{out}(\kappa(x)) p_B^\pm\frac{\partial \psi}{\partial n}(V(0)\cdot n)\\
&\;-\int_\gamma\kappa(V(0)\cdot n) (-D_t p_B^\pm \Delta_\gamma\psi\mp v_0 \partial_s \psi p_B^\pm+r_\mathrm{out}(\kappa(x))p_B^\pm\psi).
\end{split}
\label{eqn: LHS of weak p_B' equation}
\end{equation}
To further simplify \eqref{eqn: LHS of weak p_B' equation}, we need the formulae for $(\Delta_\gamma \psi|_\gamma)'$ and $(\partial_s\psi|_\gamma)'$ derived in Section \ref{section: derivation of shape derivative of Laplace psi and grad psi on gamma},
\begin{align*}
(\partial_s \psi)' = &\;\kappa(V(0)\cdot n)\partial_s\psi+\partial_s\left(\frac{\partial\psi}{\partial n}(V(0)\cdot n)\right),\\
(\Delta_\gamma\psi|_\gamma)' =&\;\partial_s [\kappa (V(0)\cdot n)]\cdot \partial_s\psi+\partial_{ss}\left(\frac{\partial\psi}{\partial n}(V(0)\cdot n)\right) +2\kappa(V(0)\cdot n) \partial_{ss}\psi.
\end{align*}
Hence, \eqref{eqn: LHS of weak p_B' equation} becomes
\begin{equation*}
\begin{split}
\eqref{eqn: LHS of weak p_B' equation}=&\; \int_\gamma -D_t p_B^\pm{}'\Delta_\gamma \psi \mp v_0\partial_s\psi p_B^\pm{}'+r_\mathrm{out}(\kappa(x)) p_B^\pm{}'\psi + \frac{dr_\mathrm{out}}{d\kappa}(\kappa(x))\kappa'(x) p_B^\pm\psi\\
&\; +\int_\gamma -D_t p_B^\pm\left(\partial_s [\kappa (V(0)\cdot n)]\cdot \partial_s\psi+\partial_{ss}\left(\frac{\partial\psi}{\partial n}(V(0)\cdot n)\right) +2\kappa(V(0)\cdot n) \partial_{ss}\psi\right)\\
&\;\mp\int_\gamma  v_0 p_B^\pm \left(\kappa(V(0)\cdot n)\partial_s\psi+\partial_s\left(\frac{\partial\psi}{\partial n}(V(0)\cdot n)\right)\right)+\int_\gamma r_\mathrm{out}(\kappa(x)) p_B^\pm\frac{\partial \psi}{\partial n}(V(0)\cdot n)\\
&\;-\int_\gamma\kappa(V(0)\cdot n) (-D_t p_B^\pm \Delta_\gamma\psi\mp v_0 \partial_s \psi p_B^\pm+r_\mathrm{out}(\kappa(x))p_B^\pm\psi).
\end{split}
\end{equation*}
We take $\psi$ such that $\frac{\partial \psi}{\partial n} = 0$ on $\gamma$ and do integration by parts
\begin{equation}
\begin{split}
\eqref{eqn: LHS of weak p_B' equation} =&\; \int_\gamma -D_t \Delta_\gamma p_B^\pm{}' \psi \pm v_0\psi \partial_s p_B^\pm{}'+r_\mathrm{out}(\kappa(x)) p_B^\pm{}'\psi + \frac{dr_\mathrm{out}}{d\kappa}(\kappa(x))\kappa'(x) p_B^\pm\psi\\
&\; -\int_\gamma D_t \partial_s(\partial_s [\kappa(V(0)\cdot n)]p_B^\pm)\cdot \psi-2D_t\int_\gamma \partial_s[\kappa(V(0)\cdot n)\partial_s p_B^\pm]\cdot \psi\\
&\;\pm\int_\gamma  v_0 \partial_s[p_B^\pm \kappa(V(0)\cdot n)]\psi\\
&\;-\int_\gamma -D_t  \Delta_\gamma[\kappa (V(0)\cdot n) p_B^\pm]\psi \pm v_0 \psi \partial_s[\kappa(V(0)\cdot n) p_B^\pm]+r_\mathrm{out}(\kappa(x))\kappa (V(0)\cdot n) p_B^\pm\psi\\
=&\; \int_\gamma -D_t \Delta_\gamma p_B^\pm{}' \psi \pm v_0\psi \partial_s p_B^\pm{}'+r_\mathrm{out}(\kappa(x)) p_B^\pm{}'\psi + \frac{dr_\mathrm{out}}{d\kappa}(\kappa(x))\kappa'(x) p_B^\pm\psi\\
&\; -D_t\int_\gamma \partial_s[\kappa(V(0)\cdot n)\partial_s p_B^\pm]\cdot \psi -\int_\gamma r_\mathrm{out}(\kappa(x))\kappa (V(0)\cdot n) p_B^\pm\psi.
\end{split}
\label{eqn: final version of LHS of weak p_B' equation}
\end{equation}
On the other hand, with $\frac{\partial \psi}{\partial n} = 0$, \eqref{eqn: RHS of weak p_B' equation} becomes
\begin{equation}
\begin{split}
\eqref{eqn: RHS of weak p_B' equation}=&\;\int_\Gamma \psi \frac{dr_\mathrm{in}}{d\kappa}(\kappa(x)) \kappa'(x)\rho_\pm(\beta) p(x,\theta)+\int_\Gamma \psi r_\mathrm{in}(\kappa(x))\frac{d\rho_\pm}{d\beta}(\beta)(-\alpha'(x)) p(x,\theta)\\
&\;+\int_\Gamma \psi r_\mathrm{in}(\kappa(x))\rho_\pm \left[p'(x,\theta)|_\Gamma+\frac{\partial p}{\partial n_\Gamma}(x,\theta)(V(0)\cdot n)\right]-\int_\Gamma \kappa(V(0)\cdot n)\psi r_\mathrm{in}\rho_\pm p(x,\theta).
\end{split}
\label{eqn: final version of RHS of weak p_B' equation}
\end{equation}
Equating \eqref{eqn: final version of LHS of weak p_B' equation} and \eqref{eqn: final version of RHS of weak p_B' equation}, we immediately find the equations for $p_B^\pm{}'$
\begin{equation}
\begin{split}
&\;-D_t \Delta_\gamma p_B^\pm{}' \pm v_0 \partial_s p_B^\pm{}'+r_\mathrm{out}(\kappa(x)) p_B^\pm{}' \\
=&\;\int_0^1  \frac{dr_\mathrm{in}}{d\kappa}(\kappa(x)) \kappa'(x)\rho_\pm p(x,\theta)+r_\mathrm{in}(\kappa(x))\frac{d\rho_\pm}{d\beta}(\beta)(-\alpha'(x)) p(x,\theta)\,d\theta\\
&\; +\int_0^1 r_\mathrm{in}(\kappa(x))\rho_\pm \left[p'(x,\theta)|_\Gamma+\frac{\partial p}{\partial n_\Gamma}(x,\theta)(V(0)\cdot n)\right]- \kappa(V(0)\cdot n) r_\mathrm{in}\rho_\pm p(x,\theta)\,d\theta\\
&\;- \frac{dr_\mathrm{out}}{d\kappa}(\kappa(x))\kappa'(x) p_B^\pm +D_t\partial_s[\kappa(V(0)\cdot n)\partial_s p_B^\pm] +r_\mathrm{out}(\kappa(x))\kappa (V(0)\cdot n) p_B^\pm.
\end{split}
\label{eqn: equation for p_B'}
\end{equation}

In summary, \eqref{eqn: equation for p'}, \eqref{eqn: boundary condition of p'} and \eqref{eqn: equation for p_B'} form the coupled system for $p'$ and $p_B^\pm{}'$, in which $\alpha'(x)$ and $\kappa'(x)$ are given by \eqref{eqn: alpha'} and \eqref{eqn: kappa'} respectively.
As before, $\beta = 2\pi\theta - \alpha(x)$ is defined for $(x,\theta)\in \Gamma$.

\subsection{Derivation of $\kappa'(x)$ and $\alpha'(x)$}\label{section: derivation of k' and alpha'}
In this section, we shall derive explicit formulae for $\kappa'(x)$ and $\alpha'(x)$, which appear in the equations of $p'$ and $p_{B}^\pm{}'$.
For the elementary differential geometry involved in the following calculation, readers are referred to \cite{do1976differential}.

Let $s$ denote the arclength parameter of $\gamma$ at $t = 0$.
Let $\tau(s) = \partial_s\gamma(s)$ denote the unit tangent vector of $\gamma$ at $t = 0$, and let $n(s)$ be the unit normal vector pointing towards $O$, the domain enclosed by $\gamma$.
We shall write $V(0,\gamma(s)) = V(0)$ whenever it is convenient.

By definition, the unit tangent vector of $\gamma_t$ at time $t$ is given by
\begin{equation}
\tau^t(s) = \frac{\frac{d}{ds}(T_t\circ \gamma)(s)}{\left|\frac{d}{ds}(T_t\circ \gamma)(s)\right|} = \frac{\nabla T_t(\gamma(s))\partial_s\gamma(s)}{\left|\nabla T_t(\gamma(s))\partial_s\gamma(s)\right|}.
\label{eqn: tau_t(s)}
\end{equation}
To derive the material derivative of the unit tangent vector, we take $t$-derivative at $t = 0$ and find
\begin{equation}
\begin{split}
\dot{\tau}(s) = &\;\frac{\nabla V(0)\partial_s\gamma(s)\left|\partial_s\gamma(s)\right| - \frac{\partial_s\gamma(s)}{\left|\partial_s\gamma(s)\right|}\partial_s\gamma(s)^T \nabla V(0) \partial_s\gamma(s)}{\left|\partial_s\gamma(s)\right|^2}\\
=&\; \nabla V(0)\partial_s\gamma(s)- \left[\partial_s\gamma(s)^T \nabla V(0) \partial_s\gamma(s)\right]\partial_s\gamma(s)\\
=&\; \left[n(s)^T\nabla_\gamma V(0) \tau(s)\right] n(s)\\
=&\; \left[n(s)\cdot\partial_s V(0)\right] n(s).
\end{split}
\label{eqn: dot tau}
\end{equation}
Here we used the fact that
\begin{equation}
\left.\nabla T_t\right|_{t=0} = Id,\quad\left.\frac{d \nabla T_t (\gamma(s))}{d t}\right|_{t=0} = \nabla V(0),
\label{eqn: T_t and grad T_t at t_0}
\end{equation}
and the following orthogonal decomposition in dimension two
\begin{equation*}
\nabla V(0)\partial_s\gamma(s) = \left[\partial_s\gamma(s)^T \nabla V(0) \partial_s\gamma(s)\right]\partial_s\gamma(s)+ \left[n(s)^T\nabla V(0) \tau(s)\right] n(s).
\end{equation*}
By definition,
\begin{equation*}
\tau'(s) = \dot{\tau}(s) - V(0)\cdot\nabla_\Gamma\tau = \dot{\tau}(s) - [V(0)\cdot \tau(s)] \partial_s\tau = \dot{\tau}(s) - [V(0)\cdot\tau(s)] \kappa(s) n(s),
\end{equation*}
where $\kappa(s)$ is the curvature of $\gamma$.

$\dot{n}(s)$ can be calculated easily with the observation
\begin{equation*}
n^t(s) = (n_1^t(s), n_2^t(s))^T = (-\tau_2^t(s), \tau_1^t(s))^T.
\end{equation*}
Combining this with \eqref{eqn: dot tau}, we find
\begin{equation}
\dot{n}(s) = -(n(s)\cdot\partial_s V(0)) \tau(s),
\label{eqn: dot n}
\end{equation}
and thus
\begin{equation*}
n'(s) = \dot{n}(s) - V(0)\cdot\nabla_\gamma n(s)=\dot{n}(s) - [V(0)\cdot \tau(s)] \partial_s n(s)=\dot{n}(s) +\kappa(s) \tau(s) [V(0)\cdot\tau(s)].
\end{equation*}

Recall that the angle $\alpha$ of the outer normal vector of $\gamma$ is determined by
\begin{equation*}
(\cos\alpha(s), \sin\alpha(s))^T = n(s).
\end{equation*}
Take material derivatives on both sides and we have
\begin{equation*}
(-\sin\alpha(s)\cdot \dot{\alpha}(s), \cos\alpha(s)\cdot \dot{\alpha}(s))^T = \dot{n}(s).
\end{equation*}
Take inner product with $-\tau(s) = (-\sin\alpha(s), \cos\alpha(s))^T$ and we find by \eqref{eqn: dot n}
\begin{equation*}
\dot{\alpha}(s) = -\dot{n}(s)\cdot \tau(s) = n(s)\cdot\partial_s V(0).
\end{equation*}
By taking derivative with respect to $s$ in the definition of $\alpha(s)$,
\begin{equation*}
\partial_s\alpha(s)(-\sin\alpha(s), \cos\alpha(s))^T = \partial_s n(s) = -\kappa(s) \tau(s).
\end{equation*}
Taking inner product with $-\tau(s)$, we obtain $\partial_s\alpha(s) = \kappa(s)$.
Hence,
\begin{equation}
\begin{split}
\alpha'(s) =&\; \dot{\alpha}(s) -V(0)\cdot\nabla_\gamma \alpha = n(s)\cdot\partial_s V(0) - [V(0)\cdot\tau(s)] \partial_s \alpha(s)\\
= &\; \partial_s V(0)\cdot n(s) - \kappa(s)V(0)\cdot\tau(s) = \partial_s[V(0)\cdot n(s)].
\end{split}
\label{eqn: alpha'}
\end{equation}

Next we are going to calculate $\kappa'(s)$. At time $t$,
\begin{equation*}
\kappa^t(s) = n^t(s)\cdot \frac{\frac{d}{ds}\tau^t(s)}{\left|\partial_s\gamma_t(s)\right|}.
\end{equation*}
Note that $s$ is not necessarily the arclength parameter at time $t$.
By using \eqref{eqn: tau_t(s)} and
\begin{equation*}
\left|\partial_s \gamma_t(s)\right| = \left|\nabla T_t(\gamma(s))\partial_s \gamma(s)\right|,
\end{equation*}
we know that
\begin{equation}
\begin{split}
\kappa^t(s) =&\;\frac{1}{\left|(\nabla T_t)\partial_s \gamma(s)\right|^4}\cdot n^t_i(s)\left[\partial_k (\nabla T_t)_{ij}\partial_s\gamma_k(s)\partial_s\gamma_j(s)\left|(\nabla T_t)\partial_s\gamma(s)\right|^2\right.\\
&\;+(\nabla T_t)_{ij}\partial_{ss}\gamma_j(s)\left|(\nabla T_t)\partial_s\gamma(s)\right|^2\\
&\;\left.-(\nabla T_t)_{ij}\partial_s \gamma_j(s)\cdot\partial_s\gamma(s)^T \nabla T_t^T\left(\nabla T_t \partial_{ss}\gamma(s)+ \partial_k (\nabla T_t)\cdot \partial_s\gamma_k(s)\partial_s\gamma(s)\right)\right].
\end{split}
\label{eqn: crude formula for curvature of gamma at t}
\end{equation}
Here we used Einstein summation convention and wrote $\nabla T_t(\gamma(s))$ as $\nabla T_t$.
Thanks to \eqref{eqn: tau_t(s)}, the last term in \eqref{eqn: crude formula for curvature of gamma at t} should vanish since
\begin{equation*}
n^t_i(s)(\nabla T_t)_{ij}\partial_s \gamma_j(s) = n^t(s)\cdot(\nabla T_t\tau(s)) = n^t(s)\cdot\tau^t(s)|\nabla T_t\tau(s)| = 0.
\end{equation*}
This gives
\begin{equation*}
\kappa^t(s) =\frac{1}{\left|\nabla T_t\partial_s \gamma(s)\right|^2}\cdot n^t_i(s)\left[\partial_k (\nabla T_t)_{ij}\partial_s\gamma_k(s)\partial_s\gamma_j(s)+(\nabla T_t)_{ij}\partial_{ss}\gamma_j(s)\right].
\end{equation*}
Taking $t$-derivative on both sides at $t = 0$, we find by \eqref{eqn: T_t and grad T_t at t_0} that
\begin{equation*}
\begin{split}
\dot{\kappa}(s) =&\; n(s)^T\left[\partial_s (\nabla V(0))\tau(s)\right] + \dot{n}(s)\cdot \partial_{ss}\gamma(s) + n(s)^T\nabla V(0)\partial_{ss}\gamma(s)\\
&\;-(n(s)\cdot \partial_{ss}\gamma(s))\cdot 2\tau(s)^T\nabla V(0)\tau(s)\\
=&\; n(s)^T \partial_s(\nabla V(0))\tau(s) - [(n(s)\cdot\partial_s V(0)) \tau(s)\cdot \kappa(s)n(s)] +\kappa(s) n(s)^T\nabla V(0) n(s)\\
&\;-2\kappa(s)\tau(s)^T\nabla V(0)\tau(s)\\
=&\; n(s)^T \partial_s(\nabla V(0))\tau(s) +\kappa(s)\left[ n(s)^T\nabla V(0) n(s)-2\tau(s)^T\nabla V(0)\tau(s)\right].
\end{split}
\end{equation*}
By definition,
\begin{equation*}
\kappa'(s) = \dot{\kappa}(s) - \nabla_\gamma \kappa(s)\cdot V(0).
\end{equation*}
It is not difficult to verify that
\begin{equation}
\kappa'(s) = \partial_{ss}( V(0)\cdot n(s))+\kappa^2(s) (V(0)\cdot n(s)).
\label{eqn: kappa'}
\end{equation}
Indeed,
\begin{equation*}
\partial_s(V(0)\cdot n(s)) = \partial_s V(0)\cdot n(s) - V(0)\cdot \kappa(s) \tau(s) = n(s)^T \nabla V(0) \tau(s) - \kappa(s)( V(0)\cdot \tau(s)),
\end{equation*}
and
\begin{equation*}
\begin{split}
\partial_{ss}(V(0)\cdot n(s)) = &\;-\kappa(s)\tau(s)^T \nabla V(0) \tau(s) +n(s)^T \partial_s\nabla V(0) \tau(s) +\kappa(s)n(s)^T \nabla V(0) n(s)\\
&\; - \partial_s\kappa(s)(V(0)\cdot \tau(s)) - \kappa(s)(\partial_s V(0)\cdot \tau(s)) - \kappa^2(s)(V(0)\cdot n(s)).
\end{split}
\end{equation*}
Hence,
\begin{equation*}
\begin{split}
&\;\partial_{ss}(V(0)\cdot n(s))+\kappa^2(s)(V(0)\cdot n(s))\\
= &\;-2\kappa(s)\tau(s)^T \nabla V(0) \tau(s) +n(s)^T \partial_s\nabla V(0) \tau(s)+\kappa(s)n(s)^T \nabla V(0) n(s)- \partial_s\kappa(s)( V(0)\cdot \tau(s))\\
= &\; \kappa(s)\left[n(s)^T \nabla V(0) n(s)-2\tau(s)^T \nabla V(0) \tau(s)\right]+n(s)^T \partial_s\nabla V(0) \tau(s)- \nabla_\gamma\kappa(s)\cdot V(0)\\
=&\; \dot{\kappa}(s)- \nabla_\gamma\kappa(s)\cdot V(0)=\kappa'(s).
\end{split}
\end{equation*}
We remark that the first term on the right hand side of \eqref{eqn: kappa'} shows how $V$ changes the curvature by bending $\gamma$, while the second term  accounts for the effect of $V$ changing the arclength of $\gamma$.

\subsection{Derivation of $(\partial_s \psi|_\gamma)\dot{}$ and $(\Delta_\gamma \psi|_\gamma)'$}\label{section: derivation of shape derivative of Laplace psi and grad psi on gamma}
This section is devoted to calculating $(\partial_s \psi|_\gamma)\dot{}$ and $(\Delta_\gamma \psi|_\gamma)'$, where $\psi\in \mathcal{D}_\#(Y)$ and the derivatives are calculated under the perturbation vector field $V$ introduced in Section \ref{section: a formal overview of the theory of shape optimization} and used before.
These two quantities are used in the derivation of equations for $p_B^\pm{}'$ in Section \ref{section: equations for p_B pm}.

We start from $(\partial_s \psi|_\gamma)\dot{}$.
By \eqref{eqn: dot tau},
\begin{equation*}
(\partial_s \psi|_\gamma)\dot{} = (\tau\cdot\partial_\gamma \psi|_\gamma)\dot{} = \dot{\tau}\cdot\partial_\gamma \psi +\tau\cdot(\partial_\gamma \psi|_\gamma)\dot{}=(n\cdot\partial_s V(0))n\cdot \partial_\gamma \psi +\tau\cdot(\partial_\gamma \psi|_\gamma)\dot{}=\tau\cdot(\partial_\gamma \psi|_\gamma)\dot{}.
\end{equation*}
Recall that for an arbitrary smooth function $f$ defined on $\gamma$, $\partial_\gamma f = (\partial_s f)\tau\in\mathds{R}^2$, and $\partial_s f = \partial_\gamma f\cdot\tau\in\mathds{R}$.
It is known that \cite[(2.137) and (2.138)]{sokolowski1992introduction}
\begin{equation}
[\partial_{\gamma_t} \psi|_{\gamma_t}]\circ T_t = (\nabla T_t)^{-T}\cdot[\nabla(\psi\circ T_t) -((B(t)n)\cdot\nabla(\psi\circ T_t)) n],
\label{eqn: transport of gradient operator on a boundary}
\end{equation}
where $n$ is the unit normal vector of $\gamma$ defined as before and
\begin{equation*}
B(t) = \left|(\nabla T_t)^{-T}n\right|^{-2}(\nabla T_t)^{-1}\cdot(\nabla T_t)^{-T}.
\end{equation*}
Take $t$-derivative at $t = 0$ on both sides of \eqref{eqn: transport of gradient operator on a boundary} and we obtain
\begin{equation*}
(\partial_\gamma \psi)\dot{} = \left.\frac{d}{dt}\right|_{t=0}(\nabla T_t)^{-T}\cdot[\nabla(\psi\circ T_t) -((B(t)n)\cdot\nabla(\psi\circ T_t)) n].
\end{equation*}
Using the fact that
\begin{align*}
\left.(\nabla T_t)^{-1}\right|_{t=0} = Id,&\;\quad\left.\frac{d}{dt}\right|_{t=0} (\nabla T_t)^{-1} = -\nabla V(0),\\
B(0) = Id,&\;\quad\left.\frac{d}{dt}\right|_{t=0} B(t) = -\nabla V(0)-\nabla V(0)^T + n^T(\nabla V(0)+\nabla V(0)^T)n\cdot Id,\\
\left.\nabla(\psi\circ T_t)\right|_{t=0} = \nabla\psi,&\;\quad\left.\frac{d}{dt}\right|_{t=0} \nabla(\psi\circ T_t) = \nabla(\nabla\psi\cdot V(0)),
\end{align*}
we find
\begin{equation*}
(\partial_\gamma \psi)\dot{} = -(\nabla V(0))^T[\nabla \psi - (n\cdot \nabla \psi) n] +[\nabla(\nabla\psi\cdot V(0))-Cn],
\end{equation*}
where $C = \left.\frac{d}{dt}\right|_{t=0}[(B(t)n)\cdot\nabla(\psi\circ T_t)] $ is a scalar. Hence,
\begin{equation*}
(\partial_s \psi)\dot{} = \tau\cdot(\partial_\gamma \psi)\dot{} = -\tau^T(\nabla V(0))^T[\nabla \psi - (n\cdot \nabla \psi) n] +\tau\cdot\nabla(\nabla\psi\cdot V(0)),
\end{equation*}
and
\begin{equation*}
\begin{split}
(\partial_s \psi)' = &\;(\partial_s \psi)\dot{}-\partial_\gamma(\partial_s\psi)\cdot V(0) \\
= &\;-\tau^T(\nabla V(0))^T[\nabla \psi - \langle n, \nabla \psi\rangle n] +\tau\cdot\nabla(\nabla\psi\cdot V(0))-\partial_\gamma(\partial_s\psi)\cdot V(0)\\
= &\;-\partial_s V(0)\cdot\partial_\gamma \psi +\partial_s(\nabla\psi\cdot V(0))-\partial_{ss}\psi\cdot (V(0)\cdot\tau)\\
= &\;-\partial_s \left[(V(0)\cdot n) n + (V(0)\cdot\tau)\tau\right]\cdot\partial_\gamma \psi\\
 &\;+\partial_s\left(\partial_s\psi (V(0)\cdot\tau)+\frac{\partial\psi}{\partial n}(V(0)\cdot n)\right)-\partial_{ss}\psi\cdot (V(0)\cdot\tau)\\
= &\;-[-\kappa(V(0)\cdot n)+\partial_s (V(0)\cdot\tau)]\partial_s\psi +\partial_s\psi \partial_s(V(0)\cdot\tau)+\partial_s\left(\frac{\partial\psi}{\partial n}(V(0)\cdot n)\right)\\
= &\;\kappa(V(0)\cdot n)\partial_s\psi+\partial_s\left(\frac{\partial\psi}{\partial n}(V(0)\cdot n)\right).
\end{split}
\label{eqn: grad psi'}
\end{equation*}
We remark that the first term above accounts for the change of $\partial_s \psi$ due to $V$ changing the arclength of $\gamma$; while the second term results from the motion of $\gamma$ by $V$ in the normal direction.

Next we turn to $(\Delta_\gamma\psi|_\gamma)'$. It is known that \cite[(2.147)]{sokolowski1992introduction}
\begin{equation}
(\Delta_{\gamma_t} \psi|_{\gamma_t})\circ T_t = \omega(t)^{-1}\mathrm{div}_\gamma\left[C(t)\cdot\left(\partial_\gamma[\psi\circ T_t]-(n^T C(t)n)^{-1}[ (C(t)n)\cdot\partial_\gamma(\psi\circ T_t)] n\right)\right],
\label{eqn: transport of laplace operator on a boundary}
\end{equation}
where
\begin{equation*}
\begin{split}
&\; \omega(t) = \det(\nabla T_t)\left|\nabla T_t^{-T}n\right|,\\
&\; C(t) = \omega(t)\nabla T_t^{-1}\cdot\nabla T_t^{-T}.
\end{split}
\end{equation*}
To derive the material derivative $(\Delta_\gamma\psi|_\gamma)\dot{}$, we take $t$-derivative on both sides of \eqref{eqn: transport of laplace operator on a boundary} and use the following facts
\begin{align*}
\omega(0) = 1,&\;\quad \left.\frac{d}{dt}\right|_{t=0}\omega(t) = \mathrm{div}V(0) -n^T\nabla V(0) n = \mathrm{div}_\gamma V(0),\\
C(0) = Id,&\;\quad  \left.\frac{d}{dt}\right|_{t=0} C(t) = \mathrm{div}_\gamma V(0)\cdot Id -\nabla V(0) - \nabla V(0)^T.
\end{align*}
Hence,
\begin{equation*}
\begin{split}
(\Delta_\gamma\psi|_\gamma)\dot{} =&\; -\mathrm{div}_\gamma V(0)\cdot\mathrm{div}_\gamma (\partial_\gamma \psi)+\mathrm{div}_\gamma\left[\left(\mathrm{div}_\gamma V(0)\cdot Id -\nabla V(0) - \nabla V(0)^T\right)\partial_\gamma\psi\right]\\
&\;+\mathrm{div}_\gamma\left[\partial_\gamma(\nabla\psi\cdot V(0))-([\mathrm{div}_\gamma V(0)\cdot Id -\nabla V(0) - \nabla V(0)^T]n\cdot \partial_\gamma\psi) n\right]\\
=&\; -\mathrm{div}_\gamma V(0)\cdot\Delta_\gamma \psi+\mathrm{div}_\gamma\left[\left(\mathrm{div}_\gamma V(0)\cdot Id -\nabla V(0) - \nabla V(0)^T\right)\partial_\gamma\psi\right]\\
&\;+\mathrm{div}_\gamma\left[\partial_\gamma(\nabla\psi\cdot V(0))+([\nabla V(0) +\nabla V(0)^T]n\cdot\partial_\gamma\psi) n\right]\\
=&\; -\mathrm{div}_\gamma V(0)\cdot\Delta_\gamma \psi+\mathrm{div}_\gamma\left[\mathrm{div}_\gamma V(0)\cdot \partial_\gamma\psi\right]+\mathrm{div}_\gamma\left[\partial_\gamma(\nabla\psi\cdot V(0))\right]\\
&\;-\mathrm{div}_\gamma\left[\left(\tau^T\left(\nabla V(0) +\nabla V(0)^T\right)\partial_\gamma\psi\right)\cdot \tau\right]\\
=&\;\partial_s\left(\mathrm{div}_\gamma V(0)\right)\cdot \partial_s\psi+\partial_{ss}\left(\frac{\partial\psi}{\partial n}(V(0)\cdot n)+\partial_s\psi( V(0)\cdot\tau)\right)-2\mathrm{div}_\gamma\left[\left(\tau^T\nabla V(0)\tau\right)\partial_s\psi \tau\right].
\end{split}
\end{equation*}
Here we used the identity that \cite[Lemma 2.63]{sokolowski1992introduction}
\begin{equation*}
\begin{split}
\mathrm{div}_\gamma[\mathrm{div}_\gamma V(0)\cdot \partial_\gamma\psi] = &\;\partial_\gamma \mathrm{div}_\gamma V(0)\cdot \partial_\gamma\psi+\mathrm{div}_\gamma V(0)\mathrm{div}_\gamma(\partial_\gamma\psi)\\
 = &\;\partial_s (\mathrm{div}_\gamma V(0))\cdot \partial_s\psi+\mathrm{div}_\gamma V(0)\Delta_\gamma\psi.
\end{split}
\end{equation*}
Since $\mathrm{div}_\gamma V(0) = \mathrm{div}_\gamma[(V(0)\cdot\tau)\tau] - \kappa(V(0)\cdot n)= \partial_s(V(0)\cdot\tau) - \kappa(V(0)\cdot n)$,
\begin{equation*}
\begin{split}
(\Delta_\gamma\psi|_\gamma)\dot{}=&\;\partial_s[\partial_s (V(0)\cdot\tau) -\kappa (V(0)\cdot n)]\cdot \partial_s\psi\\
&\;+\partial_{ss}\left(\frac{\partial\psi}{\partial n}(V(0)\cdot n)+\partial_s\psi(V(0)\cdot\tau)\right)-2\partial_s[(\tau\cdot\partial_s V(0))\cdot \partial_s\psi]\\
=&\;\partial_{ss}(V(0)\cdot\tau) \cdot\partial_s\psi -\partial_s [\kappa (V(0)\cdot n)]\cdot \partial_s\psi+\partial_{ss}\left(\frac{\partial\psi}{\partial n}(V(0)\cdot n)\right)\\
&\;+\partial_{s}[\partial_{ss}\psi(V(0)\cdot\tau)+\partial_s\psi\partial_s(V(0)\cdot\tau)]-2\partial_s[\partial_s(\tau\cdot V(0))\partial_s\psi]+2\partial_s[(\partial_s\tau\cdot V(0)) \partial_s\psi]\\
=&\;-\partial_{s}(V(0)\cdot\tau)\cdot\partial_{ss}\psi -\partial_s [\kappa (V(0)\cdot n)]\cdot \partial_s\psi+\partial_{ss}\left(\frac{\partial\psi}{\partial n}(V(0)\cdot n)\right)+\partial_{s}(\partial_{ss}\psi(V(0)\cdot\tau))\\
&\;+2\partial_s[\kappa (n\cdot V(0))\partial_s\psi]\\
=&\;-\partial_s[\kappa (V(0)\cdot n)]\cdot \partial_s\psi+\partial_{ss}\left(\frac{\partial\psi}{\partial n}(V(0)\cdot n)\right)+\partial_{sss}\psi(V(0)\cdot\tau)+2\partial_s[\kappa(V(0)\cdot n) \partial_s\psi].\\
\end{split}
\end{equation*}
Therefore, by definition,
\begin{equation}
\begin{split}
(\Delta_\gamma\psi|_\gamma)' =&\; (\Delta_\gamma\psi|_\gamma)\dot{} - V(0)\cdot\partial_\gamma \Delta_\gamma\psi\\
=&\;-\partial_s [\kappa (V(0)\cdot n)]\cdot \partial_s\psi+\partial_{ss}\left(\frac{\partial\psi}{\partial n}(V(0)\cdot n)\right) +2\partial_s[\kappa(V(0)\cdot n) \partial_s\psi]\\
=&\;\partial_s [\kappa (V(0)\cdot n)]\cdot \partial_s\psi+\partial_{ss}\left(\frac{\partial\psi}{\partial n}(V(0)\cdot n)\right) +2\kappa(V(0)\cdot n) \partial_{ss}\psi.
\label{eqn: lap psi'}
\end{split}
\end{equation}
We remark that the first term in \eqref{eqn: lap psi'} results from the motion of $\gamma$ by $V$ in the normal direction; the second term comes from the change of $\partial_s\psi$ due to the change of arclength of $\gamma$; the last term is the directly consequence of the change of arclength.

\section{Shape functionals and auxiliary functions}\label{section: shape functional and auxiliary functions}

Recall that in Section \ref{section: modeling}, we define the normalized net flux induced by the post to be
\begin{equation*}
E(\Omega) = \frac{F(\Omega)}{N(\Omega)},
\label{eqn: normalized flux definition}
\end{equation*}
where
\begin{align*}
N(\Omega) =&\;\int_{\Omega} p(x,\theta)\,\mathrm{d}x\mathrm{d}\theta+\int_\gamma [p_B^+(x)+p_B^-(x)]\,\mathrm{d}\gamma,\\
F(\Omega) =&\;\int_{\partial\Omega \cap \{x_2 = b/2\}} -D_t\frac{\partial p}{\partial x_2} + v_0p\sin2\pi\theta\,dA.
\end{align*}

Suppose we deform the domain $\Omega$ (or equivalently $\Gamma$) by the vector field $\tilde{V}(t,x,\theta)$ (or equivalently by the family of flow maps $\{\tilde{T}_t\}_{t\in[0,\varepsilon)}$).
The Eulerian derivative of $E(\Omega)$ with respect to $\tilde{V}$ is that
\begin{equation}
dE(\Omega;\tilde{V}) = \frac{N(\Omega)\cdot dF(\Omega;\tilde{V}) - F(\Omega)\cdot dN(\Omega;\tilde{V})}{N(\Omega)^2},
\label{eqn: dE the very first form}
\end{equation}
where
\begin{equation}
dN(\Omega;\tilde{V}) = \int_{\Omega}p'+\int_{\Gamma} p(\tilde{V}(0)\cdot n_\Gamma) +\int_\gamma \left(p_B^+{}'+p_B^-{}'\right)- \left( p_B^+{}+p_B^-{}\right)\kappa( V(0)\cdot n),
\label{eqn: dN the very first form}
\end{equation}
and
\begin{equation}
\begin{split}
dF(\Omega;\tilde{V}) =&\; \int_{\partial\Omega \cap \{x_2 = b/2\}}-D_t \left(\left.\frac{\partial p}{\partial x_2}\right|_{\partial\Omega \cap \{x_2 = b/2\}}\right)' + \left(p|_{\partial\Omega \cap \{x_2 = b/2\}}\right)' v_0\sin 2\pi\theta\,dA\\
=&\; \int_{\partial\Omega \cap \{x_2 = b/2\}}-D_t \left.\frac{\partial p'}{\partial x_2}\right|_{\partial\Omega \cap \{x_2 = b/2\}} + p'|_{\partial\Omega \cap \{x_2 = b/2\}} v_0\sin 2\pi\theta\,dA
\end{split}
\label{eqn: dF the very first form}
\end{equation}
Here we use the assumption that $V$ is zero on $\partial Y$.

According to Hadamard formula (Theorem \ref{thm: Hadamard formula}), we formally write $dE(\Omega; \tilde{V})$ as
\begin{equation}
dE(\Omega;\tilde{V}) = \int_\gamma G_E(x)(V(0)\cdot n) \,d\gamma,
\label{eqn: Hadamard structure theorem}
\end{equation}
for some distribution $G_E(x)$ defined on $\gamma$.
To find out $G_E(x)$, we need some auxiliary functions.

Let $\beta = 2\pi\theta-\alpha(x)$ as before.
Define $f_1(x,\theta)$ in $\Omega$ to be the solution of
\begin{align}
&\; \mathrm{div}(D\nabla f_1) + v_0(\cos 2\pi\theta, \sin 2\pi\theta, 0)\cdot \nabla f_1 = 1,\quad(x,\theta)\in\Omega,\label{eqn: equation for f_1}\\
&\; D_t\frac{\partial f_1}{\partial n_\Gamma} + r_\mathrm{in}(\kappa(x)) [\rho_+(\beta)+ \rho_-(\beta)]f_1 = 0,\quad(x,\theta)\in\Gamma,\label{eqn: boundary condition for f_1}\\
&\; f_1\mbox{ satisfies periodic boundary condition on }\partial\Omega\backslash\Gamma.\label{eqn: periodic boundary condition for f_1}
\end{align}
Then by \eqref{eqn: equation for p'}, \eqref{eqn: boundary condition of p'}, \eqref{eqn: equation for f_1}-\eqref{eqn: periodic boundary condition for f_1} and integration by parts,
\begin{equation}
\begin{split}
\int_{\Omega} p'=&\; \int_{\Omega}p'[\mathrm{div}(D\nabla f_1) + v_0(\cos 2\pi\theta, \sin 2\pi\theta, 0)\cdot \nabla f_1]\\
=&\; \int_{\Omega}[\mathrm{div}(D\nabla p') - v_0(\cos 2\pi\theta, \sin 2\pi\theta,0)\cdot\nabla p']\cdot f_1\\
&\; +\int_\Gamma D_t\left(p'\frac{\partial f_1}{\partial n_\Gamma} - f_1\frac{\partial p'}{\partial n_\Gamma}\right) +v_0(\cos 2\pi\theta, \sin 2\pi\theta, 0)\cdot n_\Gamma\cdot p'f_1\\
= &\;0+\int_\Gamma -f_1\left[D_t\frac{\partial p'}{\partial n_\Gamma} - v_0(\cos2\pi\theta, \sin 2\pi\theta, 0)\cdot n_\Gamma p' +p'r_\mathrm{in}(\rho_++\rho_-)\right]\\
= &\;\int_\Gamma -f_1\cdot \left[\mbox{RHS of \eqref{eqn: boundary condition of p'}}\right].
\end{split}
\label{eqn: simplification of integral of p'}
\end{equation}

Define $f_2^\pm(x)$ on $\gamma$ to be solutions of
\begin{equation}
-D_t \Delta_\gamma f_2^\pm\mp v_0 \partial_s f_2^\pm+r_\mathrm{out}(\kappa(x)) f_2^\pm = 1,\quad x\in\gamma.
\label{eqn: equations for f_2 pm}
\end{equation}
By \eqref{eqn: equation for p_B'}, \eqref{eqn: equations for f_2 pm} and integration by parts,
\begin{equation}
\begin{split}
\int_\gamma p_B^+{}'+p_B^-{}' = &\; \int_\gamma p_B^+{}'[-D_t \Delta_\gamma f_2^+ - v_0 \partial_s f_2^++r_\mathrm{out}(\kappa(x)) f_2^+]\\
&\; +\int_\gamma p_B^-{}'[-D_t \Delta_\gamma f_2^- + v_0 \partial_s f_2^-+r_\mathrm{out}(\kappa(x)) f_2^-]\\
=&\;\int_\gamma f_2^+[-D_t \Delta_\gamma p_B^+{}' + v_0 \partial_s p_B^+{}'+r_\mathrm{out}(\kappa(x)) p_B^+{}']\\
&\; +\int_\gamma f_2^-[-D_t \Delta_\gamma p_B^-{}' - v_0 \partial_s p_B^-{}'+r_\mathrm{out}(\kappa(x)) p_B^-{}']\\
= &\;\int_\gamma f_2^+\cdot\left[\mbox{RHS of \eqref{eqn: equation for p_B'}, plus case}\right]+ f_2^-\cdot\left[\mbox{RHS of \eqref{eqn: equation for p_B'}, minus case}\right].
\end{split}
\label{eqn: simplification of integral of p_B'}
\end{equation}

In order to handle ingredients of $dF(\Omega;\tilde{V})$ in \eqref{eqn: dF the very first form}, we start from the following calculation based on \eqref{eqn: equation for p'},
\begin{equation}
\begin{split}
&\;\int_{\Omega} v_0 p'\sin 2\pi\theta\\
= &\;\int_{\Omega} p'\cdot[\mathrm{div}(D\nabla x_2) + v_0(\cos 2\pi\theta, \sin 2\pi\theta, 0)\cdot \nabla x_2]\\
&\;-\int_{\Omega} x_2\cdot[\mathrm{div}(D\nabla p') - v_0(\cos 2\pi\theta, \sin 2\pi\theta, 0)\cdot \nabla p']\\
= &\;\int_{\partial \Omega} p'(D\nabla x_2)\cdot n_{\partial\Omega} - x_2(D\nabla p')\cdot n_{\partial\Omega} + v_0\int_{\partial\Omega} x_2 p'(\cos 2\pi\theta, \sin 2\pi\theta, 0)\cdot n_{\partial\Omega}\\
=&\;\int_\Gamma D_t \left(p'\frac{\partial x_2}{\partial n_\Gamma} - x_2\frac{\partial p'}{\partial n_\Gamma}\right) - b\int_{\partial\Omega \cap \{x_2 = b/2\}} D_t \frac{\partial p'}{\partial x_2} +v_0b\int_{\partial\Omega \cap \{x_2 = b/2\}} p'\sin 2\pi\theta \\
&\;+ v_0\int_\Gamma x_2 p'(\cos 2\pi\theta, \sin 2\pi\theta, 0)\cdot n_\Gamma\\
= &\;b \cdot dF(\Omega;\tilde{V}) + \int_\Gamma D_t \left(p'\frac{\partial x_2}{\partial n_\Gamma} - x_2\frac{\partial p'}{\partial n_\Gamma}\right) + v_0 x_2 p'\cos\beta\\
\end{split}
\label{eqn: a representation of dF}
\end{equation}
Hence, we define $f_3$ on $\Omega$ to be the solution of
\begin{align}
&\; \mathrm{div}(D\nabla f_3) + v_0(\cos 2\pi\theta, \sin 2\pi\theta, 0)\cdot \nabla f_3 = v_0\sin2\pi\theta,\quad(x,\theta)\in\Omega,\label{eqn: equation for f_3}\\
&\; D_t\frac{\partial (f_3-x_2)}{\partial n_\Gamma} + r_\mathrm{in}(\kappa(x)) [\rho_++ \rho_-](f_3-x_2) =  0,\quad(x,\theta)\in\Gamma,\label{eqn: boundary condition for f_3}\\
&\; f_3\mbox{ satisfies periodic boundary condition on }\partial\Omega\backslash \Gamma.\label{eqn: periodic boundary condition for f_3}
\end{align}
Then
\begin{equation}
\begin{split}
b\cdot dF(\Omega;\tilde{V}) = &\;\int_{\Omega} v_0 p'\sin 2\pi\theta - \int_\Gamma D_t \left(p'\frac{\partial x_2}{\partial n_\Gamma} - x_2\frac{\partial p'}{\partial n_\Gamma}\right) + v_0 x_2p'\cos\beta\\
=&\;\int_{\Omega}  p'[\mathrm{div}(D\nabla f_3) + v_0(\cos 2\pi\theta, \sin 2\pi\theta, 0)\cdot \nabla f_3]\\
&\;- \int_\Gamma D_t \left(p'\frac{\partial x_2}{\partial n_\Gamma} - x_2\frac{\partial p'}{\partial n_\Gamma}\right) + v_0 x_2 p'\cos\beta\\
=&\;\int_{\Omega}  f_3[\mathrm{div}(D\nabla p') - v_0(\cos 2\pi\theta, \sin 2\pi\theta, 0)\cdot \nabla p']\\
&\;+\int_\Gamma D_t \left(p'\frac{\partial f_3}{\partial n_\Gamma} - f_3\frac{\partial p'}{\partial n_\Gamma}\right) + v_0 f_3p'\cos\beta\\
&\;- \int_\Gamma D_t \left(p'\frac{\partial x_2}{\partial n_\Gamma} - x_2\frac{\partial p'}{\partial n_\Gamma}\right) + v_0 x_2p'\cos\beta\\
=&\;0+\int_\Gamma D_t \left(p'\frac{\partial (f_3-x_2)}{\partial n_\Gamma} - (f_3-x_2)\frac{\partial p'}{\partial n_\Gamma}\right) + v_0 (f_3-x_2)p'\cos\beta\\
=&\;\int_\Gamma -(f_3-x_2) \left(r_\mathrm{in}(\kappa(x))[\rho_++\rho_-]p' +D_t \frac{\partial p'}{\partial n_\Gamma}- v_0p'\cos\beta\right)\\
=&\;\int_\Gamma -(f_3-x_2)\cdot\left[\mbox{RHS of \eqref{eqn: boundary condition of p'}}\right].
\end{split}
\label{eqn: simplification of dF}
\end{equation}

In summary, by \eqref{eqn: dE the very first form}, \eqref{eqn: dN the very first form}, \eqref{eqn: dF the very first form}, \eqref{eqn: simplification of integral of p'}, \eqref{eqn: simplification of integral of p_B'} and \eqref{eqn: simplification of dF},
\begin{equation}
\begin{split}
dE(\Omega;\tilde{V}) = &\;\int_\Gamma \left(\frac{1}{bN(\Omega)}(x_2-f_3)+\frac{F(\Omega)}{N(\Omega)^2}f_1\right)\cdot\left[\mbox{RHS of \eqref{eqn: boundary condition of p'}}\right]  \\
&\;- \frac{F(\Omega)}{N(\Omega)^2} \left[\int_\gamma f_2^+\cdot\left[\mbox{RHS of \eqref{eqn: equation for p_B'}, plus case}\right]+ f_2^-\cdot\left[\mbox{RHS of \eqref{eqn: equation for p_B'}, minus case}\right]\right.\\
&\; \left.+ \int_{\Gamma} p( \tilde{V}(0)\cdot n_\Gamma)-\int_\gamma \left( p_B^+{}+p_B^-{}\right)\kappa(V(0)\cdot n) \right].
\end{split}
\label{eqn: final formula for dE}
\end{equation}
We remark that once the linear dependence of $(p', p_B^+{}', p_B^-{}')$ on $V(0)\cdot n$ (or equivalently $\tilde{V}(0)\cdot n_\Gamma$) is established, by the formulae for $\alpha'(x)$ and $\kappa'(x)$ (\eqref{eqn: alpha'} and \eqref{eqn: kappa'} respectively), $dE(\Omega;\tilde{V})$ is a linear functional of $V(0)\cdot n$.
Then the distribution $G_E(x)$ in \eqref{eqn: Hadamard structure theorem} is well-defined.

\section{Linear dependence of $(p', p_B^+{}', p_B^-{}')$ on $V(0)\cdot n$}\label{section: linear dependence of p' p_B' on Vn}
In this section, we shall show that we can indeed define a linear map from $V(0)\cdot n$ (or equivalent $\tilde{V}(0)\cdot n_\Gamma$) to $(p', p_B^+{}', p_B^-{}')$. 

Once $(p,p_B^+, p_B^-)$ has been solved via the method in Section \ref{subsection: numerical method to solve the original coupled system}, they can be viewed as known functions in the coupled system \eqref{eqn: equation for p'}, \eqref{eqn: boundary condition of p'} and \eqref{eqn: equation for p_B'} for $p'$ and $p_B^\pm{}'$.
We simply rewrite the equations for $p'$ and $p_B^\pm{}'$ as follows
\begin{align}
&\; \mathrm{div}(D\nabla p') = v_0(\cos2\pi\theta, \sin 2\pi\theta,0)\cdot\nabla p',\quad(x,\theta)\in\Omega,\label{eqn: simplified form of the coupled system for shape derivatives p' equation}\\
&\; D_t\frac{\partial p'}{\partial n_\Gamma} - v_0p'\cos\beta +r_\mathrm{in} (\rho_++\rho_-)p'=r_\mathrm{out}\cdot 2\pi(\tau_+ p_B^+{}'+\tau_-p_B^-{}')+\mathcal{L}_1(\tilde{V}(0)\cdot n_\Gamma),\quad(x,\theta)\in \Gamma,\label{eqn: simplified form of the coupled system for shape derivatives p' boundary condition}\\
&\;-D_t \Delta_\gamma p_B^\pm{}' \pm v_0 \partial_s p_B^\pm{}'+r_\mathrm{out} p_B^\pm{}'= \mathcal{L}^\pm_2(V(0)\cdot n) +\int_0^1 r_\mathrm{in}\rho_\pm p'\,d\theta,\quad x\in \gamma,\label{eqn: simplified form of the coupled system for shape derivatives p_B' equation}\\
&\; p'\mbox{ satisfies periodic boundary condition on }\partial\Omega\backslash \Gamma.\label{eqn: periodic boundary condition for p'}
\end{align}
Here $\mathcal{L}_1$ and $\mathcal{L}_2^\pm$ are linear operators depending on all the known functions including $p$ and $p_B^\pm$.
To be more precise,
\begin{equation}
\begin{split}
\mathcal{L}_1(\tilde{V}(0)\cdot n_\Gamma) = &\;\frac{dr_\mathrm{out}}{d\kappa}(\kappa(x))\kappa'(x)[2\pi\tau_+(\beta)p_B^+(x)+2\pi\tau_-(\beta)p_B^-(x)]\\
    &\;+r_\mathrm{out}(\kappa(x))\cdot 2\pi(-\alpha'(x))\cdot\left[\frac{d\tau_+}{d\beta}(\beta) p_B^+(x) +\frac{d\tau_-}{d\beta}(\beta) p_B^-(x)\right]\\
    &\;-\frac{dr_\mathrm{in}}{d\kappa}(\kappa(x))\kappa'(x)(\rho_+(\beta)+\rho_-(\beta))p\\
    &\;-r_\mathrm{in}(\kappa(x))(-\alpha'(x))\cdot\left[\frac{d\rho_+}{d\beta}(\beta)+\frac{d\rho_-}{d\beta}(\beta)\right]p\\
    &\;-r_\mathrm{in} (\rho_++\rho_-)\frac{\partial p}{\partial n_\Gamma}(\tilde{V}(0)\cdot n_\Gamma) -\kappa(\tilde{V}(0)\cdot n_\Gamma)(g-r_\mathrm{in}p(\rho_++\rho_-))\\
    &\;+\mathrm{div}_\Gamma[(\tilde{V}(0)\cdot n_\Gamma) D\nabla_\Gamma p - v_0 (\tilde{V}(0)\cdot n_\Gamma) p \mathds{P}_\Gamma(\cos2\pi\theta, \sin 2\pi\theta, 0)],
\end{split}
\label{eqn: definition of L_1}
\end{equation}
and
\begin{equation}
\begin{split}
\mathcal{L}_2^\pm(V(0)\cdot n) = &\;\int_0^1  \frac{dr_\mathrm{in}}{d\kappa}(\kappa(x)) \kappa'(x)\rho_\pm p(x,\theta)+r_\mathrm{in}(\kappa(x))\frac{d\rho_\pm}{d\beta}(\beta)(-\alpha'(x)) p(x,\theta)\,d\theta\\
&\; +\int_0^1 r_\mathrm{in}(\kappa(x))\rho_\pm \frac{\partial p}{\partial n_\Gamma}(x,\theta)(V(0)\cdot n)- \kappa(V(0)\cdot n) r_\mathrm{in}\rho_\pm p(x,\theta)\,d\theta\\
&\;- \frac{dr_\mathrm{out}}{d\kappa}(\kappa(x))\kappa'(x) p_B^\pm +D_t\partial_s[\kappa(V(0)\cdot n)\partial_s p_B^\pm] +r_\mathrm{out}(\kappa(x))\kappa (V(0)\cdot n) p_B^\pm.
\end{split}
\label{eqn: definition of L_2 pm}
\end{equation}
Note that by \eqref{eqn: alpha'} and \eqref{eqn: kappa'}, $\alpha'(x)$ and $\kappa'(x)$ are linear in $V(0)\cdot n$.

\begin{remark}\label{rmk: sum of integrals of L1 and L2pm is zero}
It could be shown that
\begin{equation}
\int_\Gamma \mathcal{L}_1(\tilde{V}(0)\cdot n_\Gamma) + \int_\gamma \mathcal{L}_2^+(V(0)\cdot n)+\mathcal{L}_2^-(V(0)\cdot n)  = 0.
\label{eqn: sum of integrals of L1 and L2pm is zero}
\end{equation}
Indeed, by \eqref{eqn: boundary condition of p'} and \eqref{eqn: equation for p_B'},
\begin{equation*}
\begin{split}
\int_\Gamma \mathcal{L}_1(\tilde{V}(0)\cdot n_\Gamma)
=   &\;\int_\gamma\frac{dr_\mathrm{out}}{d\kappa}(\kappa(x))\kappa'(x)(p_B^++p_B^-)-\kappa(V(0)\cdot n)r_\mathrm{out}(\kappa(x))(p_B^++p_B^-)\\
    &\;+\int_\Gamma\left\{-\frac{dr_\mathrm{in}}{d\kappa}(\kappa(x))\kappa'(x)(\rho_+(\beta)+\rho_-(\beta))p\right.\\
    &\;\left.-r_\mathrm{in} (\rho_++\rho_-)\frac{\partial p}{\partial n_\Gamma}(\tilde{V}(0)\cdot n_\Gamma)+\kappa(\tilde{V}(0)\cdot n_\Gamma)r_\mathrm{in}p(\rho_++\rho_-)\right\},
\end{split}
\end{equation*}
and
\begin{equation*}
\begin{split}
\int_\gamma \mathcal{L}_2^\pm(V(0)\cdot n)
= &\;\int_\Gamma\left\{ \frac{dr_\mathrm{in}}{d\kappa}(\kappa(x)) \kappa'(x)\rho_\pm p(x,\theta)\right.\\
&\; +\left.r_\mathrm{in}(\kappa(x))\rho_\pm \frac{\partial p}{\partial n_\Gamma}(x,\theta)(V(0)\cdot n)- \kappa(V(0)\cdot n) r_\mathrm{in}\rho_\pm p(x,\theta)\right\}\\
&\;+\int_\gamma- \frac{dr_\mathrm{out}}{d\kappa}(\kappa(x))\kappa'(x) p_B^\pm +r_\mathrm{out}(\kappa(x))\kappa (V(0)\cdot n) p_B^\pm.
\end{split}
\end{equation*}
Then \eqref{eqn: sum of integrals of L1 and L2pm is zero} follows immediately.
\end{remark}

We may follow exactly the same scheme as in Section \ref{subsection: numerical method to solve the original coupled system} to solve the system \eqref{eqn: simplified form of the coupled system for shape derivatives p' equation}-\eqref{eqn: periodic boundary condition for p'}.
To be more precise, we consider the following equations with given $p_B^\pm{}'$ and $V(0)\cdot n$
\begin{align}
&\; \mathrm{div}(D\nabla \tilde{p}') = v_0(\cos2\pi\theta, \sin 2\pi\theta,0)\cdot\nabla \tilde{p}',\quad(x,\theta)\in\Omega,\label{eqn: sub-problem of the coupled system for shape derivatives p' equation}\\
&\; D_t\frac{\partial \tilde{p}'}{\partial n_\Gamma} - v_0\tilde{p}'\cos\beta +r_\mathrm{in} (\rho_++\rho_-)\tilde{p}' =r_\mathrm{out}\cdot 2\pi(\tau_+ p_B^+{}'+\tau_-p_B^-{}')+\mathcal{L}_1(\tilde{V}(0)\cdot n_\Gamma),\quad(x,\theta)\in\Gamma,\label{eqn: sub-problem of the coupled system for shape derivatives p' boundary condition}\\
&\;-D_t \Delta_\gamma \tilde{p}_B^\pm{}' \pm v_0 \partial_s \tilde{p}_B^\pm{}'+r_\mathrm{out} \tilde{p}_B^\pm{}'= \mathcal{L}_2^\pm(V(0)\cdot n) +\int_0^1 r_\mathrm{in}\rho_\pm \tilde{p}'\,d\theta, \quad x\in \gamma,\label{eqn: sub-problem of the coupled system for shape derivatives p_b' equations}\\
&\; \tilde{p}'\mbox{ satisfies periodic boundary condition on }\partial\Omega\backslash \Gamma.\label{eqn: sub-problem of the coupled system for shape derivatives p' periodic boundary condition}
\end{align}
By solving the above equations, we can establish the linear map $K':\,(p_B^+, p_B^-, \tilde{V}(0)\cdot n_\Gamma)\mapsto (\tilde{p}_B^+, \tilde{p}_B^-)$.
We already know that
\begin{equation*}
K'[p_B^+{}',p_B^-{}',0] = (\tilde{p}_B^+{}',\tilde{p}_B^-{}') = K[p_B^+{}',p_B^-{}'],
\end{equation*}
where $K$ is defined in Section \ref{subsection: numerical method to solve the original coupled system}.
Hence, by linear superposition principle, it suffices to study
\begin{equation}
S(V(0)\cdot n) \triangleq K'[0,0,V(0)\cdot n],
\label{eqn: definition of the operator S}
\end{equation}
which is a linear map.
Once $S$ is found out, solving \eqref{eqn: equation for p'}, \eqref{eqn: boundary condition of p'} and \eqref{eqn: equation for p_B'} is equivalent to solving
\begin{equation}
(K-Id)[p_B^+{}', p_B^-{}'] = -S (V(0)\cdot n),
\label{eqn: fixed point equation for p_B'pm}
\end{equation}
\begin{remark}
By Remark \ref{rmk: r_out r_out is in the orthogonal compliment of K-Id}, a necessary condition for \eqref{eqn: fixed point equation for p_B'pm} to be solvable is that
\begin{equation*}
\langle (r_\mathrm{out},r_\mathrm{out}),S (V(0)\cdot n)\rangle =0,
\end{equation*}
where $\langle\cdot,\cdot\rangle$ again denotes the inner product of $L^2(\gamma)\times L^2(\gamma)$.
This is automatically satisfied.
Indeed, by \eqref{eqn: sub-problem of the coupled system for shape derivatives p' equation}-\eqref{eqn: sub-problem of the coupled system for shape derivatives p' periodic boundary condition}, we proceed as in \eqref{eqn: proof of r_out r_out is orthogonal to the image of K-Id},
\begin{equation*}
\begin{split}
&\;\langle (r_\mathrm{out},r_\mathrm{out}),S (V(0)\cdot n)\rangle \\
=&\;\int_\gamma r_\mathrm{out}(\tilde{p}_B^++\tilde{p}_B^-)\\
=&\;\int_\gamma \mathcal{L}_2^+(V(0)\cdot n)+\mathcal{L}_2^-(V(0)\cdot n) +\int_\Gamma r_\mathrm{in}\tilde{p}'(\rho_++\rho_-)\\
=&\;\int_\gamma \mathcal{L}_2^+(V(0)\cdot n)+\mathcal{L}_2^-(V(0)\cdot n)+\int_\Gamma -D_t\frac{\partial \tilde{p}'}{\partial n_\Gamma} + v_0\tilde{p}'\cos\beta +\mathcal{L}_1(\tilde{V}(0)\cdot n_\Gamma)\\
=&\;\int_\gamma \mathcal{L}_2^+(V(0)\cdot n)+\mathcal{L}_2^-(V(0)\cdot n) +\int_\Gamma \mathcal{L}_1(\tilde{V}(0)\cdot n_\Gamma)\\
&\;-\int_\Omega \mathrm{div}(D\nabla \tilde{p}')-v_0(\cos2\pi\theta, \sin 2\pi\theta,0)\cdot\nabla \tilde{p}'\\
=&\;0
\end{split}
\end{equation*}
In the last line, we used Remark \ref{rmk: sum of integrals of L1 and L2pm is zero} and \eqref{eqn: sub-problem of the coupled system for shape derivatives p' equation}.
\end{remark}


Suppose \eqref{eqn: fixed point equation for p_B'pm} is solvable, i.e.\;$S(V(0)\cdot n)$ is in the range of $(K-Id)$.
It will have infinitely many solutions since $(K-Id)$ has a nontrivial kernel.
Recall that by \eqref{equation: the linear system leading to the solution}, any element in the kernel of $(K-Id)$ is a solution of the coupled system \eqref{equation: model in the steady state with drift equation in the bulk}-\eqref{equation: model in the steady state with drift definition of boundary fluxes} for $p$ and $p_B^\pm$.
However, we shall show that any two solutions of \eqref{eqn: fixed point equation for p_B'pm}, whose difference is in the kernel of $(K-Id)$, will give the same $dE(\Omega;\tilde{V})$ in \eqref{eqn: final formula for dE}.
For given $V(0)\cdot n$ and given the unique positive solution $(p,p_B^+,p_B^-)$ of \eqref{equation: model in the steady state with drift equation in the bulk}-\eqref{equation: model in the steady state with drift definition of boundary fluxes} satisfying \eqref{equation: normalization condition}, assume $(p_k',p_{B,k}^+{}',p_{B,k}^-{}')$, $k = 1,2$, to be two solutions of \eqref{eqn: fixed point equation for p_B'pm}.
By \eqref{equation: the linear system leading to the solution} and \eqref{eqn: fixed point equation for p_B'pm}, $(q,q_{B}^+{},q_{B}^-{})\triangleq (p_1'-p_2',p_{B,1}^+{}'-p_{B,2}^+{}',p_{B,1}^-{}'-p_{B,2}^-{}')$ is a solution of \eqref{equation: model in the steady state with drift equation in the bulk}-\eqref{equation: model in the steady state with drift definition of boundary fluxes}.
In particular, it is a multiple of $(p,p_B^+,p_B^-)$ by the assumption that $(K-Id)$ has only one-dimensional kernel (see Remark \ref{rmk: existence and uniqueness of positive solution of the coupled system}).
Let $dE_i(\Omega;\tilde{V})$ be the Eulerian derivative of $E(\Omega)$ represented in terms of the solutions $(p_k',p_{B,k}^+{}',p_{B,k}^-{}')$ respectively.
By \eqref{eqn: final formula for dE},
\begin{equation}
\begin{split}
&\;dE_1(\Omega;\tilde{V})-dE_2(\Omega;\tilde{V}) \\
= &\;\int_\Gamma \left(\frac{1}{bN(\Omega)}(x_2-f_3)+\frac{F(\Omega)}{N(\Omega)^2}f_1\right)\cdot r_\mathrm{out}(\kappa)\cdot 2\pi[\tau_+(\beta) q_B^+{}(x)+\tau_-(\beta)q_B^-{}(x)]  \\
&\;- \frac{F(\Omega)}{N(\Omega)^2}\int_\Gamma  q|_\Gamma r_\mathrm{in}(\kappa)[f_2^+(x)\rho_+(\beta) + f_2^-\rho_-(\beta)]\\
= &\;\int_\Gamma \left(\frac{1}{bN(\Omega)}(x_2-f_3)+\frac{F(\Omega)}{N(\Omega)^2}f_1\right)\left(r_\mathrm{in}(\kappa(x))[\rho_++\rho_-]q +D_t \frac{\partial q}{\partial n_\Gamma}- v_0q\cos\beta\right)\\
&\;- \frac{F(\Omega)}{N(\Omega)^2}\int_\gamma  f_2^+(x) (-D_t\Delta_\gamma q_B^++v_0 \partial_\gamma p_B^+ +r_\mathrm{out}p_B^+)\\
&\;- \frac{F(\Omega)}{N(\Omega)^2}\int_\gamma  f_2^-(x) (-D_t\Delta_\gamma q_B^--v_0 \partial_\gamma p_B^- +r_\mathrm{out}p_B^-).
\end{split}
\label{eqn: first formula for dE_1-dE_2}
\end{equation}
Here we used the assumption that $(q,q_{B}^+{},q_{B}^-{})$ satisfies \eqref{equation: model in the steady state with drift equation in the bulk}-\eqref{equation: model in the steady state with drift definition of boundary fluxes}.
Following the derivation of \eqref{eqn: a representation of dF} and \eqref{eqn: simplification of dF} in the reverse direction, with $p'$ replaced by $q$, we find that
\begin{equation}
\begin{split}
&\;\int_\Gamma (x_2-f_3)\cdot r_\mathrm{out}(\kappa)\cdot 2\pi[\tau_+(\beta) q_B^+{}(x)+\tau_-(\beta)q_B^-{}(x)]\\
= &\; - b\int_{\partial\Omega \cap \{x_2 = b/2\}} D_t \frac{\partial q}{\partial x_2} +v_0b\int_{\partial\Omega \cap \{x_2 = b/2\}} q\sin 2\pi\theta  = bF(\Omega;q),
\end{split}
\label{eqn: dE_1-dE_2 part 1}
\end{equation}
where $F(\Omega;q)$ is the net flux generated by the distribution $q$ in $\Omega$.
Similarly, proceeding as in \eqref{eqn: simplification of integral of p'} and \eqref{eqn: simplification of integral of p_B'} in the reverse direction, with $(p',p_B^+{}',p_B^-{}')$ replaced by $(q,q_B^+{},q_B^-{})$, we find that
\begin{equation}
\begin{split}
&\;\int_\Gamma f_1\left(r_\mathrm{in}(\kappa(x))[\rho_++\rho_-]q +D_t \frac{\partial q}{\partial n_\Gamma}- v_0q\cos\beta\right)\\
&\;- \int_\gamma  f_2^+(x) (-D_t\Delta_\gamma q_B^++v_0 \partial_\gamma p_B^+ +r_\mathrm{out}p_B^+)- \int_\gamma  f_2^-(x) (-D_t\Delta_\gamma q_B^--v_0 \partial_\gamma p_B^- +r_\mathrm{out}p_B^-)\\
=&\; -\int_\Omega q-\int_\gamma (q_B^++q_B^-) = N(\Omega;(q,q_B^\pm)).
\end{split}
\label{eqn: dE_1-dE_2 part 2}
\end{equation}
Hence, combining \eqref{eqn: first formula for dE_1-dE_2}, \eqref{eqn: dE_1-dE_2 part 1} and \eqref{eqn: dE_1-dE_2 part 2}, we find
\begin{equation*}
dE_1(\Omega;\tilde{V})-dE_2(\Omega;\tilde{V}) = \frac{F(\Omega;q)}{N(\Omega)} - \frac{F(\Omega)N(\Omega;(q,q_B^\pm))}{N(\Omega)^2} = \frac{N(\Omega;(q,q_B^\pm))}{N(\Omega)}[E(\Omega;(q,q_B^\pm))-E(\Omega)].
\end{equation*}
Since $(q,q_B^+,q_B^-)$ is a multiple of $(p,p_B^+,p_B^-)$, then $N(\Omega;(q,q_B^\pm)) = 0$ or $E(\Omega;(q,q_B^\pm))=E(\Omega)$, which implies that $dE_1(\Omega;\tilde{V})=dE_2(\Omega;\tilde{V})$.

Therefore, it suffices to consider the solution $(p_B^+,p_B^-)$ as an element in the quotient space $\mathrm{dom}(K-Id)/\ker(K-Id)$;
it is well-defined and is linear in $V(0)\cdot n$.
We are thus able to define a linear map from $V(0)\cdot n$ to $(p',p_B^+{}',p_B^-{}')$.
By \eqref{eqn: final formula for dE}, $dE(\Omega;\tilde{V})$ is indeed a linear functional of $V(0)\cdot n$.
Therefore, $G_E(x)$ in \eqref{eqn: Hadamard structure theorem} is well-defined.

\section{Numerical methods for the shape optimization}\label{section: numerical method for shape optimization}
As is discussed in Remark \ref{rmk: shape gradient as a measure on gamma instead of on Gamma}, $E(\Omega)$ increases fastest if $\gamma$ evolves in the direction of $V$ such that $V\cdot n = G_E(x)$ for $x\in \gamma$.
Note that only the value of $V(t,x)$ for $x\in\gamma$ and $t=0$ will be used in representing $dE(\Omega, \tilde{V})$.
We shall omit the $t$-dependence of $V$ and $\tilde{V}$ in the sequel.

In order to apply steepest ascent method, the explicit form of $G_E(x)$ is needed, which means a vectorial representation of $G_E$ in the discretized case.
Therefore, the numerical method below aims at first establishing a discretized representation of the linear operator $S$ defined in \eqref{eqn: definition of the operator S}, and then finding out the map from $V\cdot n$ to $(p',p_B^+{}',p_B^-{}')$ and $dE(\Omega;\tilde{V})$ in its discrete form.
The vectorial representation of $G_E$ then follows.
It goes as follows:
\begin{enumerate}
  \item Fix $\gamma$. We represent $\gamma$ using the same $N$ points $\{x_1,\cdots, x_{N}\}\triangleq X$ as in Section \ref{subsection: numerical method to solve the original coupled system}.
      We also use \eqref{eqn: central difference scheme to calculate alpha}-\eqref{eqn: central difference scheme to calculate kappa} to evaluate $\alpha(x)$, $n(x)$ and $\kappa(x)$ on the grid points in $X$.
      In the $\theta$-direction, $[0,1]$ is again discretized evenly using $M$ points $\{\theta_1,\cdots,\theta_M\}$ defined in Section \ref{subsection: numerical method to solve the original coupled system}.
      Let $X_\Gamma$ be defined as in \eqref{eqn: defintion of X_Gamma}.
  \item Through the recipe in Section \ref{subsection: numerical method to solve the original coupled system}, matrices $U$ and $W$ can be constructed and a normalized solution $(p,p_B^+,p_B^-)$ can be found out.
      In particular, values of $p$ have been determined on the boundary grid $X_\Gamma$ on $\Gamma$, while $p_B^\pm$ have been solved on $X$.
      Unnormalized net flux $F(\Omega)$ and the normalizing factor $N(\Omega) = 1$ could be evaluated.

      Calculate the pseudo-inverse of $(W-U)$, denoted by $(W-U)^\dag$, by singular value decomposition \cite{golub2012matrix}.
      Recall that the pseudo-inverse have the following properties: $(W-U)^\dag(W-U)\xi = \xi$ for $\forall\, \xi\bot\ker(W-U)\in\mathds{R}^{2N}$; and $\mathrm{range}(W-U)^\dag \bot \ker(W-U)$.

  \item Let $\{\mathbf{y}_1, \cdots,\mathbf{y}_{N}\}$ be an orthogonal basis of $\mathds{R}^{N}$, which are column vectors.
  They form a basis of all possible vectorial representations of $V\cdot n$ on the grid points in $X$.
  Denote $Y = (\mathbf{y}_1, \cdots,\mathbf{y}_{N})$.
  We take 
   $Y = U_0^T$ defined in \eqref{eqn: definition of U_0 sin modes} and \eqref{eqn: definition of U_0 cos modes}.

  \item \label{numerical scheme step: set the initial values of p_B' s}
  Let $V_i$ be a perturbation vector field defined on $\gamma$, whose values on the grid points in $X$ are given by
  \begin{equation*}
  (V_i(x_1),\cdots, V_i(x_{N}))^T = (y_{i,1}\cdot n(x_1),\cdots,y_{i,N}\cdot n(x_N))^T.
  \end{equation*}
  Here $y_{i,j}\in\mathds{R}$ is the $j$-th component of $\mathbf{y}_i$.
  Hence,
  \begin{equation*}
  ([V_i\cdot n](x_1),\cdots, [V_i\cdot n](x_{N}))^T = \mathbf{y}_i.
  \end{equation*}
  We use the central difference scheme to calculate $\partial_{s}[V_i\cdot n]$ and $\partial_{ss}[V_i\cdot n]$ on points in $X$,
  \begin{align}
  \partial_{s}[V_i\cdot n](x_j) = &\;\frac{1}{2\Delta s}(y_{i,j+1}-y_{i,j-1}),\nonumber\\
  \partial_{ss}[V_i\cdot n](x_j) = &\;\frac{1}{(\Delta s)^2}(y_{i,j+1}-2y_{i,j}+y_{i,j-1}).\label{eqn: central difference scheme to calculate second derivative of test Vn}
  \end{align}
  Then $\alpha'(x_j)$'s and $\kappa'(x_j)$'s are evaluated using \eqref{eqn: alpha' in the main body} and \eqref{eqn: kappa' in the main body} respectively.
  \item  \label{numerical scheme step: calculate L1 and L2}
  Given values of $p$ on $X_\Gamma$ and values of $p_B^\pm$ on $X$, we evaluate $\mathcal{L}_1(\tilde{V}_i\cdot n_\Gamma)$ on $X_\Gamma$ by \eqref{eqn: definition of L_1}.
  The tangential divergence $\mathrm{div}_\Gamma$ is calculated by central difference scheme in $\theta$- and $s$-coordinates, where $s$ is the arclength parameter of $\gamma$.
  Periodic boundary condition of $p$ in the $\theta$-direction on $\Gamma$ is used here.
  Similarly, we calculate $\mathcal{L}_2^\pm(V_i\cdot n)$ on $X$ by \eqref{eqn: definition of L_2 pm}.

  \item \label{numerical scheme step: solve for p' s}
  With $\mathcal{L}_1(\tilde{V}_i\cdot n_\Gamma)$ specified on $X_\Gamma$, we solve the following subproblem by COMSOL
  \begin{align*}
&\; \mathrm{div}(D\nabla q'_i) = v_0(\cos2\pi\theta, \sin 2\pi\theta,0)\cdot\nabla q'_i,\quad(x,\theta)\in\Omega,\\
&\; D_t\frac{\partial q'_i}{\partial n_\Gamma} - v_0q'_i\cos\beta +r_\mathrm{in} (\rho_++\rho_-)q'_i=\mathcal{L}_1(\tilde{V}_i\cdot n_\Gamma),\quad(x,\theta)\in \Gamma,\\
&\; q'_i\mbox{ satisfies periodic boundary condition on }\partial\Omega\backslash \Gamma.
\end{align*}
  We obtain the Dirichlet boundary data of $q'_i$ on the set $X_\Gamma$.

  \item \label{numerical scheme step: solve for new p'B s}
  Now consider the equation for $q_{B,i}^\pm{}'$
  \begin{equation}
  -D_t \Delta_\gamma q_{B,i}^\pm{}' \pm v_0 \partial_s q_{B,i}^\pm{}'+r_\mathrm{out} q_{B,i}^\pm{}'= \mathcal{L}_2^\pm(V_i\cdot n) +\int_0^1 r_\mathrm{in}\rho_\pm q_i'\,d\theta, \quad x\in \gamma.
  \label{eqn: equation for p_Bi' in the numerical method}
  \end{equation}
  The integral on the right hand side is again evaluated on $X$ using the trapezoidal rule as in \eqref{eqn: trapezoidal rule}, with $\tilde{p}_i$ replaced by $q_i'$.
  Then we solve \eqref{eqn: equation for p_Bi' in the numerical method} for $q_{B,i}^\pm{}'$ by the finite difference scheme with grid points $X$.
  We denote
  \begin{equation*}
  (q_{B,i}^+{}'(x_1),\cdots, q_{B,i}^+{}'(x_{N}), q_{B,i}^-{}'(x_1),\cdots, q_{B,i}^-{}'(x_{N}))^T \triangleq  \mathbf{z}_i.
  \end{equation*}
  In this way, we obtain $\mathbf{z}_i = S\mathbf{y}_i$, where $S$, with abuse of notations, is a discrete representation of the operator $S$ defined in \eqref{eqn: definition of the operator S}.
  $S$ is not explicitly represented here.
  \item Repeating Step \ref{numerical scheme step: set the initial values of p_B' s} - Step \ref{numerical scheme step: solve for new p'B s} for $i = 1,\cdots, N$, we form an $2N\times N$ matrix $Z=(\mathbf{z}_1,\cdots,\mathbf{z}_{N})= SY$.
  \item
  Given $V_i\cdot n$, values of the corresponding $(p_{B,i}^+{}',p_{B,i}^-{}')$ on $X$ are given by
  \begin{equation}
  (p_{B,i}^+{}'(x_1),\cdots, p_{B,i}^+{}'(x_{N}), p_{B,i}^-{}'(x_1),\cdots, p_{B,i}^-{}'(x_{N}))^T = -U(W-U)^\dag \mathbf{z}_i.
  \label{eqn: formula for values of p_Bi'}
  \end{equation}
  Indeed, since $U$ is invertible, we assume $(p_{B,i}^+{}'(x_1),\cdots, p_{B,i}^+{}'(x_{N}), p_{B,i}^-{}'(x_1),\cdots, p_{B,i}^-{}'(x_{N}))^T = U\eta_i$ for some $\eta_i\in\mathds{R}^{2N}$.
  By \eqref{eqn: fixed point equation for p_B'pm},  $(K-I_{2N})U\eta_i = -S \mathbf{y}_i$, i.e. $(W-U)\eta_i = -S\mathbf{y}_i$.
  By the discussion in Section \ref{section: linear dependence of p' p_B' on Vn}, it suffices to solve for $U\eta_i$ such that $U\eta_i\bot \ker(K-I_{2N})$, i.e.\;$\eta_i\bot \ker((K-I_{2N})U) = \ker(W-U)$.
  Hence, $\eta_i = -(W-U)^\dag S\mathbf{y}_i$, which justifies \eqref{eqn: formula for values of p_Bi'}. 

  \item
  To this end, we solve for $p$ through \eqref{eqn: simplified form of the coupled system for shape derivatives p' equation}, \eqref{eqn: simplified form of the coupled system for shape derivatives p' boundary condition} and \eqref{eqn: periodic boundary condition for p'} by COMSOL.
  The values of $V_i(0)\cdot n$ and $(p_{B,i}^+{}',p_{B,i}^-{}')$ on $X$ are used in \eqref{eqn: simplified form of the coupled system for shape derivatives p' boundary condition}.
  In this way, the Dirichlet data of $p'$ on $X_\Gamma$ is obtained.
  \item
  Use COMSOL to solve for $f_1$ and $f_3$  by \eqref{eqn: equation for f_1}-\eqref{eqn: periodic boundary condition for f_1} and \eqref{eqn: equation for f_3}-\eqref{eqn: periodic boundary condition for f_3}, respectively. Solve for $f_2^\pm$ by applying finite difference scheme to \eqref{eqn: equations for f_2 pm} with grid points in $X$.
  \item
  Finally, we evaluate $dE(\Omega;\tilde{V}_i)$ using \eqref{eqn: final formula for dE} for $i = 1,\cdots, N$.
  The integrals in \eqref{eqn: final formula for dE} are evaluated by the trapezoidal rule.
  Then under discretization, the vectorial representation of the shape gradient $G_E(x)$ in \eqref{eqn: Hadamard structure theorem}, denoted by $\mathbf{G}\in\mathds{R}^N$, is given by
  \begin{equation}
  \mathbf{G} = \sum_{i=1}^N\frac{dE(\Omega;\tilde{V}_i)}{(\mathbf{y}_i\cdot \mathbf{y}_i)\Delta s}\mathbf{y}_i.
  \label{eqn: representation of the shape gradient}
  \end{equation}
  We derive this from $dE(\Omega;\tilde{V}_i) = (\mathbf{G}\cdot \mathbf{y}_i)\Delta s$ and the orthogonality of $\{\mathbf{y}_i\}$.
  \item
  We apply steepest ascent method with fixed step-size to evolve $\gamma$.
  Choose $\varepsilon$ to be suitably small, and calculate
  \begin{equation}
  \tilde{x}_j = x_j+\varepsilon n(x_j)\cdot G_j,
  \label{eqn: evolution of the boundary}
  \end{equation}
  where ${G}_j$ is the $j$-th component of $\mathbf{G}$.
  Note that by Hadamard formula \eqref{eqn: Hadamard structure theorem}, only the normal component of the perturbation vector field matters.
  The new configuration of the post boundary, denoted by $\tilde{\gamma}$, is then given by $\{\tilde{x}_j\}$ and a spline interpolation.
  In the Remark \ref{rmk: preconditioning in the evolution of gamma} below, we shall propose a better way of evolving $\gamma$.
  \item
  Repeat all the above steps to find posts that induce larger and larger normalized net flux.
  Note that $\{\tilde{x}_j\}$ are not equally-spaced along $\tilde{\gamma}$; a reparameterization is needed before starting a new iteration.
\end{enumerate}

\begin{remark}\label{rmk: preconditioning in the evolution of gamma}
Since $\mathbf{y}_i$ is a discrete Fourier mode, we use $k_i$ to denote its wave number.
To be more precise,
\begin{align*}
k_i = &\;i,\quad i\in\{1,\cdots, N/2-1\},\\
k_i = &\;i-N/2,\quad i\in\{N/2,\cdots, N\}.
\end{align*}
By definition, $\mathbf{y}_i\sim O(1)$.
Then \eqref{eqn: central difference scheme to calculate second derivative of test Vn} implies $\partial_{ss} [V_i\cdot n] \sim O(k_i^2)$.
By \eqref{eqn: kappa'}, it follows that $\kappa'(x)$ corresponding to the perturbation vector field $V_i$ is of order $1+k_i^2$.
Hence, $\mathcal{L}_1(\tilde{V}_i\cdot n)$ and $\mathcal{L}_2(V_i\cdot n)$ are going to be huge and highly oscillatory when $k_i$ is large, leading to very strong stiffness when solving for $(q_i',q_{B,i}^+{}',q_{B,i}^-{}')$.
On the other hand, it also implies that $dE(\Omega;\tilde{V}_i)$ is of order $1+k_i^2$, i.e.\;$E(\Omega)$ is sensitive to perturbations on $\gamma$ with higher frequencies.
Combining this with \eqref{eqn: representation of the shape gradient}, we see that the evolution of $\gamma$ in \eqref{eqn: evolution of the boundary} will be mostly governed by the high-frequency modes, and thus the step-size has to be extremely small.
Therefore, to remove the stiffness in the steepest ascent method \eqref{eqn: evolution of the boundary}, we make a modification of $\mathbf{G}$, denoted by $\mathbf{G}'$.

Let $N'\leq N/2$ be a positive integer.
Define
\begin{equation*}
\mathbf{G}' = \sum_{\{i:\;k_i \leq N'\}}\max\{1,k_i\}^{-2} \frac{dE(\Omega;\tilde{V}_i)}{(\mathbf{y}_i\cdot \mathbf{y}_i)\Delta s}\mathbf{y}_i,
\end{equation*}
which also gives an ascent direction since $\mathbf{G}\cdot \mathbf{G}'>0$. Then we replace \eqref{eqn: evolution of the boundary} by
\begin{equation*}
\tilde{x}_j = x_j+\varepsilon n(x_j)\cdot \mathbf{G}'_j.
\label{eqn: modified evolution of the boundary}
\end{equation*}
In other words, high-frequency modes in $\mathbf{G}$ get suppressed or even filtered out.
The coefficients $\max\{1,k_i\}^{-2}$ should be understood as preconditioning in the steepest ascent method.
This enforces the smoothness of $V$, and also reduces computational costs if $N'\ll N/2$.
\end{remark}

\end{document}